\documentclass[nofootinbib]{revtex4-2}
\usepackage{amsfonts}
\usepackage{amsmath}
\usepackage{amssymb}
\usepackage{graphicx}

\renewcommand{\theequation}{\thesection.\arabic{equation}}
\input{tcilatex}
\begin{document}

\title{\textbf{Neutrino phenomenology in\ the flavored NMSSM without domain
wall problems}}
\author{M. A. Ouahid}
\thanks{mohamedamine\_ouahid@um5.ac.ma}
\author{M. A. Loualidi}
\thanks{mr.medamin@gmail.com}
\author{R. Ahl Laamara}
\thanks{ahllaamara@gmail.com}
\author{E. H. Saidi}
\thanks{h-saidi@fsr.ac.ma}
\affiliation{LPHE-Modeling and Simulations, Faculty of \ Science, Mohammed V University
in Rabat, 10090 Rabat, Morocco}
\pacs{}
\affiliation{Centre of Physics and Mathematics, CPM, Faculty of Science, Rabat 10090,
Morocco}

\begin{abstract}
We propose a next-to-minimal supersymmetric Standard Model (NMSSM) extended
by an $\mathbb{A}_{4}\times \boldsymbol{Z}_{3}$ flavor symmetry and three
right-handed neutrinos providing a detailed study of the neutrino sector and
a solution to the domain wall problem. In this proposal, neutrino masses are
generated through Type I seesaw mechanism while the mixing angles are
described by the\textrm{\ }trimaximal mixing realized using the NMSSM singlet%
\textrm{\ }$S$ and only two flavon fields. The phenomenology of neutrino
parameters is studied for normal and inverted mass hierarchies. In
particular, we numerically evaluated the observables related to neutrino
masses and mixing, namely, $\sum m_{i}$, $m_{ee}$, $m_{\nu _{e}}$, and $%
\delta _{CP}$ where we find that the ranges of $m_{ee}$\ and $m_{\nu _{e}}$
are accessible by current and future experiments while the obtained ranges
of $\sum m_{i}$ and $\delta _{CP}$ lie within the current experimental data.
Another attractive feature we discussed in this paper is the circumvention
of the\textrm{\ }domain wall problem induced by the spontaneous breaking of
the\textrm{\ }$\mathbb{A}_{4}\times \boldsymbol{Z}_{3}$\textrm{\ }discrete
symmetry. We first showed that the domain walls in the charged lepton sector
occur at high energy scale leading to unproblematic domain walls, while in
the neutrino sector they are inevitable. Then, to solve this problem, we
reconsidered the well-known approach that relies on the explicit breaking of
the discrete symmetry through the insertion of Planck-suppressed operators
induced by supergravity.\newline
\emph{Keywords}: Neutrino physics, Discrete flavor symmetry, Trimaximal
mixing, Domain walls
\end{abstract}

\maketitle

\typeout{Filename: reftest4-2.tex for revtex 4.X 2014/12/31 (AO)}

\makeatletter

%\tableofcontents

\section{Introduction}

Physics beyond the standard model (SM) of elementary particles has been
widely explored after the experimental discovery of neutrino oscillations
\cite{R1,A1,B1}, and of the Higgs boson with a mass of about $125$ $\mathrm{%
GeV}$ \cite{R2,A2}. One of the most studied extensions of the SM is the
minimal supersymmetric Standard Model \cite{R3,A3,A4} (MSSM) which predicts
the lightest Higgs boson---among the five it contains---to be lighter than
the $Z$-boson. Thus, large quantum corrections are required in order to
reach the mass of the discovered Higgs boson. Although the MSSM solves many
problems encountered in the SM such as the hierarchy problem and the gauge
coupling unification, it undergoes a naturalness concern related to the mass
term involving the up and down Higgs doublets $\mu H_{u}H_{d}$. In order to
generate a convenient electroweak symmetry breaking, the $\mu $-parameter
which has a dimension of mass must be of the order of the SUSY breaking
scale $M_{SUSY}$. However, since the MSSM might be an effective theory that
originates from a more fundamental high energy theory with some cutoff scale
such as the GUT scale $M_{GUT}$ or the Planck scale $M_{Pl}$, the question
that arises is why is the scale of $\mu $ far below these scales? This is
the so-called the $\mu $-problem \cite{R4}.\ \newline
The most simple way to deal with the $\mu $-problem as well as some of the
other shortcomings of the MSSM is by introducing an extra hyperchargeless
singlet chiral superfield $S$ which gives rise to the well known
next-to-minimal supersymmetric Standard Model \cite{R5,A5,A6,A7}. In the
latter, the $\mu $-term is replaced by the Yukawa coupling $\lambda
SH_{u}H_{d}$, where after the scalar component of the chiral superfield $S$
acquires a vacuum expectation value (VEV) of the order of $M_{SUSY}$, an
effective $\mu $-term---$\mu _{eff}=\lambda \left\langle S\right\rangle $%
---with the appropriate order is generated, thus solving the $\mu $-problem
of the MSSM.\textrm{\ }Moreover, since the superpotential of the NMSSM is
scale invariant, the old $\mu $-term is excluded by an accidental $\mathcal{Z%
}_{3}$ discrete symmetry under which all the chiral superfields $\hat{\Phi}$
of the NMSSM transform as $\hat{\Phi}\rightarrow e^{2\pi i/3}\hat{\Phi}$
\cite{R6}. However, the rise of this discrete symmetry group $\mathcal{Z}%
_{3} $ is not without weakness from the cosmological perspective; indeed,
the spontaneous breaking of the $\mathcal{Z}_{3}$ symmetry leads to the
creation of cosmological domain walls (DWs)\textrm{\ }in the early Universe\
\cite{R7}; that is different\textrm{\ }$\mathcal{Z}_{3}$\textrm{\ }ground
states created during the electroweak phase transition \cite{R8,A8}.\
Nevertheless, there are different ways suggested in the literature to avoid
the DW problem among which we mention: (\textbf{a}) the breaking of $%
\mathcal{Z}_{3}$ symmetry taking place before the end of inflation \cite%
{R9,A9}; and (\textbf{b}) taking into account the supergravitational effect
by adding nonrenormalizable terms to the superpotential to break explicitly
the $\mathcal{Z}_{3}$ leading to favor one of the three $\mathcal{Z}_{3}$
vacua over the others \cite{R6,R10}. Notice by the way that the annihilation
of DWs has been also suggested as a plausible source of the gravitational
waves (GWs) as discussed in different extensions of the SM, see for instance%
\textrm{\ }\cite{B2,B3,B4}.\textrm{\ }This annihilation would constitute an
interesting aspect of DWs especially after the recent observation of these
GWs at the LIGO and the Virgo experiments\textrm{\ }\cite{B5}, which could
offer a way for investigating the existence of DWs in the early Universe.\ \
\newline
Aside from these features related to the scalar sector, the SM as well as
its supersymmetric extensions predict massless neutrinos, thus further
extensions are required to explain the results of the neutrino oscillation
experiments and hence, the observed mixing angles and the tiny neutrino
masses. Recent data from these experiments give hints on the mass hierarchy,
the $CP$ violating ($CPV$) phase $\delta _{CP}$\ and nonmaximal atmospheric
mixing $\theta _{23}$ where two different octants of $\theta _{23}$ are
allowed; the lower octant (LO) with $\theta _{23}<\pi /4$ and the higher
octant (HO) with $\theta _{23}>\pi /4$. The values of neutrino oscillation
parameters $\theta _{ij}$, $\Delta m_{ij}^{2}$\ and $\delta _{CP}$\ can be
found in the latest global fit analyses \cite{R11,R12,R13,C0,C1}. From the
theoretical point of view, one of the most well-known ways to generate such
tiny neutrino masses is through type-I seesaw mechanism which implies the
introduction of heavy right-handed Majorana neutrinos to the SM or any of
its supersymmetric extensions \cite{R14,A10,A11,A12,A13,A14}. However,\ the
seesaw mechanism does not allow us to explain the neutrino mixing nor to
determine the neutrino mass hierarchy. In this regard, non-Abelian discrete
flavor symmetries have been used extensively in recent years to describe the
family structure of both leptons and quarks, in particular, these symmetries
are known to produce mixing patterns compatible with large mixing angles of
the lepton sector such as the tribimaximal mixing (TBM) matrix which
corresponds to a unitary matrix of the form \cite{R15}%
\begin{equation}
U_{TBM}=\left(
\begin{array}{ccc}
\sqrt{\frac{2}{3}} & \frac{1}{\sqrt{3}} & 0 \\
-\frac{1}{\sqrt{6}} & \frac{1}{\sqrt{3}} & \frac{1}{\sqrt{2}} \\
-\frac{1}{\sqrt{6}} & \frac{1}{\sqrt{3}} & -\frac{1}{\sqrt{2}}%
\end{array}%
\right)  \label{e1}
\end{equation}%
However, while the TBM matrix is consistent with the large solar $\theta
_{12}$\ and atmospheric $\theta _{23}$ angles, it is in conflict with the
nonzero reactor $\theta _{13}$\ angle discovered in 2012 \cite{R16,A15,A16}.
Nevertheless, the TBM matrix is still considered as a good first order
approximation while small deviations are required to accommodate the small
value of $\theta _{13}$. These deviations are usually realized by
introducing higher dimensional effective operators while the leading TBM
contribution is produced by the seesaw mechanism. In the same manner,
several non-Abelian discrete groups have been used in the literature to
generate the deviation from TBM and provided a successful description of all
the neutrino mixing angles; see for instance Table 3 of Ref. \cite{R17} and
Refs. \cite{R18,R19,R20,R21,R22,R23,R24,R25,R26,R27,R28,R29,R30,R31}.
Moreover, there are additional unanswered questions relevant to neutrino
physics such as whether neutrinos are Dirac or Majorana particles and the
issue concerning the absolute neutrino mass scale. In this regard, many
nonoscillation experiments such as beta decay \cite{R32,A17} and
neutrinoless double beta decay \cite{R33,A18,A19,A20}\textrm{\ }experiments
have been proposed to probe the mass scale of neutrinos. The latter would
also prove lepton number violation and establish the Majorana nature of
neutrinos; for a recent review, see \cite{R34}. In addition, cosmological
observations can probe the sum of neutrino masses $\sum
m_{i}=m_{1}+m_{2}+m_{3}$ where the current upper bound given by the Planck
collaboration is $\sum m_{i}<0.17$ $\mathrm{eV}$ \cite{R35}.\ \ \newline
In this paper, we aim to achieve two main objectives: First, we study the
neutrino masses and mixing in the context of a flavored\textrm{\ }NMSSM%
\textrm{\ }prototype (\emph{FNMSSM}) with discrete symmetry group\textrm{\ }$%
G_{\mathrm{f}}=\mathbb{A}_{4}\times \boldsymbol{Z}_{3}$.\textrm{\ }We
compute among others the $\Delta m_{ij}^{2}$ and $\sin \theta _{ij}^{2}$
oscillation parameters for both normal (NH) and inverted (IH) mass
hierarchies with numerical estimations that agree with advanced theoretical
modeling in this matter; and which fit well with recent experimental
measurements on neutrinos. Second, we use results on the perturbation of the
scalar potential of the theory by higher dimensional operators suppressed by
powers of the Planck scale $M_{Pl}$; and on the effective field action
approach in the flavon sector of\textrm{\ \emph{FNMSSM} }to propose a
scenario preventing the domain wall problem created by the spontaneous
breaking of the full flavor group $G_{\mathrm{f}}$ in \emph{FNMSSM}\textrm{.}%
\newline
The building blocks of \emph{FNMSSM}\textrm{\ }are given by the usual\textrm{%
\ }NMSSM\textrm{\ }ones supplemented by two other kinds of chiral
superfields:\textrm{\ }$\left( \mathbf{\alpha }\right) $\textrm{\ }three
right-handed neutrinos $N_{i}^{c}$ in order to have massive neutrinos, and $%
\left( \mathbf{\beta }\right) $ three sets of flavon superfields $\Phi ,$ $%
\Omega ,$ $\mathbf{\chi }$\textrm{, }with quantum numbers as in Tables \ref%
{A}-\ref{C}, allowing us to generate suitable neutrino masses and mixing as
well as avoiding DWs. To build the chiral superpotential of the\textrm{\
\emph{FNMSSM} }prototype, we extend the \emph{SM} gauge symmetry group by
the global discrete flavor symmetry $\mathbb{A}_{4}\times \boldsymbol{Z}_{3}$%
. Notice that a supersymmetric model based on\textrm{\ }$\mathbb{A}%
_{4}\times \boldsymbol{Z}_{3}$ group was first proposed by Altarelli and
Feruglio in \cite{R43} and predicted a mixing of TBM type. Here, by using
the same flavor group we work in the context of the NMSSM which is motivated
by the several implications carried by its singlet superfield $S$\textrm{\ }%
where besides its phenomenological advantages mentioned above it also
contributes to the Higgs mass and to the neutrino\textrm{\ }masses.\textrm{\
}The non-Abelian alternating $\mathbb{A}_{4}$\ group is introduced to
explain the recent experimental data on neutrino oscillation experiments,%
\textrm{\ }while the $\boldsymbol{Z}_{3}$\ symmetry is imposed to avoid the
interchange between the flavon superfields $\Phi $ and $\Omega $
respectively appearing in the charged and chargeless lepton sectors. In the
neutrino sector, the tiny neutrinos masses are generated through type I
seesaw mechanism\textrm{\ }where besides the usual NMSSM singlet $S$, two
flavons are required to achieve a neutrino mass matrix compatible with the
TBM matrix. Notice that\ the contribution of the singlet $S$\ is naturally
small\textrm{\ }\cite{K0}---it must be at the SUSY scale (few
TeVs)---compared to the flavon fields added in the lepton sector; thus,
providing another motivation of working in the framework of the NMSSM.
Accordingly, $S$\textrm{\ }may be viewed as a small perturbation of the
Majorana neutrino mass matrix---breaking the\textrm{\ }$\mu -\tau $\textrm{\
}symmetry \cite{K1,K2,R36}---leading to a deviation of TBM; for a recent
review on this manner of doing, see \cite{R37} and references therein. The
mixing matrix in this scenario is obtained by rotating the TBM in the 1-3
plane, and thus leading to the so-called trimaximal mixing (TM$_{2}$) \cite%
{R21,R23,R38,R39,A21,R40,C2}.\emph{\ }Next, we perform a numerical study
concerning the neutrino mixing angles for both NH and IH where we find that
they all fit the current experimental data. We have also studied\textrm{\ }%
the phenomenological consequences of the model where we obtained interesting
predictions regarding the effective Majorana neutrino mass $m_{ee}$ measured
in neutrinoless double beta decay experiments, the effective electron
neutrino mass $m_{\nu _{e}}$ measured in tritium beta decay experiments and
the Dirac CPV phase. In particular, the obtained ranges of $m_{ee}$\ and $%
m_{\nu _{e}}$\ may be tested in current and near future experiments, and
which will first start to explore the regions corresponding to the inverted
mass hierarchy.\newline
As mentioned above, the\textrm{\ }NMSSM suffers from the DW problem coming
from the spontaneous breaking of the discrete $\mathcal{Z}_{3}$\ symmetry at
the electroweak phase transition\textrm{\ }\cite{R6,R10,R41}.\textrm{\ }In
our \textrm{\emph{FNMSSM} }prototype, the DW problem arises from the
spontaneous breaking of the $\mathbb{A}_{4}\times \boldsymbol{Z}_{3}$\
flavor symmetry bearing in mind that the charge assignments under the extra $%
\boldsymbol{Z}_{3}$ discrete group is different from those in the usual
NMSSM. In particular, two\textrm{\ }$\mathbb{A}_{4}$\textrm{\ }breaking
patterns are analysed: \textbf{(i)} the first one driven by an $\mathbb{A}%
_{4}$ triplet leads to the spontaneous breaking of $G_{\mathrm{f}}$ down to
its subgroup $\mathbb{Z}_{3}\times \boldsymbol{Z}_{3}$\ in the charged
lepton sector,\textrm{\ }where the domain walls appear as the boundaries
separating the degenerate vacua generated by this breaking. We find that
these DWs are remarkably described by the Klein-four\textrm{\ }$\mathbb{V}%
_{4}\cong \mathbb{Z}_{2}\times \mathbb{Z}_{2}$\textrm{\ }flavor subsymmetry
group, and are formed around the inflationary scale; and thus, present no
danger from the cosmological view. \textbf{(ii)} the second breaking is
given by $G_{\mathrm{f}}$\textrm{\ }to a subgroup $\mathbb{Z}_{2}$ in the
neutrino sector, it is generated by the VEV of $\Omega $\textrm{\ }and $S$
and creates\textrm{\ }domain walls that expand between the boundaries of
degenerate vacua which are characterized by the $\mathbb{Z}_{2}\rtimes
\mathbb{Z}_{3}\times \boldsymbol{Z}_{3}$ subgroup of $G_{\mathrm{f}}$. The
standard cosmology requires that these domain\ walls disappear at least
before nucleosynthesis \cite{R42}, and one of the first proposed solution
suggests that, after breaking the degeneracy of the vacua, the true vacuum
dominates \cite{R8}. Here, we show that breaking explicitly the\textrm{\ }$%
\mathbb{Z}_{2}\rtimes \mathbb{Z}_{3}\times \boldsymbol{Z}_{3}$\textrm{\ }%
subsymmetry of $G_{\mathrm{f}}$ at high energy by means of additional
nonrenormalizable terms---Planck-suppressed operators---is capable to remove
the degeneracy between the different domains.\emph{\ }In addition, the
theory at low energy will not undergo a remarkable change as long as these
operators are suppressed by powers of the Planck scale.\textrm{\newline
}This paper is organized as follows. In Sec. II, we present our \textrm{%
\emph{FNMSSM}} proposal giving an extension of NMSSM with right-handed
neutrinos and a global discrete $\mathbb{A}_{4}\times \boldsymbol{Z}_{3}$\
flavor symmetry. Within this section\ we give the superfield properties
describing the flavored NMSSM, then we describe the $\mathbb{A}_{4}\times
\boldsymbol{Z}_{3}$ invariant Yukawa couplings in the charged lepton sector.
In Sec. III, we study the neutrino masses and mixing as well as the
deviation from TBM matrix, and we use the\textrm{\ }$3\sigma $\textrm{\ }%
values of\ the neutrino oscillation parameters to extract the allowed ranges
of our model parameters.\textrm{\ }In Sec. IV, we perform a phenomenological
study of our model and provide the predictions concerning the parameters $%
m_{ee}$, $m_{\nu _{e}}$\ and the Dirac CPV phase $\delta _{CP}$\ along with
the Jarlskog parameter and its impact on the $\theta _{23}$\ octant
degeneracy. In Sec. V, we discuss the domain wall problem in our \emph{FNMSSM%
} created via the spontaneous breaking of $G_{\mathrm{f}}$\textrm{\ }and we
provide a viable solution to this cosmological problem in the neutrino
sector by adding a Planck-suppressed operator that breaks explicitly $G_{%
\mathrm{f}}$.\ In Sec. IV, we give our conclusion and then we give four
appendices A, B, C and D.\textrm{\ }In Appendix A, we provide some useful
details on the discrete $\mathbb{A}_{4}$ group and its representations.%
\textrm{\ }In Appendix B, we determine the vacuum alignments of the $\mathbb{%
A}_{4}$ flavon triplets $\Phi $\ and $\Omega $.\ In Appendix C, we give some
comments on additional nonrenormalizable operators that break explicitly the
$\mathbb{A}_{4}\times \boldsymbol{Z}_{3}$ group and we illustrate the
contribution of one of these operators to the effective potential of the
theory. In Appendix D, we describe the case where the breaking pattern in
the neutrino sector is driven by the VEV\ of the NMSSM singlet $S$.

\section{Building blocks in flavon extended NMSSM}

In this section, we introduce the flavon extended NMSSM that we propose to
produce neutrino masses, their mixing and bypassing the domain wall problem
created by spontaneous breaking of the flavor symmetry. To get straight to
the point, we will mainly focus on leptons by first giving the chiral
superfield content of the model; then describing the $\mathbb{A}_{4}\times
\boldsymbol{Z}_{3}$ invariant Yukawa coupling in the charged lepton sector.

\subsection{Superfield content of the model}

In addition to the usual chiral and gauge superfields of the MSSM, our model
contains extra chiral superfields carrying quantum numbers under the $%
\mathbb{A}_{4}\times \boldsymbol{Z}_{3}$ flavor symmetry group:

\begin{description}
\item[ $\left( i\right) $] the chiral singlet superfield $S$ of NMSSM that
couples to the usual Higgs doublets $H_{u}$ and $H_{d}$. This complex
superfield carries a negative unit charge under $\boldsymbol{Z}_{3}$; and
carry nontrivial quantum charge under $\mathbb{A}_{4}$.

\item[ $\left( ii\right) $] three right-handed neutrinos $N_{i=1,2,3}^{c}$\
needed to generate tiny masses for neutrinos through type I seesaw
mechanism. They carry a negative charge under $\boldsymbol{Z}_{3}$ exactly
like $S$; but form altogether an irreducible triplet under $\mathbb{A}_{4}$.

\item[ $\left( iii\right) $] three flavon superfields $\{ \Phi ,\Omega ,%
\mathbf{\chi }\}$ which are gauge singlets but carry non-trivial quantum
numbers under $\mathbb{A}_{4}\times \boldsymbol{Z}_{3}$ symmetry.
\end{description}

The $\mathbb{A}_{4}\times \boldsymbol{Z}_{3}$ representations hosting these
chiral superfields and the role they play in the construction are described
below.

\subsubsection{Quantum charges}

The gauge quantum numbers under $SU(2)_{L}\times U(1)_{Y}$ of the lepton and
the Higgs superfields of the NMSSM as well as the right-handed neutrinos $%
N_{i}^{c}$ are as shown in Table \ref{A}. The lepton doublets of the three
generations $L_{i}=\left( L_{e},L_{\mu },L_{\mu }\right) $\ and the three
right-handed neutrinos $N_{i}^{c}=(\nu _{e}^{c},\nu _{\mu }^{c},\nu _{\tau
}^{c})$ are assigned to irreducible triplets $\mathbf{3}_{(-1,0)}$ of the
discrete group $\mathbb{A}_{4}$. The three right-handed leptons $%
E_{i}^{c}=(e^{c},\mu ^{c},\tau ^{c})$\ sit however in the three different $%
\mathbb{A}_{4}$ singlets $\mathbf{1}_{(1,\omega ^{2})},\mathbf{1}_{(1,\omega
)},\mathbf{1}_{(1,1)}$ where $\omega =e^{i\frac{2\pi }{3}}$ while $(1,1),$ $%
(1,\omega )$ and $(1,\omega ^{2})$ denote the group characters used to
discriminate the one- dimensional representations of the $\mathbb{A}_{4}$
group. Regarding the Higgs superfields of the NMSSM namely the usual three ($%
H_{u},H_{d},S)$ are hosted by ($\mathbf{1}_{(1,1)},\mathbf{1}_{(1,\omega )},%
\mathbf{1}_{(1,\omega ^{2})})$; and carry charges of the extra discrete $%
\boldsymbol{Z}_{3}$.
\begin{table}[h]
\centering \renewcommand{\arraystretch}{1.2}
\begin{tabular}{|c|c|c|c|c|c|c|c|c|}
\hline
Superfields & $L_{i}$ & $e^{c}$ & $\mu ^{c}$ & $\tau ^{c}$ & $N_{i}^{c}$ & $%
H_{u}$ & $H_{d}$ & $S$ \\ \hline
$SU(2)_{L}$ & $2$ & $1$ & $1$ & $1$ & $1$ & $2$ & $2$ & $1$ \\ \hline
$U(1)_{Y}$ & $-1$ & $2$ & $2$ & $2$ & $0$ & $1$ & $-1$ & $0$ \\ \hline
\end{tabular}%
\caption{Gauge charges of lepton and Higgs superfields of the model.}
\label{A}
\end{table}
In addition to the superfields in Table \ref{A}, the three flavon
superfields $\{\Phi ,\Omega ,\mathbf{\chi }\}$ carry nontrivial quantum
numbers under $\mathbb{A}_{4}$ symmetry as shown in Table \ref{C}. The
flavon $\Phi $, transforming as $\mathbf{3}_{(-1,0)}$ under $\mathbb{A}_{4}$%
, is added in the charged lepton sector; while the $\Omega $\ and $\mathbf{%
\chi }$\ are both added in the neutrino sector. We will show later that the
flavon $\mathbb{A}_{4}$- triplet $\Omega $ with the $\mathbb{A}_{4}$-
singlet superfield $\mathbf{\chi }$\ are required to produce a neutrino mass
matrix compatible with the tribimaximal mixing; while $S$, hosted by the $%
\mathbf{1}_{(1,\omega ^{2})}$ of the $\mathbb{A}_{4}$ symmetry is needed for
the deviation from TBM matrix.
\begin{table}[h]
\centering \renewcommand{\arraystretch}{1.2}
\begin{tabular}{|c|c|c|c|c|c|}
\hline
Superfields & $L_{i}$ & $e^{c}$ & $\mu ^{c}$ & $\tau ^{c}$ & $N_{i}^{c}$ \\
\hline
$\mathbb{A}_{4}$ & $\mathbf{3}_{(-1,0)}$ & $1_{(1,\omega ^{2})}$ & $%
1_{(1,\omega )}$ & $1_{(1,1)}$ & $\mathbf{3}_{(-1,0)}$ \\ \hline
$\boldsymbol{Z}_{3}$ & $\mathbf{1}_{Q}$ & $\mathbf{1}_{Q}$ & $\mathbf{1}_{Q}$
& $\mathbf{1}_{Q}$ & $\mathbf{1}_{Q^{2}}$ \\ \hline
\end{tabular}%
\caption{Lepton and right-handed neutrino superfields and their quantum
numbers under $\mathbb{A}_{4}\times \boldsymbol{Z}_{3}$.}
\label{B}
\end{table}
Moreover, to obtain the right vacuum alignment of the $\mathbb{A}_{4}$-
triplets $\Phi $ and $\Omega $ that lead to the desired masses and mixing,
we need to avoid the communication between the charged lepton and neutrino
sectors. This is achieved by introducing the $\boldsymbol{Z}_{3}$ discrete
symmetry under which the flavons triplets $\Phi $ and $\Omega $ act
differently. The $\boldsymbol{Z}_{3}$ quantum numbers for all the
superfields are as given in Tables \ref{B} and \ref{C} with $Q=e^{i\frac{%
2\pi }{3}}$.
\begin{table}[h]
\centering \renewcommand{\arraystretch}{1.2}
\begin{tabular}{|c|c|c|c|c|c|c|}
\hline
Superfields & $H_{u}$ & $H_{d}$ & $\Phi $ & $\Omega $ & $S$ & $\mathbf{\chi }
$ \\ \hline
$\mathbb{A}_{4}$ & $\mathbf{1}_{(1,1)}$ & $\mathbf{1}_{(1,\omega )}$ & $%
\mathbf{3}_{(-1,0)}$ & $\mathbf{3}_{(-1,0)}$ & $\mathbf{1}_{(1,\omega ^{2})}$
& $\mathbf{1}_{(1,1)}$ \\ \hline
$\boldsymbol{Z}_{3}$ & $\mathbf{1}_{1}$ & $\mathbf{1}_{Q}$ & $\mathbf{1}_{1}$
& $\mathbf{1}_{Q^{2}}$ & $\mathbf{1}_{Q^{2}}$ & $\mathbf{1}_{Q^{2}}$ \\
\hline
\end{tabular}%
\caption{Higgs and flavon superfields and their quantum numbers under $%
\mathbb{A}_{4}\times \boldsymbol{Z}_{3}$.}
\label{C}
\end{table}
To fix ideas on some useful features of the $\mathbb{A}_{4}\times
\boldsymbol{Z}_{3}$ flavor symmetry of the proposed model, recall that the
subgroup $\mathbb{A}_{4}$ has four irreducible representations $\mathbf{1}$,
$\mathbf{1}^{\prime }$, $\mathbf{1}^{\prime \prime }$ and $\mathbf{3}$ used
to host the chiral superfield of the model; while the $\boldsymbol{Z}_{3}$
subsymmetry has three irreducible representations $\mathbf{\tilde{1}}$, $%
\mathbf{\tilde{1}}^{\prime }$, $\mathbf{\tilde{1}}^{\prime \prime }$. The
properties of these representations will play an important role in this
study; for this reason, and also for the study of the breaking of $\mathbb{A}%
_{4}$ down to its $\mathbb{Z}_{3}$ and $\mathbb{Z}_{2}$ subgroups, we
thought it is useful to recall briefly some useful properties on the $%
\mathbb{A}_{4}$ group to better illustrate the construction.

\subsubsection{Flavor symmetry $G_{\mathrm{f}}=\mathbb{A}_{4}\times
\boldsymbol{Z}_{3}$}

The alternating $\mathbb{A}_{4}$ subsymmetry of $G_{\mathrm{f}}$ is a
non-Abelian discrete group living inside of the well known permutation group
$\mathbb{S}_{4}$. It is generated by two noncommuting operators $\mathcal{S}$
and $\mathcal{T}$ ($\mathcal{ST}\neq \mathcal{TS}$) with the property $%
\mathcal{S}^{2}=\mathcal{T}^{3}=I$, these $\mathcal{S}$ and $\mathcal{T}$
cannot be diagonalized simultaneously; and we will use later the basis where
$\mathcal{T}$ is represented by a diagonal matrix \cite{R43}. The order of
the $\mathbb{A}_{4}$ group is $12$; it is related to the dimensions of its
four irreducible representations through the character relation
\begin{equation}
\mathbb{A}_{4}:12=\left( 1\right) ^{2}+\left( 1^{\prime }\right) ^{2}+\left(
1^{\prime \prime }\right) ^{2}+\left( 3\right) ^{2}  \label{ch}
\end{equation}%
Throughout this work, we will denote these four irreducible representations
by their basis characters as $\mathbf{3}_{(-1,0)}$, $\mathbf{1}_{(1,1)}$, $%
\mathbf{1}_{(1,\omega )}$ and $\mathbf{1}_{(1,\omega ^{2})}$ with $\omega
=\exp \left( 2i\pi /3\right) $; this notation gives a tricky manner to
distinguish the three one- dimensional representations by the characters of
the generators of $\mathbb{A}_{4}$. Because of the property $\omega ^{2}=%
\bar{\omega}$, we can also denote the one- dimensional representations as $%
\mathbf{1}_{(1,1)},$ $\mathbf{1}_{(1,\omega )},$ $\mathbf{1}_{(1,\bar{\omega}%
)}$; for other aspects, see Appendix A and Refs.\textrm{\ }\cite{R44,S}.
\newline
The $\mathbb{A}_{4}$ has ten subgroups that we want to describe briefly;
they are needed in Sec. IV when we study the breaking of $\mathbb{A}_{4}$.
All the proper subgroups of $\mathbb{A}_{4}$ are Abelian and so have one-
dimensional irreducible representations that can be determined by using the
analogue of the character formula (\ref{ch}). In addition to the set $%
\mathbb{A}_{4}$ itself and to the identity $I_{id}$, the subgroups of $%
\mathbb{A}_{4}$ are as follows: (\textbf{a)} the Klein-four group $\mathbb{V}%
_{4}$ with non trivial elements given by double transpositions $\left(
ij\right) \left( kl\right) $ as reported hereinbelow
\begin{equation}
\mathbb{V}_{4}=\left\{ I_{id},\left( 12\right) \left( 34\right) ,\left(
13\right) \left( 24\right) ,\left( 14\right) \left( 23\right) \right\}
\label{v4}
\end{equation}%
where $\left\{ 1,2,3,4\right\} $ are numbers indexing some four given
objects $\left\{ X_{1},X_{2},X_{3},X_{4}\right\} $. In our case, the $\Phi
_{i}$, $\Omega _{i}$ and $\mathbf{\chi }$\ flavon chiral superfields are
given by a subsets of $X_{i}$'s or also by linear combinations of them.
\textbf{(b)} four kinds of cyclic groups $\mathbb{Z}_{3}$, one of them is
given by the set $\left\{ I_{id},\left( 123\right) ,\left( 132\right)
\right\} $ where the fourth point $X_{4}$ is fixed; a second example of a $%
\mathbb{Z}_{3}$\ subgroup is given by $\left\{ I_{id},\left( 124\right)
,\left( 142\right) \right\} $ where now the point $X_{3}$ which is fixed.
\textbf{(c)} three cyclic groups $\mathbb{Z}_{2}$ generated by double
transpositions; they are subgroups of $\mathbb{V}_{4}$; one of these three $%
\mathbb{Z}_{2}$'s is given by $\left\{ I_{id},\left( 12\right) \left(
34\right) \right\} $; a second one is given by $\left\{ I_{id},\left(
13\right) \left( 24\right) \right\} $. The multiplicities of these subgroups
are as in the following Table%
\begin{equation}
\begin{tabular}{|l|l|l|l|l|l|}
\hline
$\text{{\small subgroups}}$ & $\mathbb{A}_{4}$ & $\mathbb{V}_{4}$ & $\
\mathbb{Z}_{3}$ & $\ \mathbb{Z}_{2}$ & $\ I_{id}$ \\ \hline
$\text{{\small order}}$ & ${\small 12}$ & ${\small 4}$ & ${\small \ 3}$ & $%
{\small 2}$ & ${\small 1}$ \\ \hline
$\text{{\small multiplicity}}$ & ${\small 1}$ & ${\small 1}$ & $\ {\small 4}$
& ${\small 3}$ & $\ {\small 1}$ \\ \hline
\end{tabular}
\label{a4}
\end{equation}%
The $\boldsymbol{Z}_{3}$ subsymmetry of $G_{\mathrm{f}}$ is an Abelian
discrete group which has not to be confused with $\mathbb{Z}_{3}$ which is a
subgroup of $\mathbb{A}_{4}$; it has three different one- dimensional
representations $\mathbf{\tilde{1}}$, $\mathbf{\tilde{1}}^{\prime }$, $%
\mathbf{\tilde{1}}^{\prime \prime }$ that can be denoted as well like $%
\mathbf{1}_{1}$, $\mathbf{1}_{Q}$, and $\mathbf{1}_{Q^{2}}$ with $Q=e^{i%
\frac{2\pi }{3}}$.\newline
To summarize, the flavor symmetry $G_{\mathrm{f}}=\mathbb{A}_{4}\times
\boldsymbol{Z}_{3}$ has three generators $\mathcal{S}$, $\mathcal{T}$, and $%
\mathcal{\tilde{T}}$ and \textrm{12} irreducible representations namely%
\begin{equation}
\begin{tabular}{llll}
$\mathbf{1}_{(1,1,1)},$ & $\mathbf{1}_{(1,\omega ,1)},$ & $\mathbf{1}_{(1,%
\bar{\omega},1)}$ & $\mathbf{3}_{(-1,0,1)}$ \\
$\mathbf{1}_{(1,1,Q)},$ & $\mathbf{1}_{(1,\omega ,Q)},$ & $\mathbf{1}_{(1,%
\bar{\omega},Q)}$ & $\mathbf{3}_{(-1,0,Q)}$ \\
$\mathbf{1}_{(1,1,\bar{Q})},$ & $\mathbf{1}_{(1,\omega ,\bar{Q})},$ & $%
\mathbf{1}_{(1,\bar{\omega},\bar{Q})}$ & $\mathbf{3}_{(-1,0,\bar{Q})}$%
\end{tabular}%
\end{equation}%
where the indices refer to the characters of the generators, the two first
entries for the $\mathcal{S}$, $\mathcal{T}$ of alternating $\mathbb{A}_{4}$
and the third for the $\mathcal{\tilde{T}}$ of $\boldsymbol{Z}_{3}$.

\subsection{Charged lepton sector}

With the extra $\mathbb{A}_{4}\times \boldsymbol{Z}_{3}$ quantum numbers; we
have a powerful tool to suppress undesired superfield couplings in building
the general lepton mass matrix. Following \textrm{\cite{R43}}, the usual
Yukawa couplings, giving the masses $\left( m_{e},m_{\mu },m_{\tau }\right) $%
\ of the charged leptons, are given by the $W_{lep^{+}}^{0}=%
\sum_{i,j}y_{0}^{ij}H_{d}E_{i}^{c}L_{j}$ tri-superfield couplings which in
general are not invariant under the flavor symmetry. To make the charged
leptons superpotential invariant under the full discrete $\mathbb{A}%
_{4}\times \boldsymbol{Z}_{3}$ symmetry group, we promote the Yukawa
coupling constants $y_{0}^{ij}$ to be flavon $\Phi $- dependent like $%
y_{0}^{ij}\rightarrow y_{0}^{ijk}\frac{\Phi _{k}}{\Lambda }$ where $\Lambda $
denotes a cutoff scale of the $G_{\mathrm{f}}$ symmetry of GUT scale order.
Thus, the desired $W_{lep^{+}}$ chiral superpotential\ for the charged
leptons is flavon dependant$\ W_{lep^{+}}\left[ \Phi \right] $ and is given
by a four- order operator generated by the monomials $H_{d}E_{i}^{c}L_{j}%
\Phi _{k}$. However, because $L_{j}$ and $\Phi _{k}$ transform into
nontrivial representations of $G_{\mathrm{f}}$, these four order monomials
carry non trivial quantum charges of the discrete flavor symmetry. In the $%
\mathbb{A}_{4}\times \boldsymbol{Z}_{3}$ representation language, the $%
H_{d}E_{i}^{c}L_{j}\Phi _{k}$ transform in the $\mathbb{A}_{4}$- tensor
product representation $\mathbf{1}_{\left( 1,\bar{\omega}^{i-1}\right)
}\otimes \left[ \mathbf{3}_{\left( -1,0\right) }\otimes \mathbf{3}_{\left(
-1,0\right) }\right] $ as it can be checked from Tables \ref{B} and \ref{C}.
From this tensor product structure, we learn that the $3^{3}=27$ possible
monomials $H_{d}E_{i}^{c}L_{j}\Phi _{k}$ are manifestly $\boldsymbol{Z}_{3}$%
- invariant ($Q^{3}=1$); while invariance under the non-Abelian $\mathbb{A}%
_{4}$ requires restricting the sum to the particular trace
\begin{equation}
Tr_{\mathbb{A}_{4}}\left( \frac{y^{ijk}}{\Lambda }E_{i}^{c}L_{j}\Phi
_{k}\right) \equiv \left. \left( \frac{y^{ijk}}{\Lambda }E_{i}^{c}L_{j}\Phi
_{k}\right) \right\vert _{\mathbf{1}_{\left( 1,\bar{\omega}\right) }}
\label{27}
\end{equation}%
This trace can be obtained by first using the group representation notations
$L\otimes \Phi \equiv \mathbf{3}_{\left( -1,0\right) }\otimes \mathbf{3}%
_{\left( -1,0\right) }$ showing that $L_{j}\Phi _{k}$, with $L_{j}=\left(
L_{e},L_{\mu },L_{\tau }\right) $ and $\Phi _{k}=\left( \Phi _{1},\Phi
_{2},\Phi _{3}\right) $, has nine components carrying in general different $%
\mathbb{A}_{4}$- quantum numbers. By using results from \textrm{\cite{S}} on
the reducibility property of tensors product of $\mathbb{A}_{4}$-
representations; in particular $\mathbf{3}_{\left( -1,0\right) }\otimes
\mathbf{3}_{\left( -1,0\right) }=\mathbf{9}_{\left( 1,0\right) }$ with%
\begin{equation}
\mathbf{9}_{\left( 1,0\right) }=\mathbf{1}_{\left( 1,1\right) }\oplus
\mathbf{1}_{\left( 1,\omega \right) }\oplus \mathbf{1}_{\left( 1,\bar{\omega}%
\right) }\oplus \mathbf{3}_{\left( -1,0\right) }\oplus \mathbf{3}_{\left(
-1,0\right) }
\end{equation}%
we learn that the relevant terms in the charged lepton superpotential $%
W_{lep^{+}}$ are given, in the $\mathbb{A}_{4}$ basis where the generator $%
\mathcal{T}$ is diagonal [$\left( \mathcal{T}.\Phi \right) _{r}=\omega
^{r-1}\Phi _{r}$ and $\left( \mathcal{T}.L\right) _{r}=\omega ^{r-1}L_{r}$],
by the following one dimensional $\mathbb{A}_{4}$- representations
\begin{equation}
\begin{tabular}{lll}
$\left. L\otimes \Phi \right\vert _{\mathbf{1}_{\left( 1,1\right) }}$ & $=$
& $L_{e}\Phi _{1}+L_{\tau }\Phi _{2}+L_{\mu }\Phi _{3}$ \\
$\left. L\otimes \Phi \right\vert _{\mathbf{1}_{\left( 1,\omega \right) }}$
& $=$ & $L_{\mu }\Phi _{1}+L_{e}\Phi _{2}+L_{\tau }\Phi _{3}$ \\
$\left. L\otimes \Phi \right\vert _{\mathbf{1}_{(1,\bar{\omega})}}$ & $=$ & $%
L_{\tau }\Phi _{1}+L_{\mu }\Phi _{2}+L_{e}\Phi _{3}$%
\end{tabular}
\label{product}
\end{equation}%
For the explicit expressions of the other irreducible representations; see
Eq. (\ref{re}) of Appendix A. Notice that the above restrictions to the
three one- dimensional $\left. L\otimes \Phi \right\vert _{\mathbf{1}%
_{\left( 1,\omega ^{r}\right) }}$ of the $\mathbb{A}_{4}$ symmetry are
because of the quantum numbers of the right handed leptons $%
E_{i}^{c}=(e^{c},\mu ^{c},\tau ^{c})$ which transform under $\mathbb{A}_{4}$
into three different one dimensional representations like $E_{i}^{c}\sim
\mathbf{1}_{\left( 1,\bar{\omega}^{i}\right) }$. Therefore, the desired $%
\mathbb{A}_{4}$- invariant quantity is given by
\begin{equation}
Tr_{\mathbb{A}_{4}}\left( \frac{y^{ijk}}{\Lambda }E_{i}^{c}L_{j}\Phi
_{k}\right) \sim \sum_{i=1}^{3}\left( \mathbf{1}_{\left( 1,\omega
^{-i}\right) }\times \left. \left( L\otimes \Phi \right) \right\vert _{%
\mathbf{1}_{_{(1,\omega ^{i-1})}}}\right)
\end{equation}%
leading to the following quartic $\mathbb{A}_{4}\times \boldsymbol{Z}_{3}$
invariant superpotential%
\begin{equation}
W_{lep^{+}}=\sum_{i=1}^{3}\frac{y_{i}}{\Lambda }H_{d}\left( \left. L\otimes
\Phi \right\vert _{\mathbf{1}_{(1,\omega ^{i-1})}}\right) E_{i}^{c}
\label{lep}
\end{equation}%
Taking the VEV of $\mathbb{A}_{4}$- triplet $\Phi $ like $\langle \Phi
\rangle =(\upsilon _{\Phi },0,0)$ and the VEV of the Higgs $H_{d}$ as usual--%
$\left\langle H_{d}\right\rangle =\upsilon _{d}$--; then substituting these
expressions back into the superpotential (\ref{lep}), we obtain the desired
supersymmetric Yukawa couplings%
\begin{equation}
W_{lep^{+}}=\frac{y_{e}\upsilon _{\Phi }}{\Lambda }H_{d}e^{c}L_{e}+\frac{%
y_{\mu }\upsilon _{\Phi }}{\Lambda }H_{d}\mu ^{c}L_{\mu }+\frac{y_{\tau
}\upsilon _{\Phi }}{\Lambda }H_{d}\tau ^{c}L_{\tau }  \label{mm}
\end{equation}%
leading to a diagonal charged lepton mass matrix $M_{lep^{+}}=diag(m_{e},m_{%
\mu },m_{\tau })$ with mass eigenvalues induced by the product of two VEVs
as follows%
\begin{equation}
m_{e}=y_{e}\frac{\upsilon _{d}\upsilon _{\Phi }}{\Lambda }~\quad ,\quad
m_{\mu }=y_{\mu }\frac{\upsilon _{d}\upsilon _{\Phi }}{\Lambda }~\quad
,\quad m_{\tau }=y_{\tau }\frac{\upsilon _{d}\upsilon _{\Phi }}{\Lambda }
\label{lm}
\end{equation}%
By using the experimental value of the tau mass as given by the Particle
Data Group \cite{R45}, namely $m_{\tau }=1776.86$ $\mathrm{MeV}$, and by
assuming that $y_{\tau }\upsilon _{d}\lesssim 246$ $\mathrm{GeV}$, we obtain
a lower bound on the ratio between the flavon VEV $\upsilon _{\Phi }$ and
the cutoff scale $\Lambda $\ given by%
\begin{equation}
\frac{\upsilon _{\Phi }}{\Lambda }>0.007  \label{cl}
\end{equation}%
This constraint on $\upsilon _{\Phi }$ will be used later on when studying
domain walls in the charged lepton sector.

\section{Neutrino masses and trimaximal mixing}

In this section,\textrm{\ }we study the\ neutrino masses and mixing in the
framework of trimaximal scheme in order to reconcile with the experimental
values of the reactor and atmospheric mixing angles\textrm{. }We use the PDG
parametrization for the mixing matrix\textrm{\ }$U_{PMNS}$\ given by \cite%
{R45}\textrm{\ }%
\begin{equation}
U_{PMNS}=\left(
\begin{array}{ccc}
c_{12}c_{13} & c_{13}s_{12} & s_{13}e^{-i\delta } \\
-c_{23}s_{12}-c_{12}s_{13}s_{23}e^{i\delta } &
c_{12}c_{23}-s_{12}s_{13}s_{23}e^{i\delta } & c_{13}s_{23} \\
s_{12}s_{23}-c_{12}c_{23}s_{13}e^{i\delta } &
-c_{12}s_{23}-c_{23}s_{12}s_{13}e^{i\delta } & c_{13}c_{23}%
\end{array}%
\right) .\mathcal{U}_{m}  \label{pmns}
\end{equation}%
where the short hands $c_{ij}\equiv \cos \theta _{ij}$ and $s_{ij}\equiv
\sin \theta _{ij}$ refer to the mixing angles; $\delta $\ refers to the
Dirac $CP$ phase \textrm{and }$\mathcal{U}_{m}=diag(1,e^{\frac{i}{2}\alpha
_{21}},e^{\frac{i}{2}\alpha _{31}})$\textrm{\ }encodes the Majorana-type $CP$
violating phases $\alpha _{21}$\ and $\alpha _{31}$\textrm{.}

\subsection{Trimaximal mixing as a deviation of TBM matrix}

By adding the three right-handed neutrinos $N_{i}^{c}=(\nu _{e}^{c},\nu
_{\mu }^{c},\nu _{\tau }^{c})$, the light neutrino masses are generated via
type I seesaw mechanism formula $m_{\nu }=m_{D}^{T}M_{R}^{-1}m_{D}$ with $%
M_{R}$ the right-handed Majorana neutrinos mass matrix. The Dirac mass
matrix $m_{D}$ emerges from the Yukawa couplings of the left and
right-handed neutrinos with the Higgs superfield $H_{u}$. The relevant
chiral superpotential $W_{D}$ respecting gauge and flavor symmetries of the
model is as follows%
\begin{equation}
W_{D}=Tr_{\mathbb{A}_{4}}\left( Y^{ij}L_{i}N_{j}^{c}H_{u}\right)
\end{equation}%
where $L_{i}=\left( L_{e},L_{\mu },L_{\tau }\right) $, $Y^{ij}$\ are Yukawa
coupling constants and where $Tr_{\mathbb{A}_{4}}\left[ \ast \right] $
refers to the restriction of the sum $\sum_{i,j}Y^{ij}L_{i}N_{j}^{c}$ to the
$\mathbb{A}_{4}$- invariant part. Following the same analysis used in the
charged lepton sector, $\mathbb{A}_{4}$-invariant $W_{D}$ is then given by $%
\left. \left( H_{u}\otimes L\otimes N^{c}\right) \right\vert _{\mathbf{1}%
_{\left( 1,1\right) }}$; but because $H_{u}\sim \mathbf{1}_{\left(
1,1\right) }$, this irreducible component reduces to $H_{u}$ times $\left.
\left( L\otimes N^{c}\right) \right\vert _{\mathbf{1}_{\left( 1,1\right) }}$
reading explicitly as
\begin{equation}
W_{D}=Y_{0}H_{u}\left( L_{1}N_{1}^{c}+L_{2}N_{3}^{c}+L_{3}N_{2}^{c}\right)
\end{equation}%
Giving to the neutral component of the $H_{u}$ Higgs doublet its VEV $%
\upsilon _{u}$, we get the Dirac mass matrix of neutrinos%
\begin{equation}
m_{D}=Y_{0}\upsilon _{u}\left(
\begin{array}{ccc}
1 & 0 & 0 \\
0 & 0 & 1 \\
0 & 1 & 0%
\end{array}%
\right)  \label{md}
\end{equation}%
On the other hand, the superpotential\textrm{\ }$W_{R}$ of the right-handed
Majorana neutrinos respecting gauge invariance, the $\mathbb{A}_{4}\times
\boldsymbol{Z}_{3}$\ flavor symmetry and leading to trimaximal mixing,
contains\textrm{\ }three tri-superfield couplings as in Eq. (\ref{ma1})
given below; the first one involves the chiral superfield $\mathbf{\chi }$
which is an $\mathbb{A}_{4}$- singlet, the second one involves an $\mathbb{A}%
_{4}$- flavon triplet $\Omega =\left( \Omega _{1},\Omega _{2},\Omega
_{3}\right) $ and the last one involves the NMSSM singlet $S$\ which
transforms as a nontrivial singlet under\textrm{\ }$\mathbb{A}_{4}$\textrm{.}
The superpotential having a quadratic dependence into the right-handed
neutrinos superfields---i.e: $W_{R}$ of the form $\sum \lambda _{i}N^{c}\Phi
_{i}N^{c}$---is expressed at the renormalizable level as\footnote{%
If dropping the condition of a superpotential quadratic in $N^{c},$ that is
releasing the form $W_{R}\sim \lambda _{i}N^{c}\Phi _{i}N^{c}$, the
superpotential may include other possible terms like $\lambda _{ij}N^{c}\Phi
_{i}\Phi _{j}$. A possible manner to eliminate these undesirable couplings
is to impose additional discrete symmetries. At the level of Eq. (\ref{ma1}%
), linear and cubic $N^{c}$- terms in $W_{R}$ may be eliminated by an extra $%
\mathcal{Z}_{2}:N^{c}\rightarrow -N^{c}$. In this case, the other matter
superfields of the model $L$ and $E^{c}$ have to carry the same $\mathcal{Z}%
_{2}$ charge as $N^{c}$ while the scalar superfields carry a trivial $%
\mathcal{Z}_{2}$ charge.}%
\begin{equation}
W_{R}=\lambda Tr_{\mathbb{A}_{4}}\left( \mathbf{\chi }N^{c}N^{c}\right)
+\lambda ^{\prime }Tr_{\mathbb{A}_{4}}\left( \Omega N^{c}N^{c}\right)
+\lambda ^{\prime \prime }Tr_{\mathbb{A}_{4}}\left( SN^{c}N^{c}\right)
\label{ma1}
\end{equation}%
where $\lambda $, $\lambda ^{\prime }$\ and $\lambda ^{\prime \prime }$\ are
Yukawa coupling constants. The first tri-superfield coupling $\mathbf{\chi }%
N^{c}N^{c}$\textrm{\ }transforms under the\textrm{\ }$\mathbb{A}_{4}$\textrm{%
\ }symmetry as $1_{\left( 1,1\right) }\otimes 3_{\left( -1,0\right) }\otimes
3_{\left( -1,0\right) }$, the second coupling\textrm{\ }$\Omega N^{c}N^{c}$%
\textrm{\ }transforms like\textrm{\ }$3_{\left( -1,0\right) }\otimes
3_{\left( -1,0\right) }\otimes 3_{\left( -1,0\right) }$\textrm{; }while the
third coupling\textrm{\ }$SN^{c}N^{c}$\textrm{\ }transforms as\textrm{\ }$%
1_{(1,\omega ^{2})}\otimes 3_{\left( -1,0\right) }\otimes 3_{\left(
-1,0\right) }$. Hence, by using the Clebsch-Gordan decomposition of\textrm{\
}$\mathbb{A}_{4}$\textrm{\ }(\ref{re}), the superpotential\textrm{\ }$W_{R}$%
\textrm{\ }develops into%
\begin{equation}
\begin{array}{ccc}
W_{R} & = & \lambda \mathbf{\chi }(\nu _{e}^{c}\nu _{e}^{c}+\nu _{\mu
}^{c}\nu _{\tau }^{c}+\nu _{\tau }^{c}\nu _{\mu }^{c})+\lambda ^{\prime
\prime }S(\nu _{e}^{c}\nu _{\mu }^{c}+\nu _{\mu }^{c}\nu _{e}^{c}+\nu _{\tau
}^{c}\nu _{\tau }^{c})+\text{\ \ \ \ \ \ } \\
&  & \frac{2\lambda ^{\prime }}{3}(\left[ \left( \nu _{e}^{c}\nu
_{e}^{c}-\nu _{\mu }^{c}\nu _{\tau }^{c}\right) \Omega _{1}+\left( \nu
_{\tau }^{c}\nu _{\tau }^{c}-\nu _{e}^{c}\nu _{\mu }^{c}\right) \Omega
_{2}+\left( \nu _{\mu }^{c}\nu _{\mu }^{c}-\nu _{e}^{c}\nu _{\tau
}^{c}\right) \Omega _{3}\right]%
\end{array}
\label{1}
\end{equation}%
By taking the flavon VEVs as $\langle \mathbf{\chi }\rangle =\upsilon _{%
\mathbf{\chi }}$, $\langle \Omega \rangle =(\upsilon _{\Omega },\upsilon
_{\Omega },\upsilon _{\Omega })$ and the NMSSM singlet as $\langle S\rangle
=\upsilon _{S}$, as well as adopting the following notations: $a=2\lambda
\upsilon _{\mathbf{\chi }}$, $b=2\lambda ^{\prime }\upsilon _{\Omega }$ and $%
\mathrm{\epsilon }=2\lambda ^{\prime \prime }\upsilon _{S}$---the parameters
$a$, $b$\ and $\mathrm{\epsilon }$ have a dimension of mass and at least one
of them should be complex as it will be shown below---the right-handed
Majorana neutrino mass matrix $M_{R}$\ is given by%
\begin{equation}
M_{R}=\left(
\begin{array}{ccc}
a+\frac{2b}{3} & -\frac{b}{3}+\mathrm{\epsilon } & -\frac{b}{3} \\
-\frac{b}{3}+\mathrm{\epsilon } & \frac{2b}{3} & a-\frac{b}{3} \\
-\frac{b}{3} & a-\frac{b}{3} & \frac{2b}{3}+\mathrm{\epsilon }%
\end{array}%
\right)  \label{mr1}
\end{equation}%
In the limit $\epsilon \rightarrow 0$, this matrix is diagonalized by the
TBM mixing matrix, given in Eq. (\ref{e1}). \ Notice by the way that the
form \textrm{of} $M_{R}\left( \mathrm{\epsilon }\rightarrow 0\right) $ was
produced before with at least two flavons fields \cite{R43,R46,B6}; the
novelty here is that we used the NMSSM singlet superfield $S$\ to deviate
from the TBM instead of a flavon field. It is well known that in this limit,
the matrix respects the $\mu -\tau $\ symmetry\footnote{%
\ This symmetry is based on the invariance of the neutrino mass terms under
the interchange of $\nu _{\mu }$ and $\nu _{\tau }$ \cite{R36}.}\textrm{\ }%
with the prediction of zero reactor angle and maximal atmospheric angle.
Currently, it is a well-established experimental fact that the value of the
reactor angle is different from zero; $\theta _{13}\neq 0$. On the other
hand, the NOvA experiment has disfavored recently\ the maximal atmospheric
angle; however, the significance of this indication, while at the $2.6\sigma
$\ level in \cite{NOV}, was reduced to less than 1$\sigma $\ in \cite{K3}
which means that the maximal value of $\theta _{23}$ remain possible.
Therefore, in order to produce the experimental values of these angles, we
need to go beyond the TBM framework. Thus, by using the NMSSM singlet
superfield $S$ (which plays the role of the deviation from TBM), a coupling $%
SN^{c}N^{c}$\ in the Majorana superpotential (\ref{1}) breaks the $\mu -\tau
$\ symmetry in a way that only\ the $\left( M_{R}\right) _{12}$, $\left(
M_{R}\right) _{21}$ and $\left( M_{R}\right) _{33}$ entries of the Majorana
neutrino mass matrix are affected; see Eq. (\ref{mr1}). The diagonalization
of this symmetric matrix can be obtained by performing a similarity
transformation like $M_{R}^{\prime }=$ $\mathcal{P}^{-1}M_{R}\mathcal{P}$
with eigenvalues%
\begin{equation}
M_{1}^{\prime }=b+\sqrt{a^{2}-a\mathrm{\epsilon }+\mathrm{\epsilon }^{2}}%
\quad ,\quad M_{2}^{\prime }=a+\mathrm{\epsilon }\quad ,\quad M_{3}^{\prime
}=b-\sqrt{a^{2}-a\mathrm{\epsilon }+\mathrm{\epsilon }^{2}}  \label{39}
\end{equation}%
and
\begin{equation}
\mathcal{P}=\left(
\begin{array}{ccc}
\gamma _{1}^{+} & 1 & \gamma _{1}^{-} \\
\gamma _{2}^{+} & 1 & \gamma _{2}^{-} \\
1 & 1 & 1%
\end{array}%
\right)
\end{equation}%
with%
\begin{eqnarray}
\gamma _{1}^{\mp } &=&\frac{\left( 3a+b\right) \mathrm{\epsilon }-3\mathrm{%
\epsilon }^{2}\mp \left( 3\mathrm{\epsilon }-2b\right) \sqrt{a^{2}-a\mathrm{%
\epsilon }+\mathrm{\epsilon }^{2}}-2ab}{ab-2b\mathrm{\epsilon }+3\mathrm{%
\epsilon }^{2}+\left( 3\mathrm{\epsilon }-b\right) \sqrt{a^{2}-a\mathrm{%
\epsilon }+\mathrm{\epsilon }^{2}}}  \notag \\
&&\text{ \ } \\
\gamma _{2}^{\mp } &=&\frac{\left( b-3a\right) \mathrm{\epsilon }\mp b\sqrt{%
a^{2}-a\mathrm{\epsilon }+\mathrm{\epsilon }^{2}}+ab}{ab-2b\mathrm{\epsilon }%
+3\mathrm{\epsilon }^{2}+\left( 3\mathrm{\epsilon }-b\right) \sqrt{a^{2}-a%
\mathrm{\epsilon }+\mathrm{\epsilon }^{2}}}  \notag
\end{eqnarray}%
These relations are nonlinear in the deviation parameter $\mathrm{\epsilon }$%
; they may be treated perturbatively up to order $\mathcal{O}(\mathrm{%
\epsilon }^{2})$\emph{\ }by following the same method as done in \cite{R44}%
.\ For example, the linearization of the $b\pm \sqrt{a^{2}-a\mathrm{\epsilon
}+\mathrm{\epsilon }^{2}}$ eigenvalues gives $b\pm a\mp \frac{\mathrm{%
\epsilon }}{2}+O(\frac{\mathrm{\epsilon }^{2}}{a})$; and then%
\begin{equation}
M_{1}^{\prime }\simeq b+a-\frac{\mathrm{\epsilon }}{2}\quad ,\quad
M_{2}^{\prime }=a+\mathrm{\epsilon }\quad ,\quad M_{3}^{\prime }\simeq b-a+%
\frac{\mathrm{\epsilon }}{2}  \label{e}
\end{equation}%
Nevertheless, to deal with (\ref{mr1}) we follow the procedure used in \cite%
{R23} instead of the above linearization. The method relies on using the
trimaximal mixing matrix $U_{TM_{2}}=U_{TBM}U_{R}$ obtained by multiplying $%
U_{TBM}$ by a matrix $U_{R}$ from the\ right while leaving one of the
columns in $U_{TBM}$ unaffected. Here, $U_{R}$ stands for a
\textquotedblleft complexified rotation\textquotedblright\ matrix. In the
(1-3) plane, we have%
\begin{equation}
U_{R}=\left(
\begin{array}{ccc}
\cos \theta & 0 & \sin \theta e^{-i\sigma } \\
0 & 1 & 0 \\
-\sin \theta e^{i\sigma } & 0 & \cos \theta%
\end{array}%
\right)
\end{equation}%
where the parameter $\theta $ parameterizes the deviation from tribimaximal
mixing and $\sigma $ is an arbitrary phase that will be related to the Dirac
$CP$ phase later. This complex rotation matrix $U_{R}$ has the properties $%
U_{R}^{\dagger }U_{R}=I$ and $\det U_{R}=1$; it preserves the second column
of the TBM matrix and leads to
\begin{equation}
U_{TM_{2}}=\left(
\begin{array}{ccc}
\sqrt{\frac{2}{3}}\cos \theta & \frac{1}{\sqrt{3}} & \sqrt{\frac{2}{3}}\sin
\theta e^{-i\sigma } \\
-\frac{\cos \theta }{\sqrt{6}}-\frac{\sin \theta }{\sqrt{2}}e^{i\sigma } &
\frac{1}{\sqrt{3}} & \frac{\cos \theta }{\sqrt{2}}-\frac{\sin \theta }{\sqrt{%
6}}e^{-i\sigma } \\
-\frac{\cos \theta }{\sqrt{6}}+\frac{\sin \theta }{\sqrt{2}}e^{i\sigma } &
\frac{1}{\sqrt{3}} & -\frac{\cos \theta }{\sqrt{2}}-\frac{\sin \theta }{%
\sqrt{6}}e^{-i\sigma }%
\end{array}%
\right)  \label{tm}
\end{equation}%
with $U_{TM_{2}}^{\dagger }U_{TM_{2}}=I$ and $\left\vert \det
U_{TM_{2}}\right\vert =1$. By using this mixing matrix, the Majorana
neutrino mass matrix $M_{R}$\ is diagonalized as
\begin{equation}
\left( U_{TM_{2}}^{\ast }\right) ^{T}M_{R}U_{TM_{2}}^{\ast }=\mathrm{diag}%
(M_{1},M_{2},M_{3})
\end{equation}%
provided the following condition holds%
\begin{equation}
\tan 2\theta =\frac{-\sqrt{3}\mathrm{\epsilon }}{\left( 2a+2b-\mathrm{%
\epsilon }\right) e^{i\sigma }+\left( 2a-2b-\mathrm{\epsilon }\right)
e^{-i\sigma }}.
\end{equation}%
By taking a, b and $\epsilon $\ complex numbers, then the reality of $\tan
2\theta $\ put a constraint on the phase $\sigma $; indeed using $%
a=\left\vert a\right\vert e^{i\phi _{a}}$, $b=\left\vert b\right\vert
e^{i\phi _{b}},$ $\epsilon =\left\vert \mathrm{\epsilon }\right\vert
e^{i\phi _{\epsilon }}$\ and setting $k_{a\epsilon }=\left\vert a/\mathrm{%
\epsilon }\right\vert $, $k_{b\epsilon }=\left\vert b/\mathrm{\epsilon }%
\right\vert ,$ $k_{ab}=\left\vert a/b\right\vert $\ as well as $\phi
_{ij}=\phi _{i}-\phi _{j}$, we end up with a somehow cumbersome relation
that can be simplified a little bit by taking $\phi _{b}=\phi _{\epsilon
}\neq \phi _{a}$. In this case, we have%
\begin{equation}
\begin{tabular}{lll}
$\tan 2\theta $ & $=$ & $\frac{-2\sqrt{3}\left( 2k_{a\epsilon }\cos \phi
_{a\epsilon }-1\right) \cos \sigma }{4\left( 2k_{a\epsilon }\cos \phi
_{a\epsilon }-1\right) ^{2}\cos ^{2}\sigma +\left( 4k_{a\epsilon }\cos
\sigma \sin \phi _{a\epsilon }+4k_{b\epsilon }\sin \sigma \right) ^{2}}%
,\quad \tan \sigma =-k_{ab}\sin \phi _{a\epsilon }.$%
\end{tabular}
\label{sigma}
\end{equation}%
For similar expressions to the above relations; see \cite{B7}. Furthermore,
the exact eigenvalues of the mass matrix $M_{R}$ are given by $b\pm \sqrt{%
a^{2}-a\mathrm{\epsilon }+\mathrm{\epsilon }^{2}}$ and $a+\mathrm{\epsilon }$%
; by assuming $\mathrm{\epsilon }/a<<1$, they can be expanded in a power
series in the reduced parameter $\mathrm{\epsilon }/a$; at the leading
order, we have:%
\begin{equation}
M_{1}=a+b-\frac{\mathrm{\epsilon }}{2}+O(\frac{\mathrm{\epsilon }^{2}}{a}%
)\quad ,\quad M_{2}=a+\mathrm{\epsilon \quad },\quad M_{3}=b-a+\frac{\mathrm{%
\epsilon }}{2}+O(\frac{\mathrm{\epsilon }^{2}}{a})
\end{equation}%
Now we turn to calculate the light neutrino masses by applying the type-I
seesaw formula $m_{\nu }=m_{D}^{T}M_{R}^{-1}m_{D}$. Since $M_{R}$ is
diagonalized by $U_{TM_{2}}$ given in Eq. (\ref{tm}), the inverse Majorana
neutrino mass is now given by%
\begin{equation}
M_{R}^{-1}=U_{TM_{2}}^{\ast }[\mathrm{diag}(M_{1},M_{2},M_{3})]^{-1}\left(
U_{TM_{2}}^{\ast }\right) ^{T}
\end{equation}%
Moreover, due to the form of the Dirac mass matrix (\ref{md}), the
diagonalization of $m_{\nu }$ remains of trimaximal form\footnote{%
\ Notice that in general the light neutrino masses are given by $\tilde{U}%
_{TM_{2}}m_{\nu }(\tilde{U}_{TM_{2}}^{T})=\mathrm{diag}(\hat{m}_{1},\hat{m}%
_{2},\hat{m}_{3})$ \textrm{where }$\hat{m}_{j}=m_{j}\times e^{i\eta _{j}}$
\textrm{are complex eigenvalues; the} $m_{j}=\left\vert \hat{m}%
_{j}\right\vert $'s \textrm{are positive eigenvalues and} the $\eta _{j}$'s
\textrm{stand for the phases. }}. Thus, the light neutrino masses are given
by $\left( \tilde{U}_{TM_{2}}\right) ^{T}m_{\nu }\tilde{U}_{TM_{2}}=\mathrm{%
diag}(m_{1},m_{2},m_{3})$ where $\tilde{U}_{TM_{2}}=\frac{m_{D}}{%
Y_{0}\upsilon _{u}}U_{TM_{2}},$\textrm{\ }%
\begin{equation}
m_{1}=\frac{Y_{0}^{2}\upsilon _{u}^{2}}{M_{1}}\quad ,\quad m_{2}=\frac{%
Y_{0}^{2}\upsilon _{u}^{2}}{M_{2}}\quad ,\quad m_{3}=\frac{Y_{0}^{2}\upsilon
_{u}^{2}}{M_{3}}  \label{ev}
\end{equation}%
$\tilde{U}_{TM_{2}}$ is the new neutrino mixing matrix which differs from $%
U_{TM_{2}}$ by the interchange of the second and the third row \cite{R47}
\begin{equation}
\tilde{U}_{TM_{2}}=\left(
\begin{array}{ccc}
\sqrt{\frac{2}{3}}\cos \theta & \frac{1}{\sqrt{3}} & \sqrt{\frac{2}{3}}\sin
\theta e^{-i\sigma } \\
-\frac{\cos \theta }{\sqrt{6}}+\frac{\sin \theta }{\sqrt{2}}e^{i\sigma } &
\frac{1}{\sqrt{3}} & -\frac{\cos \theta }{\sqrt{2}}-\frac{\sin \theta }{%
\sqrt{6}}e^{-i\sigma } \\
-\frac{\cos \theta }{\sqrt{6}}-\frac{\sin \theta }{\sqrt{2}}e^{i\sigma } &
\frac{1}{\sqrt{3}} & \frac{\cos \theta }{\sqrt{2}}-\frac{\sin \theta }{\sqrt{%
6}}e^{-i\sigma }%
\end{array}%
\right)  \label{TM2}
\end{equation}%
where in the PDG parametrization, we have to add the Majorana phase matrix%
\textrm{\ }$\mathcal{U}_{m}=diag(1,e^{\frac{i}{2}\alpha _{21}},e^{\frac{i}{2}%
\alpha _{31}})$ and where the\textrm{\ }$\alpha $'s are related to the $\eta
$'s of footnote 4 like $\alpha _{21}=\eta _{1}-\eta _{2}$ and\textrm{\ }$%
\alpha _{31}=\eta _{1}-\eta _{3}$\textrm{. }Therefore, using this matrix $%
\tilde{U}_{TM_{2}}$ and the definitions%
\begin{equation}
\sin \theta _{13}=\left\vert \tilde{U}_{e3}\right\vert \quad ,\quad \sin
^{2}\theta _{23}=\frac{\left\vert \tilde{U}_{\mu 3}\right\vert ^{2}}{%
1-\left\vert \tilde{U}_{e3}\right\vert ^{2}}\quad ,\quad \sin ^{2}\theta
_{12}=\frac{\left\vert \tilde{U}_{e2}\right\vert ^{2}}{1-\left\vert \tilde{U}%
_{e3}\right\vert ^{2}}
\end{equation}%
we obtain the following expressions of the three $\theta _{13}$, $\theta
_{23}$, and $\theta _{12}$ mixing angles%
\begin{eqnarray}
\sin ^{2}\theta _{12} &=&\frac{1}{3(1-\sin ^{2}\theta _{13})}\quad ,\quad
\sin ^{2}\theta _{13}=\frac{2}{3}\sin ^{2}\theta  \notag \\
\sin ^{2}\theta _{23} &=&\frac{1}{2}\pm \frac{\sin \theta _{13}\left( 1-%
\frac{3}{2}\sin ^{2}\theta _{13}\right) ^{1/2}}{\sqrt{2}\left( 1-\sin
^{2}\theta _{13}\right) }\cos \sigma  \label{mixi}
\end{eqnarray}%
Recall that the phase $\sigma $\ is related to the Dirac phase as\textrm{\ }$%
\sin \sigma =-\sin 2\theta _{23}\sin \delta _{CP}$\textrm{\ }with \cite{K4}%
\begin{equation}
\cos \delta _{CP}=\frac{\cos 2\theta _{13}\cos 2\theta _{23}}{\sin 2\theta
_{23}\sin \theta _{13}\left( 2-3\sin ^{2}\theta _{13}\right) ^{1/2}}.
\label{novo}
\end{equation}%
We can see from these equations that the atmospheric angle $\theta _{23}$\
is related to the reactor angle $\theta _{13}$,\textrm{\ }thus, it is clear
that for non zero $\theta _{13}$,\ the atmospheric neutrino mixing angle $%
\theta _{23}$\ oscillates around its maximal value---$\theta _{23}=\pi /4$%
---which is in agreement with the recent result from the NOvA experiment
\cite{NOV}. By setting $m_{0}=Y_{0}^{2}\upsilon _{u}^{2}/\left\vert
b\right\vert ,$The light neutrino masses in Eq. (\ref{ev}) can be rewritten
as follows%
\begin{equation}
m_{1}=m_{0}\frac{k_{b\varepsilon }}{k_{b\varepsilon }+k_{a\varepsilon }-%
\frac{1}{2}}\quad ,\quad m_{2}=m_{0}\frac{k_{b\varepsilon }}{k_{a\varepsilon
}+1}\quad ,\quad m_{3}=m_{0}\frac{k_{b\varepsilon }}{k_{b\varepsilon
}-k_{a\varepsilon }+\frac{1}{2}}  \label{MA}
\end{equation}%
The parameters $m_{0}$, $k_{a\varepsilon }$\ and $k_{b\varepsilon }$\ are
defined as three different ratios of the scalar field VEVs.\textrm{\ }A
discussion concerning the ranges of these parameters will be performed in
the next subsection.\textrm{\ }By using these masses, we calculate the solar
$\Delta m_{21}^{2}$\ and atmospheric $\Delta m_{31}^{2}$\ mass-squared
differences, we obtain%
\begin{equation}
\begin{array}{ccc}
\Delta m_{21}^{2} & = & m_{0}^{2}k_{b\varepsilon }^{2}\left[ \frac{1}{\left(
k_{a\varepsilon }+1\right) ^{2}}-\frac{1}{\left( k_{b\varepsilon
}+k_{a\varepsilon }-\frac{1}{2}\right) ^{2}}\right] \\
\left\vert \Delta m_{31}^{2}\right\vert & = & m_{0}^{2}k_{b\varepsilon
}^{2}\left\vert \frac{1}{\left( k_{b\varepsilon }-k_{a\varepsilon }+\frac{1}{%
2}\right) ^{2}}-\frac{1}{\left( k_{b\varepsilon }+k_{a\varepsilon }-\frac{1}{%
2}\right) ^{2}}\right\vert%
\end{array}
\label{dm}
\end{equation}%
and where the two types of neutrino spectrum NH and IH are associated
directly to the sign of\textrm{\ }$\Delta m_{31}^{2}$ which remains unknown.
\begin{figure}[th]
\begin{center}
\hspace{2.5em} \includegraphics[width=.55\textwidth]{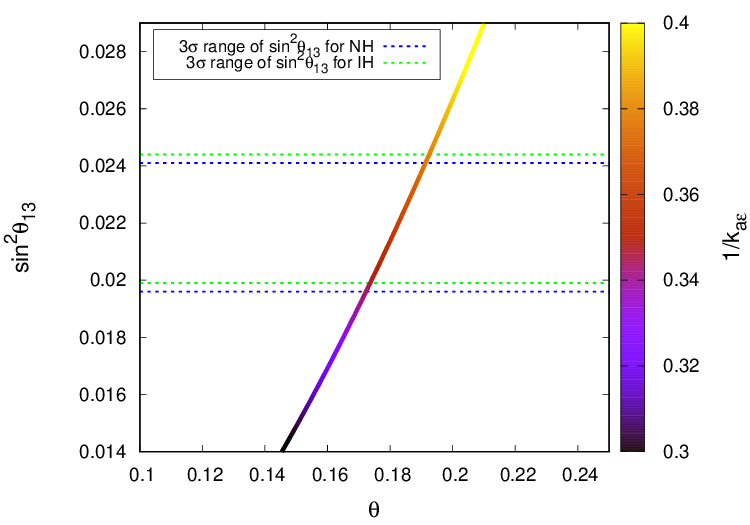}\quad
\end{center}
\par
\vspace{0.5cm}
\caption{The variation of $\sin ^{2}\protect\theta _{13}$ as a function of
the deviation parameter $\protect\theta $ with the parameter $k_{a\protect%
\epsilon }$ shown in the palette.}
\label{f0}
\end{figure}
Moreover, in order to reduce the intervals of $k_{a\epsilon }=\left\vert a/%
\mathrm{\epsilon }\right\vert $ and $\theta $, we plot in Fig. \ref{f0} $%
\sin ^{2}\theta _{13}$\ as a function of the deviation parameter $\theta $\
and $1/k_{a\epsilon }$\ induced by the VEV of the singlet $S$. The allowed
regions of $\theta $\ and $1/k_{a\epsilon }$\ are constrained by the $%
3\sigma $ experimental range of $\sin ^{2}\theta _{13}$, see Table \ref{C41}%
. Thus, the restricted regions of the parameters $\theta $\ and $%
1/k_{a\epsilon }$\ for NH (IH) are as follows%
\begin{equation}
\begin{tabular}{lllll}
$0.1726(0.1738)$ & $\lesssim $ & $\theta $ & $\lesssim $ & $0.1912(0.1923)$
\\
$0.344(0.346)$ & $\lesssim $ & $1/k_{a\epsilon }$ & $\lesssim $ & $%
0.377(0.379)$%
\end{tabular}%
\end{equation}

\subsection{Numerical analysis}

\begin{figure}[th]
\begin{center}
\hspace{2.5em} \includegraphics[width=.44\textwidth]{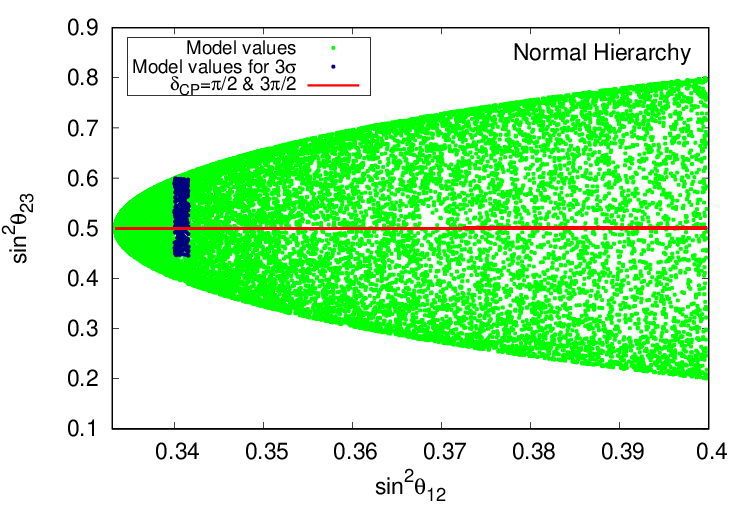}\quad %
\includegraphics[width=.44\textwidth]{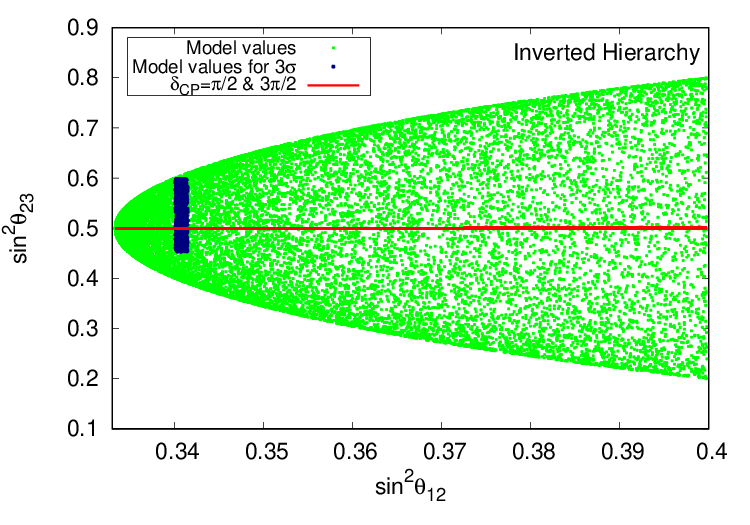}
\end{center}
\par
\vspace{0.5cm}
\caption{Correlation plots between $\sin ^{2}\protect\theta _{23}$ and $\sin
^{2}\protect\theta _{12}$ for NH (left panel) and IH (right panel). The
green region stands for all model values, while the blue region concerns the
allowed model values for 3$\protect\sigma $ range taken from Table \protect
\ref{C41}.}
\label{f2}
\end{figure}

\begin{figure}[th]
\begin{center}
\hspace{2.5em} \includegraphics[width=.44\textwidth]{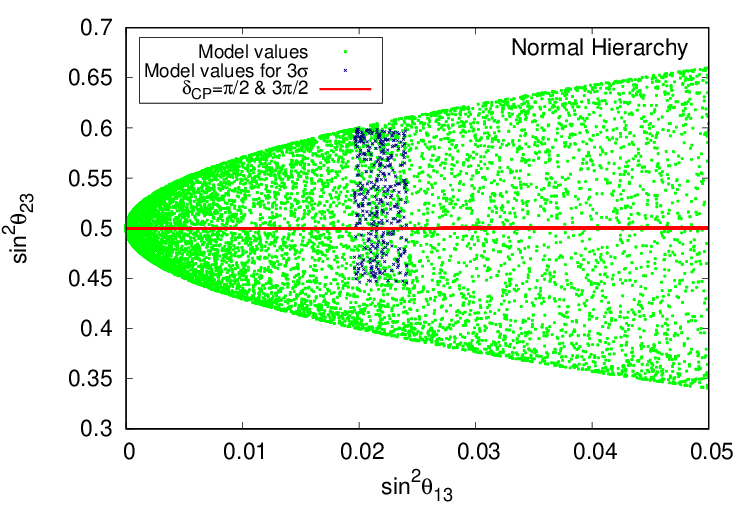}\quad %
\includegraphics[width=.44\textwidth]{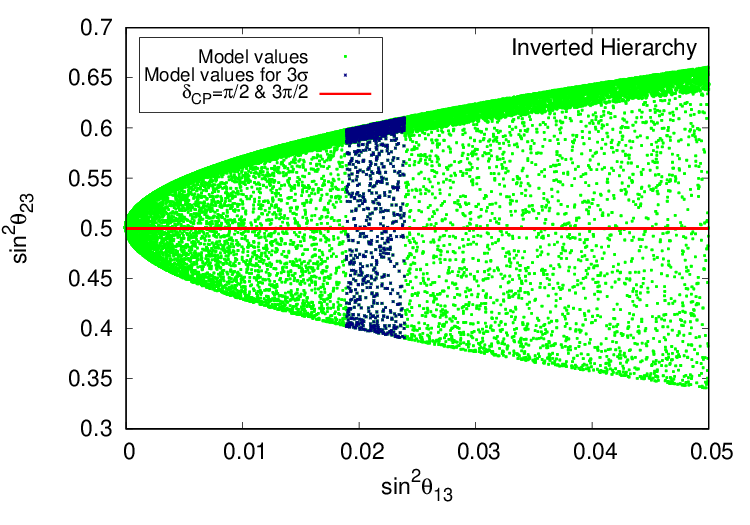}
\end{center}
\par
\vspace{0.5cm}
\caption{Same as in Fig. \protect\ref{f2} but for $\sin ^{2}\protect\theta %
_{13}$ instead of $\sin ^{2}\protect\theta _{12}$.}
\label{f3}
\end{figure}

The neutrino oscillation experiments are known to be sensitive to the three
mixing angles $\theta _{12}$, $\theta _{23}$, $\theta _{13}$\ and to the
neutrino mass-squared differences $\Delta m_{31}^{2}$\ for atmospheric
neutrinos and $\Delta m_{21}^{2}$\ for solar neutrinos.\textrm{\ }The
measurements of these angles as well as the\textrm{\ }mass-squared
differences were reported by several global fits of neutrino data \cite%
{R11,R12,R13}; see Table \ref{C41}.
\begin{table}[h]
\centering \renewcommand{\arraystretch}{1.2}
\begin{tabular}{|c|c|c|c|}
\hline
Oscillation parameters & Ordering & Best fit & $3\sigma $ \\ \hline
$\sin ^{2}\theta _{12}/10^{-1}$ & NH \& IH & $3.20$ & $2.73-3.79$ \\ \hline
$\sin ^{2}\theta _{23}/10^{-1}$ & $%
\begin{array}{c}
\text{NH} \\
\text{IH}%
\end{array}%
$ & $%
\begin{array}{c}
5.47 \\
5.51%
\end{array}%
$ & $%
\begin{array}{c}
4.45-5.99 \\
4.53-5.98%
\end{array}%
$ \\ \hline
$\sin ^{2}\theta _{13}/10^{-2}$ & $%
\begin{array}{c}
\text{NH} \\
\text{IH}%
\end{array}%
$ & $%
\begin{array}{c}
2.16 \\
2.22%
\end{array}%
$ & $%
\begin{array}{c}
1.96-2.41 \\
1.99-2.44%
\end{array}%
$ \\ \hline
$\Delta m_{21}^{2}$ $\mathrm{[10^{-5}eV}^{2}\mathrm{]}$ & NH \& IH & $7.55$
& $7.05-8.14$ \\ \hline
$\left \vert \Delta m_{31}^{2}\right \vert $ $\mathrm{[10}^{-3}\mathrm{eV}%
^{2}\mathrm{]}$ & $%
\begin{array}{c}
\text{NH} \\
\text{IH}%
\end{array}%
$ & $%
\begin{array}{c}
2.50 \\
2.42%
\end{array}%
$ & $%
\begin{array}{c}
2.41-2.60 \\
2.31-2.51%
\end{array}%
$ \\ \hline
$\delta ^{\circ }/\pi $ & $%
\begin{array}{c}
\text{NH} \\
\text{IH}%
\end{array}%
$ & $%
\begin{array}{c}
1.21 \\
1.56%
\end{array}%
$ & $%
\begin{array}{c}
0.87-1.94 \\
1.12-1.94%
\end{array}%
$ \\ \hline
\end{tabular}%
\caption{The best-fit values and the 3$\protect\sigma $ ranges of neutrino
oscillation parameters as reported by ref. \protect\cite{R13} for both
hierarchies.}
\label{C41}
\end{table}
In Fig. \ref{f2}, we show the correlation between the atmospheric neutrino
angle $\sin ^{2}\theta _{23}$\ and the solar neutrino angle $\sin ^{2}\theta
_{12}$ for the normal (left) and inverted (right) mass hierarchies where the
green points present all the possible model values while the blue points are
the allowed model values restricted by the $3\sigma $\ ranges of the mixing
angles given in Table \ref{C41}. The red line in both panels corresponds to
the values $\delta _{CP}=\frac{\pi }{2},\frac{3\pi }{2}$ which is consistent
with the maximal value of the atmospheric angle. We observe that the allowed
intervals for the mixing angles $\theta _{12}$ and $\theta _{23}$ are
\textrm{\ }%
\begin{equation}
0.34\lesssim \sin ^{2}\theta _{12}\lesssim \mathrm{\ }0.3415\mathrm{\quad },%
\mathrm{\quad }0.445\lesssim \sin ^{2}\theta _{23}\lesssim \mathrm{\ }0.598
\label{i2}
\end{equation}%
for NH; and for IH
\begin{equation}
0.34\lesssim \sin ^{2}\theta _{12}\lesssim \mathrm{\ }0.3415\mathrm{\quad },%
\mathrm{\quad }0.453\lesssim \sin ^{2}\theta _{23}\lesssim \mathrm{\ }0.597
\end{equation}%
We see that the interval of $\sin ^{2}\theta _{12}$\ gets more restrained
compared to its $3\sigma $ region while the intervals of $\sin ^{2}\theta
_{23}$ are slightly restricted compared to their $3\sigma $ region reported
in Table \ref{C41}.
The same discussion applies to the correlation between the atmospheric angle
$\sin ^{2}\theta _{23}$\ and the reactor angle $\sin ^{2}\theta _{13}$\
shown in Fig. \ref{f3}. In this case, the obtained intervals for $\theta
_{23}$ are the same as in Eq. (\ref{i2}), while the angle $\theta _{13}$
coincides with its $3\sigma $\ allowed range for NH and gets more restricted
in the case of IH; the range for the later case is $0.0199\lesssim \sin
^{2}\theta _{13}\lesssim \mathrm{\ }0.0237.$

\begin{figure}[th]
\begin{center}
\hspace{2.5em} \includegraphics[width=.44\textwidth]{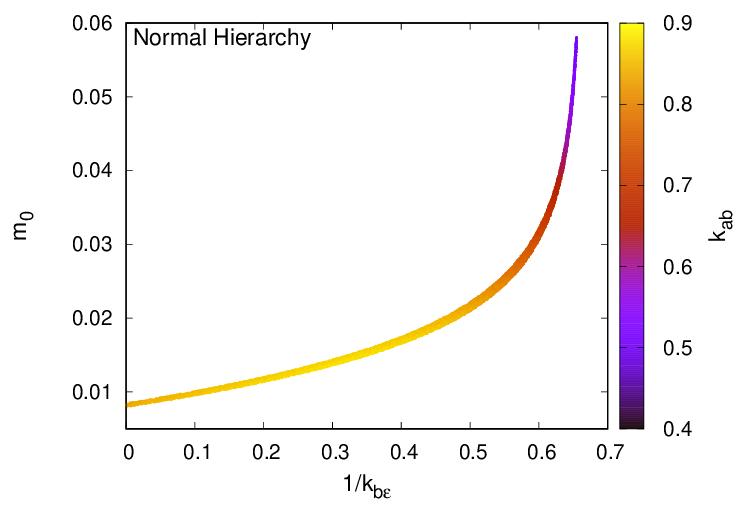}\quad %
\includegraphics[width=.44\textwidth]{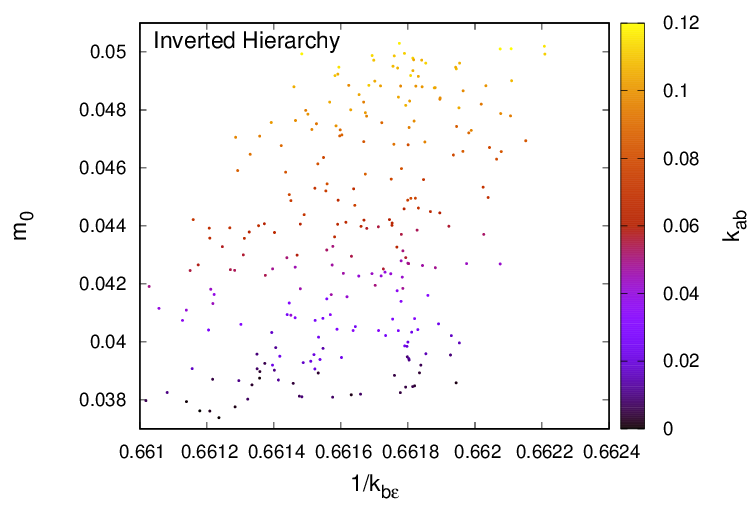}
\end{center}
\par
\vspace{0.5cm}
\caption{Scatter plot showing the allowed regions of the $m_{0}$-$d$-$c$
parameters for the normal (left panel) and inverted (right panel) mass
hierarchy.}
\label{f33}
\end{figure}
In Fig. \ref{f33}, we show for both mass hierarchies the correlation among
the parameters $m_{0}$, $k_{ab}$ and $1/k_{b\epsilon }$ where we used as
input parameters the current $3\sigma $\ allowed range of the mass squared
differences. As a result, we find that the obtained ranges are more
restricted in the IH case than the NH case. In order to get estimations\ on\
these parameters, we take the best fit value of $\Delta m_{31}^{2}$\ and $%
\Delta m_{21}^{2}$ and extract the corresponding values for $m_{0},$ $k_{ab}$
and $1/k_{b\epsilon }$. Their intervals as well as their best fit values are
as summarized\ in Table \ref{C42}.
\begin{table}[h]
\centering\renewcommand{\arraystretch}{1.2}
\begin{tabular}{|r|r|r|}
\hline
Model parameters & NH \ \ \ \ \ \ \  & IH \ \ \ \ \ \ \ \ \ \  \\ \hline
$%
\begin{array}{c}
m_{0}\text{[eV]} \\
\text{Best fit of }m_{0}\text{[eV]}%
\end{array}%
$ \  & $\ \ \
\begin{array}{c}
\left[ 0.8\times 10^{-2},0.058\right] \\
\simeq 0.0259%
\end{array}%
$ \ \ \  & $%
\begin{array}{c}
\left[ 0.037,0.05\right] \\
\simeq 0.044%
\end{array}%
$ \ \ \ \ \ \  \\ \hline
$\ \ \
\begin{array}{c}
k_{ab} \\
\text{Best fit of }k_{ab}%
\end{array}%
$ \ \  & $%
\begin{array}{c}
\left[ 0.49,0.88\right] \\
\simeq 0.788%
\end{array}%
$ \ \ \ \ \ \ \ \  & $\ \ \
\begin{array}{c}
\left[ 0.37\times 10^{-4},0.12\right] \\
\simeq 0.0617%
\end{array}%
$\ \  \\ \hline
$%
\begin{array}{c}
1/k_{b\epsilon } \\
\text{Best fit of }1/k_{b\epsilon }%
\end{array}%
$\ \ \  & $\
\begin{array}{c}
\left[ 0.16\times 10^{-3},0.65\right] \\
\simeq 0.558%
\end{array}%
$ \ \ \  & $%
\begin{array}{c}
\left[ 0.66,0.6622\right] \\
\simeq 0.6617%
\end{array}%
$ \ \ \  \\ \hline
\end{tabular}%
\caption{Allowed ranges of our model parameters as well as their best fit
values respecting the current $3\protect\sigma $ ranges of the mass squared
differences.}
\label{C42}
\end{table}
In what follows, we use these intervals as input parameters to plot the
correlations between masses.

\section{Phenomenological implications}

In this section, we first focus on the physical observables related to
experiments other than those of the neutrino oscillation experiments. These
are the Majorana neutrino mass $m_{ee}$ measured in neutrinoless double $%
\beta $-decay experiments and the neutrino electron mass $m_{\nu _{e}}$
measured in tritium beta decay experiments. Then, we provide predictions
concerning the Dirac $CPV$ phase $\delta _{CP}$ and its correlation with the
Jarlskog parameter $J_{CP}$ and the atmospheric angle $\theta _{23}$.

\subsection{Neutrino mass spectrum\emph{\ }}

\begin{figure}[th]
\begin{center}
\hspace{2.5em} \includegraphics[width=.44\textwidth]{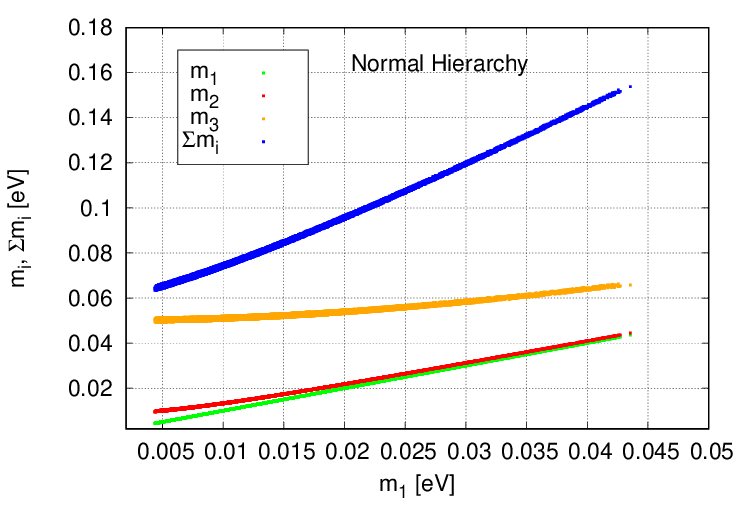}\quad %
\includegraphics[width=.44\textwidth]{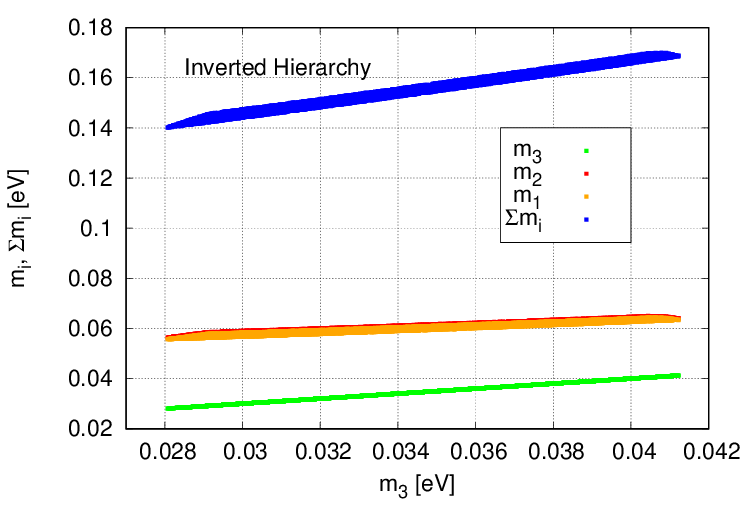}
\end{center}
\par
\vspace{0.5cm}
\caption{Prediction for the absolute neutrino masses and their sum $\sum
m_{i}$ as a function of $m_{1}(m_{3})$ for NH (IH) in the left (right)
panels. In both plots, $m_{1},$ $m_{2},$ $m_{3}$ and $\sum m_{i}$ described
by green, red, orange and blue respectively.}
\label{f6}
\end{figure}

We begin by investigating through scatter plots the neutrino mass spectrum
where we use the recent upper bound on the sum of neutrino masses from the
Planck collaboration;\textrm{\ }$\sum m_{i}\leq 0.17$ $\mathrm{eV}$ \cite%
{R35}, and the $3\sigma $\ allowed ranges of\ the mass squared differences $%
\Delta m_{21}^{2}$\ and $\Delta m_{31}^{2}$\ given in Table \ref{C41}.
Moreover, in order to fix all three neutrino masses, we express the masses $%
m_{2}$\ and $m_{3}$\ in terms of the lightest neutrino mass $m_{1}$\textrm{\
}as\textrm{\ }$\sqrt{m_{1}^{2}+\Delta m_{21}^{2}}$\textrm{\ }and\textrm{\ }$%
\sqrt{m_{1}^{2}+\Delta m_{31}^{2}}$\textrm{\ }respectively in the NH case,
while in the\textrm{\ }IH case we express $m_{1}$\ and $m_{2}$\ in terms of
the lightest neutrino mass $m_{3}$ as\textrm{\ }$\sqrt{m_{3}^{2}-\Delta
m_{31}^{2}}$\textrm{\ }and\textrm{\ }$\sqrt{m_{3}^{2}+\Delta
m_{21}^{2}-\Delta m_{31}^{2}}$\textrm{\ }respectively. Hence, we plot in
Fig. \ref{f6} the exact neutrino masses $m_{1},$\ $m_{2}$\ and $m_{3}$\
(denoted by the green, red and orange dots respectively) given in Eq. (\ref%
{MA}) and their sum $\sum m_{i}$\ (denoted by the blue dots) as a function
of the lightest neutrino mass ($m_{1}$ for NH and $m_{3}$ for IH). The
neutrino oscillation parameters are taken within their currently allowed $%
3\sigma $ ranges. For NH (left panel), the allowed range for the lightest
neutrino mass is given by $m_{1}\approx 0.0043-0.043$ $\mathrm{eV}$, while
for IH (right panel) the lightest neutrino mass lies in the range $%
m_{3}\approx 0.0282-0.041$ $\mathrm{eV}$. The allowed ranges of the
remaining masses for both mass hierarchies are as summarized in Table \ref%
{MSS}.
As for the sum of neutrino masses, in the normal hierarchy case---blue dots
in the left panel of Fig. \ref{f6}---we have $0.063$\emph{\ }$\mathrm{eV}%
\lesssim \sum m_{i}\lesssim 0.153$ $\mathrm{eV}$, while in the inverted
hierarchy case---blue dots in the right panel of Fig. \ref{f6}---we have%
\textrm{\ }$0.14\emph{\ }\mathrm{eV}\lesssim \sum m_{i}\lesssim 0.17$ $%
\mathrm{eV}$.\ From these ranges,\textrm{\ }one can suggest easily the
possibility to obtain more restricted regions of the inverted hierarchy, or
even exclude it if the sum of the light neutrino masses\ is less than $0.14$
\textrm{eV} in the future experiments. Indeed, strong cosmological studies%
\footnote{%
For more studies on constraining the sum of the light neutrino masses in
cosmological models, see Ref. \cite{R48} and the references therein.} on $%
\sum m_{i}$ provided the bound $\sum m_{i}<0.12$ $\mathrm{eV}$ at 95\% C.L.
\cite{R49,A22,A23,A24} (A more recent paper provided the bound $\sum
m_{i}=0.11\pm 0.03$ $\mathrm{eV}$ \cite{R50}).
\begin{table}[h]
\centering \renewcommand{\arraystretch}{1.2}
\begin{tabular}{|c|c|c|}
\hline
Neutrino masses & NH & IH \\ \hline
$m_{1}$ & $[0.0043-0.043]$ & $[0.0556-0.064]$ \\ \hline
$m_{2}$ & $[0.0094-0.044]$ & $[0.0562-0.065]$ \\ \hline
$m_{3}$ & $[0.0493-0.066]$ & $[0.0282-0.041]$ \\ \hline
\end{tabular}%
\caption{Our model predictions for the three light neutrino masses.}
\label{MSS}
\end{table}
Furthermore, to determine the best estimates on the neutrino masses and
their sum, we use the best fit values of the neutrino oscillation parameters
given in Table \ref{C41} as well as the best fit values of the parameters $%
m_{0}$,\ $c$, and $d$ given in Table \ref{C42}. We find for NH$\mathrm{\ }$%
\begin{equation}
m_{1}\simeq 0.0055\mathrm{\ eV\quad },\mathrm{\ \quad }m_{2}\simeq 0.0102%
\mathrm{\ eV\mathrm{\quad },\mathrm{\ \quad }\ }m_{3}\simeq 0.0503\mathrm{\
eV}
\end{equation}%
while for IH we find$\mathrm{\ \qquad }$%
\begin{equation}
m_{1}\simeq 0.0606\mathrm{\ eV\mathrm{\quad },\mathrm{\ \quad }\ }%
m_{2}\simeq 0.0612\mathrm{\ eV\mathrm{\quad },\mathrm{\ \quad }\ }%
m_{3}\simeq 0.035\mathrm{\ eV.}
\end{equation}%
Consequently, we remark that the masses in the normal hierarchy case are
lighter compared to the inverted hierarchy case, and thus our \emph{FNMSSM}
tends to favor the normal mass hierarchy over the inverted one.\textrm{\ }%
This can be seen also by calculating the best fit of\textrm{\ }$\sum m_{i}$
where we find $\sum m_{i}=0.06614$ $\mathrm{eV}$ for NH and $\sum
m_{i}=0.1573$ $\mathrm{eV}$ for IH. Thus, the later case\textrm{\ }is
excluded if we consider the aforementioned bounds from cosmological studies.

\subsection{Absolute mass scale and the nature of neutrinos}

The neutrino oscillation experiments are insensitive to the absolute mass of
neutrinos which is still unknown to this day. However, there are several
nonoscillation parameters that are sensitive to the absolute neutrino mass
scale, in particular, the effective neutrino electron mass $m_{\nu _{e}}$
that can be measured in tritium beta decay experiments, and the effective
Majorana mass $m_{ee}$ measured in neutrinoless double beta decay
experiments which will provide information on the nature of neutrinos as
well.

\begin{figure}[th]
\begin{center}
\includegraphics[width=.5\textwidth]{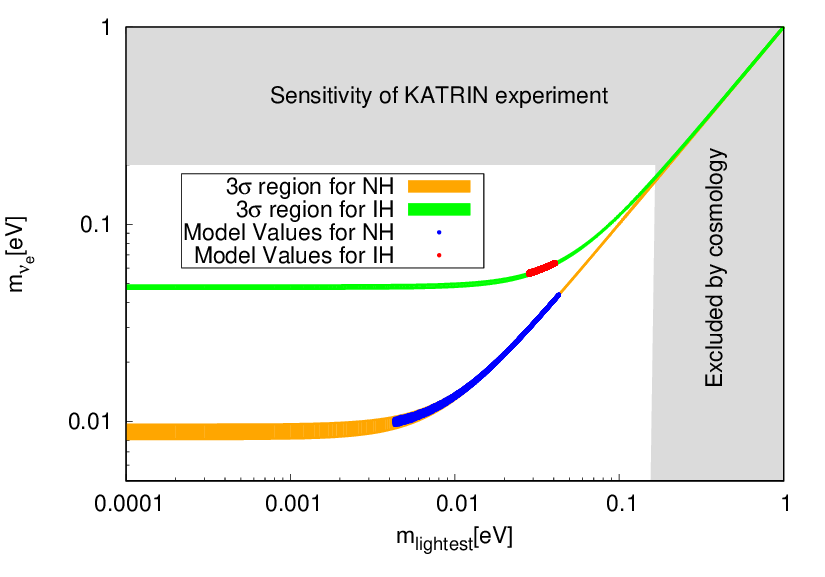}\quad
\end{center}
\caption{{}The effective neutrino mass $m_{\protect\nu _{e}}$\ as function
of the lightest neutrino mass $m_{i}$. The 3$\protect\sigma $ allowed
regions for $m_{\protect\nu _{e}}$ are represented by the orange (green)
colors for NH (IH). The blue and red points refer to the predicted model
values for NH and IH, respectively.}
\label{f7}
\end{figure}

\textbf{(i)}\emph{\ Tritium beta decay:} The study of electron spectrum near
its endpoint in tritium beta decay is one of the best methods to probe the
neutrino mass scale. The KATRIN experiment is the current generation of
direct neutrino mass measurement which is designed to measure the effective
electron neutrino mass with a sensitivity of $m_{\nu _{e}}<0.2$ $\mathrm{eV}$
(at 90 \% C.L.) \cite{R51}. The effective electron neutrino mass is defined
as
\begin{equation}
m_{\nu _{e}}^{2}=\sum_{i}\left\vert U_{ei}\right\vert ^{2}m_{i}^{2}
\end{equation}%
where $m_{i}$ are the three neutrino masses and $U_{ei}$ are the elements of
the first row of the PMNS matrix \cite{R52}. In our \emph{FNMSSM }prototype
where the mixing matrix is given by Eq. (\ref{TM2}), $m_{\nu _{e}}^{2}$ is
given by
\begin{equation}
m_{\nu _{e}}^{2}=\frac{1}{3}\left( 2m_{1}^{2}\cos ^{2}\theta
+m_{2}^{2}+2m_{3}^{2}\sin ^{2}\theta \right) .
\end{equation}%
By using this equation and the mass relations for both hierarchies given
after Table \ref{C42}, we plot in Fig. \ref{f7} the parameter $m_{\nu _{e}}$
as a function of the lightest neutrino mass. The orange region (green
region) is obtained by varying all the input parameters in their $3\sigma $
ranges for NH (IH) while our model values are presented by the blue points
(the red points). As a result, we find that the effective electron neutrino
mass lie in the ranges%
\begin{equation}
\begin{tabular}{lll}
NH: $m_{\nu _{e}}(\mathrm{eV})\in \left[ 0.00959-0.0439\right] $ & $;$ & IH:
$m_{\nu _{e}}(\mathrm{eV})\in \left[ 0.0554-0.0638\right] .$%
\end{tabular}%
\end{equation}%
Although the obtained intervals of $m_{\nu _{e}}$ are compatible with
current data, the anticipated future sensitivity from Project 8 experiment
is as low as $0.04$ $\mathrm{eV}$ \cite{R53}, which means that if no signal
is observed around this value in the future there is a good chance to
exclude the inverted mass hierarchy (the smallest value of $m_{\nu _{e}}$ in
our model is $0.0554$ $\mathrm{eV}$).

\textbf{(ii)}\emph{\ Neutrinoless double beta decay:} The most sensitive
probe of the Majorana nature of neutrinos is provided by the neutrinoless
double beta decay experiments $\left( 0\nu \beta \beta \right) $ which would
also provide a measurement of the absolute neutrino mass scale if it were
observed. Furthermore, since the lepton number $L$ is violated in $0\nu
\beta \beta $ processes, its discovery would provide a theoretical evidence
of physics beyond the SM---which conserves $L$---and would also provide a
sign of the seesaw mechanisms which imply the existence of a Majorana
neutrino mass term that violates $L$. The decay rate\ of $0\nu \beta \beta $%
\ is proportional to the square of the effective Majorana mass $\left\vert
m_{ee}\right\vert $ defined as $\left\vert m_{ee}\right\vert =\left\vert
\sum_{i}U_{ei}^{2}m_{i}\right\vert $ where $m_{i}$ and $U_{ei}$ are as
defined in the case of the effective electron mass $m_{\nu _{e}}$.
\begin{figure}[th]
\begin{center}
\hspace{2.5em} \includegraphics[width=.44\textwidth]{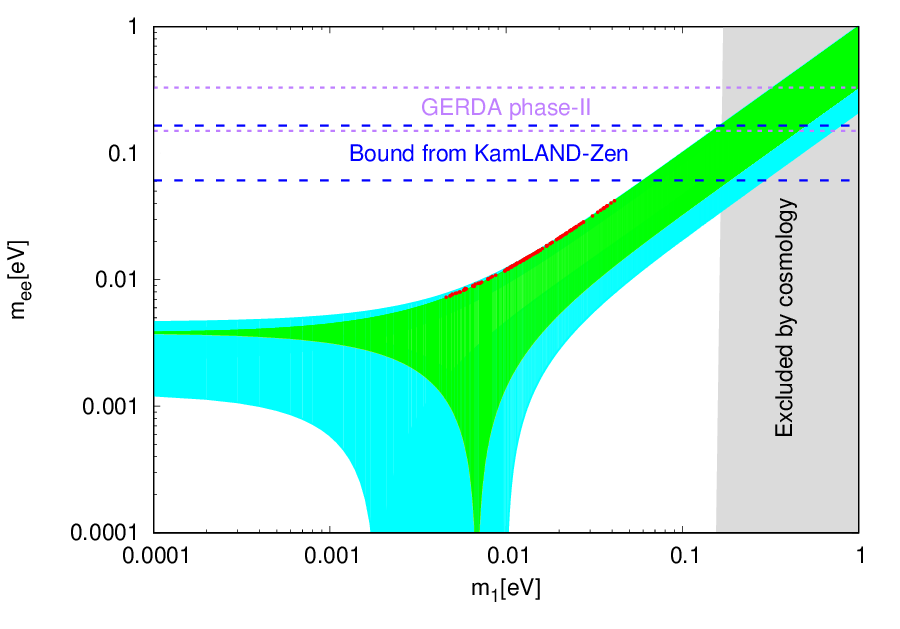}\quad %
\includegraphics[width=.44\textwidth]{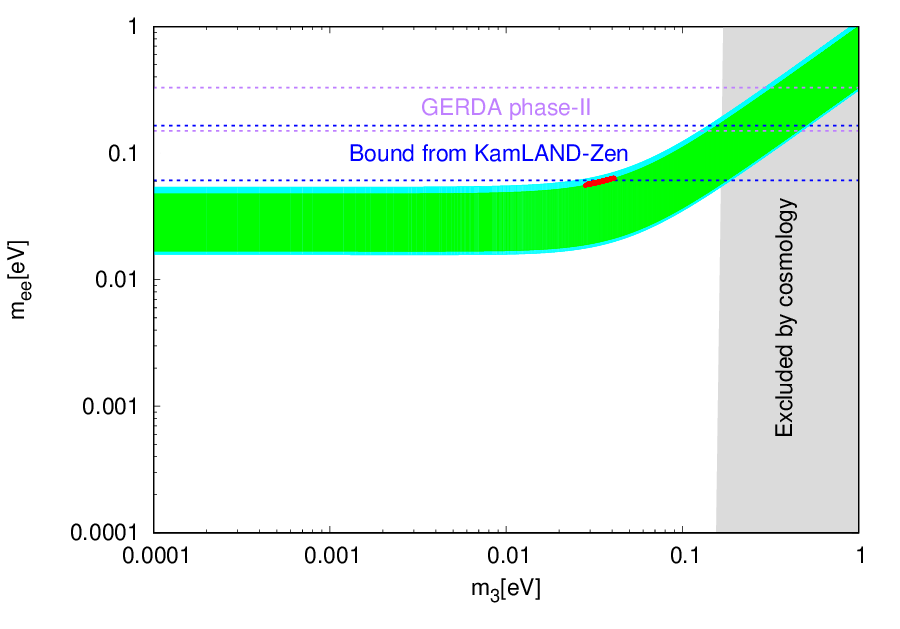}
\end{center}
\par
\vspace{0.5cm}
\caption{$m_{ee}$ as a function the lightest neutrino mass for NH (left
panel) and IH (right panel). The colored cyan regions stand for the $3%
\protect\sigma $ allowed ranges of $\left\vert m_{ee}\right\vert $ by
neutrino oscillation data given in Table \protect\ref{C41} while the green
colored region presents the best fit values. The horizontal dashed lines
refer to the upper bound on $\left\vert m_{ee}\right\vert $ from GERDA
\protect\cite{R58} and KamLAND-ZEN \protect\cite{R59}, while the vertical
gray band present the upper limit bound on the sum of the neutrino masses
from Planck collaboration. The red points are our predicted model values.}
\label{f8}
\end{figure}
In our \emph{FNMSSM}, we substitute $U_{ei}^{2}$ in $\left\vert
m_{ee}\right\vert $ by their expressions in the mixing matrix $\tilde{U}%
_{TM_{2}}\times \mathrm{diag}(1,e^{\frac{i}{2}\alpha _{21}},e^{\frac{i}{2}%
\alpha _{31}})$ (\ref{TM2}) where $\alpha _{21}$\textrm{\ }and\textrm{\ }$%
\alpha _{31}$\textrm{\ }are the Majorana $CP$ violating phases, we find%
\begin{equation}
\left\vert m_{ee}\right\vert =\frac{1}{3}\left\vert 2m_{1}\cos ^{2}\theta
+m_{2}e^{i\alpha _{21}}+2m_{3}\sin ^{2}\theta e^{i\left( \alpha
_{31}-2\sigma \right) }\right\vert .
\end{equation}%
Several experiments such as CUORE \cite{R54}, EXO \cite{R55}, NEXT \cite{R56}%
, SuperNEMO \cite{R57}, GERDA \cite{R58}, and KamLAND-ZEN \cite{R59} are
searching for signatures of $0\nu \beta \beta $. The most recent bounds of $%
\left\vert m_{ee}\right\vert $ come from the KamLAND-Zen and GERDA
experiments which are given respectively by%
\begin{equation}
\left\vert m_{ee}\right\vert <0.061-0.165\text{ }\mathrm{eV}\quad ,\quad
\left\vert m_{ee}\right\vert <0.15-0.33\text{ }\mathrm{eV.}
\end{equation}%
Similar to the above discussion on $m_{\nu _{e}}$, we plot in Fig. \ref{f8} $%
\left\vert m_{ee}\right\vert $ as a function of the lightest neutrino mass
for both mass hierarchies and we allow the Majorana phases\textrm{\ }$\alpha
_{21}$\ and\textrm{\ }$\alpha _{31}$\textrm{\ }to vary in the range $[0-2\pi
]$. The cyan regions are obtained by varying all the input parameters in
their $3\sigma $ ranges, the green regions correspond to the best fit values
while our model values are presented by the red points. Consequently, we
find in the normal (inverted) hierarchy case that the effective Majorana
mass lie in the ranges%
\begin{equation}
\begin{tabular}{lll}
NH: $\left\vert m_{ee}(\mathrm{eV})\right\vert \in \left[ 0.007271-0.04202%
\right] $ & $;$ & IH: $\left\vert m_{ee}(\mathrm{eV})\right\vert \in \left[
0.05519-0.0641\right] $%
\end{tabular}%
\end{equation}%
We see that in the normal hierarchy case the allowed range of $\left\vert
m_{ee}\right\vert $ as well as its corresponding lightest neutrino mass $%
m_{1}$ are smaller than the ranges in the inverted hierarchy case.
Furthermore, the obtained region of the effective Majorana mass in the IH
case can be reached in future experiments like KamLAND-Zen\textrm{\ }which
plans to reach a sensitivity below $0.04$ $\mathrm{eV}$ on $\left\vert
m_{ee}\right\vert $ \cite{R60}. As for the NH case, the obtained lower value
($\sim 0.007$ $\mathrm{eV}$) is far from any current or future planed
sensitivity, however, its higher values (around $0.03$ $\mathrm{eV}$) can be
reached in future experiments such as EXO-200, KamLAND2-Zen and GERDA Phase
II with the expected sensitivity given respectively by\textrm{\ }$\left(
0.015-0.025\right) $ $\mathrm{eV}$ \cite{R61},\emph{\ }$0.02$ $\mathrm{eV}$
\cite{R62} and $\left( 0.01-0.02\right) $ \textrm{eV }\cite{R63}.

\subsection{$CPV$ phase, Jarlskog invariant, and octant degeneracy}

\begin{figure}[th]
\begin{center}
\includegraphics[width=.44\textwidth]{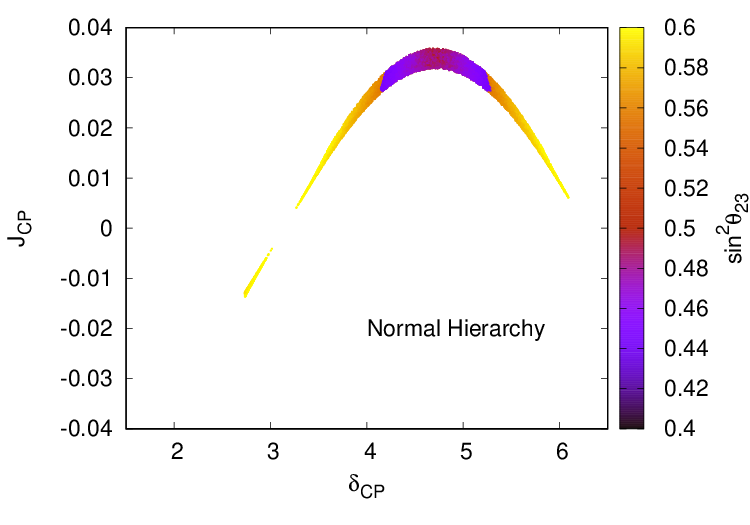} \includegraphics[width=.44%
\textwidth]{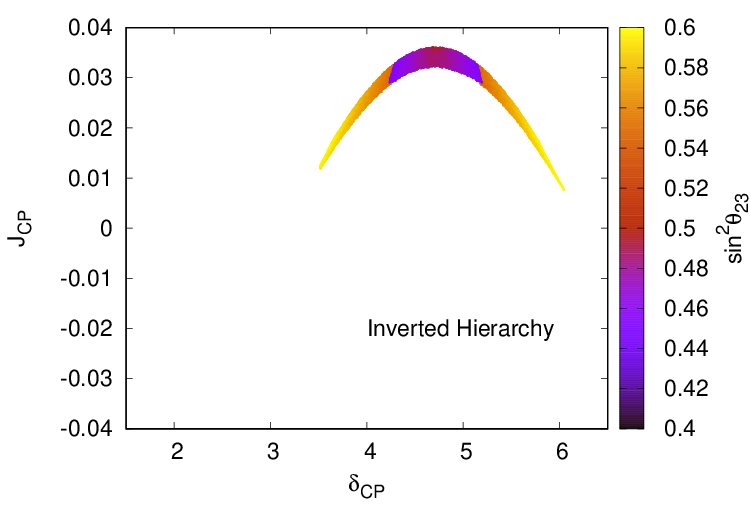} \includegraphics[width=.44\textwidth]{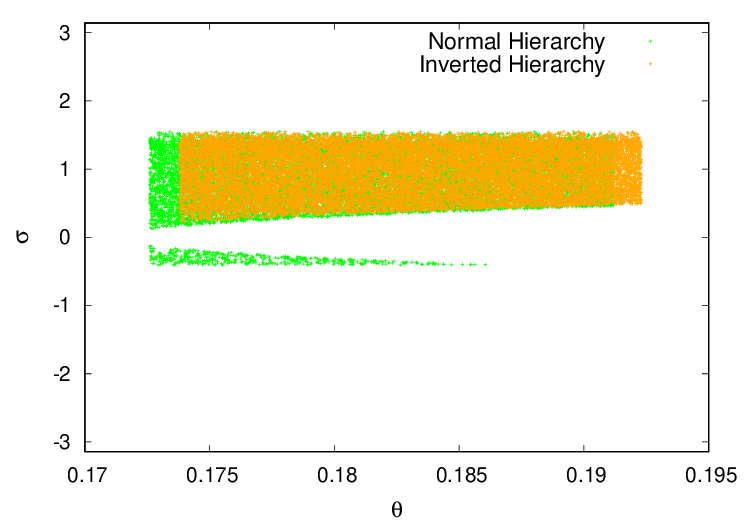}
\end{center}
\caption{This figure contains three plots; On the top, the left (right)
panel represents the predicted model values for Jarlskog invariant $J_{CP}$
as a function of the $CP$-violating phase $\protect\delta _{CP}$ with the
atmospheric angle $\sin ^{2}\protect\theta _{23}$ in the palette for NH (IH)
respectively. For NH, we have both positive and negative values for $J_{CP};$
while this invariant takes only positive ones for IH. On the bottom panel,
we plot the correlation between $\protect\sigma $\ and $\protect\theta $\
for NH (IH) represented by green (orange) points.}
\label{f15}
\end{figure}

After the discovery of the non-vanishing reactor angle, the observation of a
$CPV$ in the lepton sector is now possible. One way to measure it is by
means of the Jarlskog invariant parameter defined as\textrm{\ }$J_{CP}=\func{%
Im}(U_{e1}U_{\mu 1}^{\ast }U_{\mu 2}U_{e2}^{\ast })$. In the PDG standard
parametrization, this parameter is expressed in terms of the three mixing
angles and the Dirac $CP$ phase as follows \cite{R45}%
\begin{equation}
J_{CP}=\frac{1}{8}\sin 2\theta _{12}\sin 2\theta _{13}\sin 2\theta _{23}\cos
\theta _{13}\sin \delta _{CP}
\end{equation}%
By substituting the elements of the neutrino mixing matrix by their
expressions\ given in Eq. (\ref{TM2}), the Jarlskog invariant takes a
simpler form given by%
\begin{equation}
J_{CP}=-\frac{1}{3}(\sin ^{2}\theta _{23}-\frac{1}{2})(1-\sin ^{2}\theta
_{13})\tan \sigma  \label{jcp}
\end{equation}%
where we have expressed $\cos \theta \sin \theta $\ in terms of $\sin
^{2}\theta _{13}$ and $\sin ^{2}\theta _{23}$\ using Eq. (\ref{mixi}). In%
\textrm{\ }the bottom panel of Fig. \ref{f15}, we show the correlation
between $\sigma $\ and $\theta $\ where the green (orange) points stand for
NH (IH). The extracted range of $\sigma $\ is given by%
\begin{equation}
\begin{tabular}{cccc}
$\sigma \in $ & $\left[ -0.12\pi ,-3.9\times 10^{-2}\pi \right] \cup \left[
5\times 10^{-2}\pi ,\frac{\pi }{2}\right[ $ & for & NH \\
$\sigma \in $ & $\left[ 8.3\times 10^{-2}\pi ,\frac{\pi }{2}\right[ $ \ \ \
\ \ \ \ \ \ \ \ \ \ \ \ \ \ \ \ \ \ \ \ \ \ \ \ \ \ \ \ \ \ \ \ \ \ \  & for
& IH%
\end{tabular}
\label{val.sig}
\end{equation}%
\textrm{\ }It is clear from these ranges that\textrm{\ }$\sigma \neq 0$%
\textrm{\ }$\func{mod}\frac{\pi }{2}$\textrm{. }Therefore, it is clear now
from Eq. (\ref{jcp}) that $J_{CP}$\ cannot vanish because $\tan \sigma \neq
0 $\ and $\sin ^{2}\theta _{23}\neq \frac{1}{2}$, indicating that the $CP$
is always violated in our model; this result may be verified also by using
the relation\textrm{\ }$\sin \sigma =-\sin 2\theta _{23}\sin \delta _{CP}$.
To illustrate this feature further, we take into account the $3\sigma $\
allowed ranges of the oscillation parameters from Table \ref{C41}; and give
in the left\ and right panels of Fig. \ref{f15} the correlations between $%
J_{CP}$\ and $\delta _{CP}$ with $\sin ^{2}\theta _{23}$\ shown in the
palette for normal and inverted mass hierarchies respectively. It is clear
from this figure that $J_{CP}\neq 0$ implying that $CP$ is always violated.\
Numerically, we find that our model predicts $\delta _{CP}$ in the ranges
\begin{equation}
\delta _{CP}\in \left[ 0.87\pi ,0.96\pi \right] \cup \lbrack 1.04\pi ,1.5\pi
\lbrack \cup ]1.5\pi ,1.94\pi ]  \label{dcp1}
\end{equation}%
for NH, while it predicts the ranges
\begin{equation}
\delta _{CP}\in \lbrack 1.12\pi ,1.5\pi \lbrack \cup ]1.5\pi ,1.93\pi ]
\label{dcp2}
\end{equation}%
for IH. On the other hand, the maximal value of the Jarlskog invariant is $%
\left\vert J_{CP}\right\vert \simeq 0.0358$ $(0.036)$\ for NH (IH) which is
compatible with the upper bound on $\left\vert J_{CP}\right\vert _{\max
}\lesssim 0.03599$ $(0.03624)$ obtained using the current data\ on\ the
neutrino oscillation parameters\textrm{\ }\cite{R13}\textrm{.}\emph{\ }
It is clear from Fig. \ref{f15} that both octants of $\theta _{23}$\ are
permitted for both mass hierarchies; moreover, a nearly maximal atmospheric
mixing $\theta _{23}\approx \frac{\pi }{4}$\ favours nearly maximal $CP$
violation $\delta _{CP}\approx \frac{3\pi }{2}$\ while the\ value of\ $%
\delta _{CP}=\pi /2$---strongly disfavored by current data \cite{R11}---is
not permitted at all. The left panel in \ref{f15} shows that for the values
around $\delta _{CP}=0.87\pi $\ and $\delta _{CP}\approx \frac{3\pi }{2}$,
the $CP$ is maximally violated when the magnitude of $J_{CP}$\ is maximal ($%
J_{CP}\cong -0,0136$\ and $J_{CP}\cong 0.0358$) while in the right panel it
is maximally violated ($J_{CP}\cong 0,0136$\ and $J_{CP}\cong 0.036$) around
the values $\delta _{CP}=1.12\pi $\ and $\delta _{CP}\approx \frac{3\pi }{2}$%
\textrm{.}\newline
On the other hand, the main difficulty in determining the octant sensitivity
of the atmospheric angle $\theta _{23}$\ arises due to the unknown value of $%
\delta _{CP}$\textrm{. }Reading from Fig. \ref{f15}, the range of $\delta
_{CP}$\ corresponding to the HO for NH and IH is the same as the obtained
ones in Eqs. (\ref{dcp1}) and (\ref{dcp2}) respectively; while for the LO
case the range of $\delta _{CP}$\ corresponding to the NH (IH) is $\left[
1.31\pi ,1.67\pi \right] $\ ($\left[ 1.34\pi ,1.65\pi \right] $).

\section{Bypassing domain walls}

In this section, we study the problem of domain walls (DWs) induced by the
spontaneous breaking of the discrete flavor symmetry $G_{\mathrm{f}}=\mathbb{%
A}_{4}\times \boldsymbol{Z}_{3}$ in the\textrm{\ }flavored NMSSM with
building blocks as in Tables \ref{A}-\ref{C}. Then, we propose a way to
bypass these domains in terms of a perturbation by higher dimensional
operators suppressed by powers of the Planck scale $M_{Pl}$, induced by
supergravitational effect, and by using effective field action $\Gamma
_{eff} $ approach. \newline
More precisely, we first study the spontaneous breaking of the non-Abelian
discrete $\mathbb{A}_{4}$ group factor of $G_{\mathrm{f}}$ down to its
particular subgroups $\mathbb{Z}_{3}$ and $\mathbb{Z}_{2}$\ in (\ref{a4})
since these breaking patterns are key points in the study of the DWs in the%
\textrm{\ }\emph{FNMSSM}. Actually, the breaking $\mathbb{A}_{4}\rightarrow
\mathbb{Z}_{3}$ analysis concerns the charged lepton sector and completes
Eq. (\ref{mm}); while the breaking $\mathbb{A}_{4}\rightarrow \mathbb{Z}_{2}$
deals with the neutrino sector and deepens the analysis of Eqs. (\ref{1})
and (\ref{mr1}). Then, we discuss the induced domain walls in both charged
and chargeless lepton sectors\textrm{.} After that, we provide a solution to
this DW issue through the explicit breaking of the full discrete flavor
group by higher dimensional operators contributing to $\Gamma _{eff}$.

\subsection{Domain walls setup}

The problem of domain walls in the framework of models based on the $\mathbb{%
A}_{4}$\ discrete symmetry has been discussed in\textrm{\ }\cite{A25,A26}\
and recently in\ \cite{A27,A28,A29}. Here we consider our $\emph{FNMSSM}$
prototype\textrm{\ }with chiral superfield content as given in Tables \ref{A}%
-\ref{C} and\textrm{\ }describe some basic ingredients to approach domain
wall problem in lepton sector by using discrete symmetry group of the flavor
vacua. In Secs. II and III, we showed that the generation of the lepton
masses and their mixing angles\textrm{\ }arises from the\ spontaneous
breaking of two kinds of symmetries: \textbf{(i)} the continuous $SU\left(
2\right) _{L}\times U\left( 1\right) _{Y}$ gauge symmetry by giving the VEV $%
\upsilon _{d}$ to the $H_{d}$\ Higgs superfield as in Eq. (\ref{lm}); and
\textbf{(ii)} the discrete flavor symmetry $G_{\mathrm{f}}=\mathbb{A}%
_{4}\times \boldsymbol{Z}_{3}$ by giving non zero VEVs to some components of
$\Phi _{i},\Omega _{i},\boldsymbol{S}$ \footnote{%
Below we will mainly focus on the properties of the non-Abelian part $%
\mathbb{A}_{4}$ of the flavor symmetry $G_{f}$; the Abelian factor $%
\boldsymbol{Z}_{3}$ can be implemented in a straightforward manner.}. To
handle these flavon superfields, we will imagine the sets $\Phi _{i},\Omega
_{i},\boldsymbol{S}$ as components of complex 4d vector space expansions
like
\begin{equation}
\Phi \sim \sum_{i=1}^{3}\Phi _{i}X_{i}\qquad ,\qquad \Omega \sim
\sum_{i=1}^{3}\Omega _{i}X_{i}\qquad ,\qquad \boldsymbol{S}\sim SX_{4}
\label{fl}
\end{equation}%
with $X_{i}$'s standing for a system of vector basis with some flavon space
metric $h_{ij}=\left\langle X_{i}|X_{j}\right\rangle $; the first three
directions $\left( X_{1},X_{2},X_{3}\right) $ are for the $\mathbb{A}_{4}$-
triplets and the fourth $X_{4}$ for singlets. As these chiral
superfields---collectively denoted below by $\Upsilon $---play an important
role in the analysis of DWs, we start by giving three interesting comments
that will be used to approach DWs in the \textquotedblleft
flavon\textquotedblright\ field space. The first comment concerns the above
formal\ 4d- vector expansions (\ref{fl}) which are complex valued
developments while we will need real expansions when looking for
representing DWs by real quivers. The two other comments concern useful
features regarding the structure of the DWs viewed from space- time as well
as their properties with respect to subgroups of $G_{\mathrm{f}}$.

\ \

$\bullet $ \emph{First comment: flavon scalar potential and real VEV quivers
}\newline
\textrm{\ }The superfields we are using here are chiral super functions $%
\Upsilon =\xi _{\Upsilon }+\theta \psi _{\Upsilon }+\theta ^{2}F_{\Upsilon }$%
\ carrying in general a quantum number of\textrm{\ }$G_{\mathrm{f}}$ but no
gauge charge restricting therefore the gauge invariant kinetic energy
density $\sum \left\vert \mathcal{D}_{\mu }\xi _{\Upsilon }\right\vert ^{2}$
just to the ordinary $\sum \left\vert \partial _{\mu }\xi _{\Upsilon
}\right\vert ^{2}$ with no couplings with vector gauge potentials A$_{\mu
}^{su_{2}\times u_{1}}$. However, there are interactions---among themselves
and with other chiral superfields of the model---governed by $G_{\mathrm{f}}$
and contributing to the full classical scalar potential $\mathcal{V}_{tot}$
of the \emph{FNMSSM} by some real function of the $\xi _{\Upsilon }$ scalar
fields to which we refer hereafter to as
\begin{equation}
\begin{tabular}{lll}
$\mathcal{\tilde{V}}$ & $=$ & $\mathcal{\tilde{V}}\left( \xi _{\Phi },\xi
_{\Omega },\xi _{S};..\right) $ \\
& $:=$ & $\mathcal{\tilde{V}}\left( \xi _{\Upsilon },\bar{\xi}_{\Upsilon
}\right) $%
\end{tabular}
\label{fg}
\end{equation}%
and whose supersymmetric part can be read from $\sum \left\vert F_{\Upsilon
}\right\vert ^{2}$. Recall that, the flavor symmetry $G_{\mathrm{f}}$ has
been implemented by adding an index to the flavon superfields like $\Upsilon
_{i}$; for example the\textrm{\ }$\Phi $ flavon is a triplet under $\mathbb{A%
}_{4}\subset G_{\mathrm{f}}$; so it involves three chiral superfields $\Phi
_{1},\Phi _{2},\Phi _{3}$ that couple to each other and to other chiral
superfield of the theory in $G_{\mathrm{f}}$-invariant manner like in (\ref%
{27}). Notice also that because of supersymmetry the superfields\ in the
expansion Eq. (\ref{fl}) are complex chiral superfields, and their scalar
components; in particular the lowest $\xi _{\Upsilon }\sim \sum \xi
_{\Upsilon _{i}}X_{i}$---which may have VEVs $\left\langle \xi _{\Upsilon
_{i}}\right\rangle $\ that can be used to break symmetries---are also
complex quantities. This complex nature of field variables is somehow a
disturbing thing for later use; especially when searching a real graphic
representation of domain walls. To work around this difficulty, it is
interesting to use $\mathbb{C}\sim \mathbb{R}^{2}$\textrm{\ }and split the
complex flavon scalars $\xi _{\Upsilon }$ and the $X_{k}$- generators like $%
\xi =\func{Re}\xi +i\func{Im}\xi $\ and \textrm{\ }%
\begin{equation}
X_{k}\sim \left(
\begin{array}{c}
U_{k} \\
0%
\end{array}%
\right) \qquad ,\qquad i\otimes X_{k}\sim \left(
\begin{array}{c}
0 \\
V_{k}%
\end{array}%
\right) \qquad ,\qquad k=1,2,3,4  \label{r8}
\end{equation}%
\textrm{\ }where $U_{k}$ and $V_{k}$ are real 4d vectors.\textrm{\ }With
this real splitting, we can put the above complex 4d expansions of $\mathbb{A%
}_{4}$- flavonic representations with respect to\textrm{\ }$X_{i}$'s as real
8-dimensional vectors like\textrm{\ }%
\begin{equation}
\xi _{\Upsilon }\sim \sum_{k}\left( \func{Re}\xi _{\Upsilon _{k}}\right)
U_{k}+\sum_{k}\left( \func{Im}\xi _{\Upsilon _{k}}\right) V_{k}  \label{uv}
\end{equation}%
For example, the real VEV $\upsilon _{\Phi }$ of the flavon $\Phi $ used in
the derivation of Eq. (\ref{mm}) may be imagined as the modulus of a real 6d
vector in $\mathbb{R}^{6}\subset \mathbb{R}^{8}$ generated by (\ref{r8}).
Then, geometrically speaking, the VEV $\langle \Phi \rangle $ can be
represented by a vector $\mathbf{\varphi }_{1}U_{1}$ pointing in the first
real $U_{1}$ direction of the $\left( U_{k},V_{k}\right) $ vector space. The
actions on $\langle \Phi \rangle \mathbf{=\varphi }_{1}U_{1}:=\mathbf{%
\varphi }$ by the two generators $\mathcal{S}$ and $\mathcal{T}$ of the
flavor group $\mathbb{A}_{4}$ generate in general a polygon in the $\left(
U_{k},V_{k}\right) $ space with \emph{12} vertices given by%
\begin{equation}
\begin{tabular}{lllllllllll}
$\mathbf{\varphi }$ & $,$ & $\mathcal{ST}\mathbf{\varphi }$ & $,$ & $%
\mathcal{TST}\mathbf{\varphi }$ & $,$ & $\mathcal{T}^{2}\mathcal{ST}\mathbf{%
\varphi }$ & $,$ & $\mathcal{S}\mathbf{\varphi }$ & $,$ & $\mathcal{TS}%
\mathbf{\varphi }$ \\
$\mathcal{STS}\mathbf{\varphi }$ & $,$ & $\mathcal{T}^{2}\mathcal{S}\mathbf{%
\varphi }$ & $,$ & $\mathcal{T}\mathbf{\varphi }$ & $,$ & $\mathcal{T}^{2}%
\mathbf{\varphi }$ & $,$ & $\mathcal{ST}^{2}\mathbf{\varphi }$ & $,$ & $%
\mathcal{TST}^{2}\mathbf{\varphi }$%
\end{tabular}%
\end{equation}%
This polygonal graph may be interpreted as the DW quiver viewed from the
flavon space $\mathfrak{F}$ with completely broken $\mathbb{A}_{4}$. Notice
that for the flavon triplet $\Phi $, this $\mathfrak{F}$ space is
parametrized by the scalar fields $\xi _{\Phi _{1}},\xi _{\Phi _{2}},\xi
_{\Phi _{3}}$ and so is isomorphic to $\mathbb{C}^{3}\simeq \mathbb{R}^{6}$.
Notice also that DW quivers we are concerned with below correspond to
partial breaking of $\mathbb{A}_{4}$ down to a subgroup $H_{f}$; and then do
not have \emph{12} vertices; but a lower number given by the order of the
broken part of the\textrm{\ }$\mathbb{A}_{4}$ symmetry denoted by\textrm{\ }$%
K_{f}$. In fact we will encounter two kinds of real quivers that are
associated with two particular spontaneous breaking patterns namely: $%
\mathbb{A}_{4}$ down to $\mathbb{Z}_{3}$, and $\mathbb{A}_{4}$ down to $%
\mathbb{Z}_{2}$; by using $\mathbb{A}_{4}\cong \mathbb{V}_{4}\rtimes \mathbb{%
Z}_{3}$, the broken parts of these breakings are respectively given by $%
\mathbb{V}_{4}$ and $\mathbb{Z}_{2}\rtimes \mathbb{Z}_{3}$ \cite{A27}. For
the first breaking we have $\mathcal{T}\langle \Phi \rangle =\langle \Phi
\rangle $, then the above \emph{12} vertices reduce to \emph{4} ones
defining the vertices of a tetrahedronal quiver in $\mathfrak{F}$ (see
subsection V B); and for the second breaking, we have $\mathcal{S}\langle
\Phi \rangle =\langle \Phi \rangle $ and then the \emph{12} vertices reduce
to \emph{6} vertices defining an octahedronal quiver (see subsection V C).

$\bullet $ \emph{Second comment: Quantum numbers and domain walls }\newline
\ The quantum numbers $n_{G_{\mathrm{f}}}$ of the flavon superfields under
the non-Abelian discrete symmetry $G_{\mathrm{f}}$\ were given in Tables \ref%
{B}-\ref{C} in terms of the characters of the representation group
generators $\mathcal{S}$ and $\mathcal{T}$. Recall that the two flavons $%
\Phi $ and $\Omega $\ transform\ in the same $3_{(-1,0)}$ representation of%
\textrm{\ }$\mathbb{A}_{4}$\textrm{, }but they are distinguished by the
extra discrete\textrm{\ }$\boldsymbol{Z}_{3}$. The\textrm{\ }$S$\textrm{\ }%
is a non trivial singlet under both\textrm{\ }$\mathbb{A}_{4}$ and $%
\boldsymbol{Z}_{3}$\textrm{, }and so a linear term of this superfield in the
effective superpotential $\mathcal{W}_{eff}\left( \Phi ,\Omega ,S\right) $
breaks completely the\textrm{\ }$G_{\mathrm{f}}$\textrm{\ }flavor symmetry.
This\textrm{\ }$S$- property will be used later on in overcoming the domain
wall problem in the neutrino sector by using effective scalar potential
approach; see subsection V D for details.\newline
Moreover, under the breaking of $G_{\mathrm{f}}$ down to a given subgroup $%
H_{f}$; these $n_{G_{\mathrm{f}}}$ quantum numbers get split to two subsets $%
n_{H_{f}}$ and $n_{K_{f}}$; the $n_{H_{f}}$ for $H_{f}$ and the $n_{K_{f}}$
for the broken subsymmetry; the second subset $n_{K_{f}}$ is important for
describing the DWs; it will be used to deal with the two following things:
\newline
\textbf{(}a\textbf{)} the indexing of the split degenerate vacua $%
\left\langle \xi _{\Upsilon }\right\rangle _{a}$ of the potential $\mathcal{%
\tilde{V}}\left( \xi _{\Upsilon },\bar{\xi}_{\Upsilon }\right) $; for
convenience, we shall denote these degenerate vacua like $\left\langle \xi
_{\Upsilon }\right\rangle _{a}:\equiv \mathbf{\varphi }_{a}$ with values for
the subindex $a=1,...,n_{K_{f}}$. For the example of $H_{f}=\mathbb{Z}_{3}$,
the index $a$ takes the values $1,2,3,4$; and for $H_{f}=\mathbb{Z}_{2},$ it
takes the values $1,..,6$.\newline
(b) the study of the properties of the network of domain walls viewed from
the flavon space $\mathfrak{F}$; in particular the $l_{ab}$ DWs
interpolating between the various $\mathbf{\varphi }_{a}$- vacua whose space
time description correspond to \cite{B2,B3,B4,R10}%
\begin{equation}
l_{ab}=x\left( \mathbf{\varphi }_{b}\right) -x\left( \mathbf{\varphi }%
_{a}\right)
\end{equation}%
This space time quantity $l_{ab}$ can be interpreted as the width\footnote{%
\ For a consistent interpretation of $l_{ab}$ as a space-time width, we need
finite values of the fields at space infinity. This requires however
introducing some input parameters in the scalar potential that controls the
finitude of value of the fields at space infinity as remarkably done in \cite%
{B4}; see also \cite{A26,A28} for other approaches. We suspect that these
methods might be applied as well to our construction provided implementation
of extra nonrenormalizable terms in the superpotential although technically
is more laborious.} of the DWs since it vanishes for $\mathbf{\varphi }_{a}=%
\mathbf{\varphi }_{b}$; i.e: $l_{aa}=0$. A simple explicit expression in
terms of the fields $\xi _{\Upsilon }$ and $\bar{\xi}_{\Upsilon }$ that
defines the above $l_{ab}$ DWs is given by the following integral formula in
the space $\mathfrak{F}$ of flavon scalars,\textrm{\ }%
\begin{equation}
l_{ab}=\pm \int_{\mathbf{\varphi }_{a}}^{\mathbf{\varphi }_{b}}dx\qquad
,\qquad dx=\frac{d\eta _{\Upsilon }}{\sqrt{\mathcal{E}+\mathcal{\tilde{V}}%
\left( \xi _{\Upsilon },\bar{\xi}_{\Upsilon }\right) }}  \label{dw}
\end{equation}%
with $\mathcal{E}$ is a constant standing for the energy density of a static
flavon configuration $\xi _{\Upsilon }$; and where $(d\eta _{\Upsilon
})^{2}=d\xi _{\Upsilon }d\bar{\xi}_{\Upsilon }$ is the metric\footnote{%
\ In complex space $\mathbb{C}^{n}$ with n coordinates $z_{l}=x_{l}+iy_{l}$,
the metric is given by $ds^{2}=\sum_{l}dz_{l}d\bar{z}_{l}$ and reads with
real variables as $\sum_{l}\left( dx_{l}\right) ^{2}+\left( dy_{l}\right)
^{2}.$}\textrm{\ }in the $\mathfrak{F}$ space parameterised by $\xi
_{\Upsilon },\bar{\xi}_{\Upsilon }$; see Eq. (\ref{dx}) below. A short way
to derive this $l_{ab}$ interpolating relation is to use the 4d space time
flavon dynamics described by the following Lagrangian density
\begin{equation}
\mathcal{L}_{4d}=\partial _{\mu }\bar{\xi}_{\Upsilon }\partial ^{\mu }\xi
_{\Upsilon }-\mathcal{\tilde{V}}\left( \xi _{\Upsilon },\bar{\xi}_{\Upsilon
}\right)
\end{equation}%
and look for a relation that links the space time differentials $dx^{\mu }$
and the flavon potential $\mathcal{\tilde{V}}\left( \xi _{\Upsilon },\bar{\xi%
}_{\Upsilon }\right) $. From the above $\mathcal{L}_{4d}$ one can derive the
flavon field equations of motion that we express like
\begin{equation}
\ddot{\xi}_{\Upsilon }-\vec{\nabla}^{2}\xi _{\Upsilon }+\frac{\partial
\mathcal{\tilde{V}}}{\partial \bar{\xi}_{\Upsilon }}=0
\end{equation}%
together with the complex conjugate. By considering time independent flavon
field configurations; i.e: $\dot{\xi}_{\Upsilon }=0$, and restricting the
static 3d- space to one space direction---planar DWs---by setting $\xi
_{\Upsilon }\left( t,x,y,z\right) \equiv \zeta _{\Upsilon }\left( x\right) $
with no $y$ nor $z$ dependence, the above $\mathcal{L}_{4d}$ Lagrangian
density reduces to $\mathcal{L}_{1d}=-\bar{\zeta}_{\Upsilon }^{\prime }\zeta
_{\Upsilon }^{\prime }-\mathcal{\tilde{V}}\left( \zeta _{\Upsilon },\bar{%
\zeta}_{\Upsilon }\right) $. In this case, the 1d- field equations of motion
resulting from $\mathcal{L}_{1d}$ read as follows
\begin{equation}
\frac{d^{2}\zeta _{\Upsilon }}{dx^{2}}-\frac{\partial \mathcal{\tilde{V}}}{%
\partial \bar{\zeta}_{\Upsilon }}=0\qquad ,\qquad \frac{d^{2}\bar{\zeta}%
_{\Upsilon }}{dx^{2}}-\frac{\partial \mathcal{\tilde{V}}}{\partial \zeta
_{\Upsilon }}=0
\end{equation}%
By multiplying the first equation by $\frac{d\bar{\zeta}_{\Upsilon }}{dx}$
and the second one by $\frac{d\zeta _{\Upsilon }}{dx}$; then, adding the two
obtained relations, we end with a conserved expression $\frac{d\mathcal{E}}{%
dx}=0$ with energy density like%
\begin{equation}
\mathcal{E}=\left( \frac{d\bar{\zeta}_{\Upsilon }}{dx}\right) \left( \frac{%
d\zeta _{\Upsilon }}{dx}\right) -\mathcal{\tilde{V}}\text{.}
\end{equation}%
This conserved expression leads in turn to a relationship between the $dx$
differential in x- space and the $d\eta _{\Upsilon }$ differential in the $%
\zeta _{\Upsilon }$- field space; it read as $\left( \mathcal{E}+\mathcal{%
\tilde{V}}\right) dx^{2}=d\eta _{\Upsilon }^{2}$ with $d\eta _{\Upsilon
}^{2}=d\bar{\zeta}_{\Upsilon }d\zeta _{\Upsilon }$ and; by taking the square
root, we then have
\begin{equation}
dx=\pm \frac{d\eta _{\Upsilon }}{\sqrt{\mathcal{E}+\mathcal{\tilde{V}}}}
\label{dx}
\end{equation}%
which by integration between two given vacua $(\mathbf{\varphi }_{a},\mathbf{%
\varphi }_{b})$ of the scalar potential, we discover (\ref{dw}). Later on,
we shall think of the simple expression $\frac{d\eta _{\Upsilon }}{\sqrt{%
\mathcal{E}+\mathcal{\tilde{V}}}}$ in (\ref{dx}) as a typical Hermitian
1-form $\varpi _{1}=\frac{1}{\sqrt{\mathcal{E}+\mathcal{\tilde{V}}}}d\eta
_{\Upsilon }$ in the flavon space $\mathfrak{F}$ that characterizes the DWs
interpolating between two vacua $\mathbf{\varphi }_{ab}=\left[ \mathbf{%
\varphi }_{a},\mathbf{\varphi }_{b}\right] $. So, we can also express (\ref%
{dw}) by using $\varpi _{1}$ language like
\begin{equation}
l_{ab}=\pm \int_{\left[ \mathbf{\varphi }_{a},\mathbf{\varphi }_{b}\right]
}\varpi _{1}
\end{equation}%
that is an integral of the 1-form $\varpi _{1}$ on a compact line $\mathbf{%
\varphi }_{ab}$ with boundaries given by the critical points of the flavon
potential. This formula is suggestive in the sense that permits to think of
generic DWs extending between p- vacua in $\mathfrak{F}$ as given by
multi-integrations over $\varpi _{p-1}$ forms on the flavon space $\mathfrak{%
F}$; for the cases of $p=3$ and $p=4$, see Eqs. (\ref{l},\ref{s},\ref{v})
given below.

\ \

$\bullet $ \emph{Third comment: }$\mathbb{V}_{4}$\emph{\ and }$\mathbb{Z}%
_{2}\rtimes \mathbb{Z}_{3}$\newline
The order of the discrete alternating $\mathbb{A}_{4}$ is equal to \emph{12}%
; the same as the product $4\times 3$ which is the order of semidirect
product of $\mathbb{V}_{4}\rtimes \mathbb{Z}_{3}$ where $\mathbb{V}_{4}$ is
the Klein group. This order four group $\mathbb{V}_{4}$ is Abelian and can
be expressed like the product of two $\mathbb{Z}_{2}$ copies; that is $%
\mathbb{V}_{4}\sim \mathbb{Z}_{2}\times \mathbb{Z}_{2}.$ To distinguish
these two $\mathbb{Z}_{2}$ copies, we shall think of $\mathcal{S}$ as the
generator of the first $\mathbb{Z}_{2}$ and use a different $\mathcal{S}%
^{\prime }$ to refer to the generator of the second copy that we denote like
$\mathbb{Z}_{2}^{\prime }$. Notice that according to Eq. (\ref{a4}), there
exists a third $\mathbb{Z}_{2}^{\prime \prime }$ subgroup inside the
alternating $\mathbb{A}_{4}$ whose generator will be denoted below as $%
\mathcal{S}^{\prime \prime }$ and is related to the generators of the two
other $\mathbb{Z}_{2}$'s like $\mathcal{S}^{\prime \prime }=\mathcal{SS}%
^{\prime }=\mathcal{S}^{\prime }\mathcal{S}.$\newline
In the basis where $\mathcal{T}$ is diagonal; i.e: $\mathcal{T}=\mathrm{diag}%
\left( 1,\bar{\omega},\omega \right) $, the matrix representations of these $%
\mathcal{S}$, $\mathcal{S}^{\prime }$ and $\mathcal{S}^{\prime \prime }$ are
given by
\begin{equation}
\begin{tabular}{lllllll}
$\mathcal{T}$ & $=$ & $\left(
\begin{array}{ccc}
1 & 0 & 0 \\
0 & \bar{\omega} & 0 \\
0 & 0 & \omega%
\end{array}%
\right) $ & \quad ,\quad & $\mathcal{S}$ & $=$ & $\frac{1}{3}\left(
\begin{array}{ccc}
-1 & 2 & 2 \\
2 & -1 & 2 \\
2 & 2 & -1%
\end{array}%
\right) $ \\
$\mathcal{S}^{\prime }$ & $=$ & $\frac{1}{3}\left(
\begin{array}{ccc}
-1 & 2\omega & 2\bar{\omega} \\
2\bar{\omega} & -1 & 2\omega \\
2\omega & 2\bar{\omega} & -1%
\end{array}%
\right) $ & \quad ,\quad & $\mathcal{S}^{\prime \prime }$ & $=$ & $\frac{1}{3%
}\left(
\begin{array}{ccc}
-1 & 2\bar{\omega} & 2\omega \\
2\omega & -1 & 2\bar{\omega} \\
2\bar{\omega} & 2\omega & -1%
\end{array}%
\right) $%
\end{tabular}
\label{sp}
\end{equation}%
satisfying the properties $\mathcal{S}^{2}=\mathcal{S}^{\prime 2}=\mathcal{S}%
^{\prime \prime 2}=I$ and
\begin{equation}
\mathcal{S}^{\prime }=\mathcal{TST}^{-1}\qquad ,\qquad \mathcal{S}^{\prime
\prime }=\mathcal{T}^{2}\mathcal{ST}^{-2}
\end{equation}%
Observe that the sum of entries of the rows of $\mathcal{S}$ and $\mathcal{S}%
^{\prime }$ matrices are equal; $\sum_{j}\mathcal{S}_{ij}=1$ and $\sum_{j}%
\mathcal{S}_{ij}^{\prime }=-1$, two properties that will be useful later on
when studying the breaking of $\mathbb{A}_{4}$ in the neutrino sector.
Observe also that by giving a VEV to the complex $\Phi $- flavon like%
\footnote{%
\ \ In the real basis (\ref{r8}), this VEV may be imagined as $\left\langle
\Phi \right\rangle ^{T}=\left( \upsilon _{\Phi },0,0,0,0,0\right) .$}
\begin{equation}
\left\langle \Phi \right\rangle =\upsilon _{\Phi }\left(
\begin{array}{c}
1 \\
0 \\
0%
\end{array}%
\right)  \label{tv}
\end{equation}%
the alternating $\mathbb{A}_{4}$ gets broken down to a subgroup $\mathbb{Z}%
_{3}$ with broken part given by the Klein group $\mathbb{V}_{4}$ that
characterizes the split vacua. The breaking $\mathbb{A}_{4}\rightarrow
\mathbb{Z}_{3}$ is explicitly exhibited on the matrix representation (\ref%
{sp}) of the generators of $\mathbb{A}_{4}$\textrm{\ }which shows that%
\textrm{\ }%
\begin{equation}
\begin{tabular}{lllllll}
$\mathcal{T}\left\langle \Phi \right\rangle $ & $=$ & $\left\langle \Phi
\right\rangle $ & $\qquad ,\qquad $ & $\mathcal{T}^{3}$ & $=$ & $I_{id}$ \\
$\mathcal{S}\left\langle \Phi \right\rangle $ & $\neq $ & $\left\langle \Phi
\right\rangle $ & $\qquad ,\qquad $ & $\mathcal{S}^{2}$ & $=$ & $I_{id}$ \\
$\mathcal{S}^{\prime }\left\langle \Phi \right\rangle $ & $\neq $ & $%
\left\langle \Phi \right\rangle $ & $\qquad ,\qquad $ & $\mathcal{S}^{\prime
2}$ & $=$ & $I_{id}$%
\end{tabular}%
\end{equation}%
where\textrm{\ $\mathcal{T}$ }generates\textrm{\ }$\mathbb{Z}_{3}$. With
this choice of VEV, one gives masses to the charged leptons that are
proportional to $\frac{\upsilon _{\Phi }}{\Lambda }$ as shown in the
superpotential (\ref{mm}); but also induces DWs to be studied with details
in next subsection. If instead of $\Phi $, we give a non zero VEV to the $%
\Omega $- flavon by choosing the three VEVs equal like\footnote{%
\ \ In the real basis (\ref{r8}), this VEV is given by $\left\langle \Omega
\right\rangle ^{T}=\upsilon _{\Omega }\left( 1,0,1,0,1,0\right) .$}
\begin{equation}
\left\langle \Omega \right\rangle =\upsilon _{\Omega }\left(
\begin{array}{c}
1 \\
1 \\
1%
\end{array}%
\right)  \label{om}
\end{equation}%
the discrete $\mathbb{A}_{4}$ symmetry group gets broken down to a $\mathbb{Z%
}_{2}$ subsymmetry with broken part behaving like $\mathbb{Z}_{3}\rtimes
\mathbb{Z}_{2}$; as there is no symmetric $\mathbb{S}_{3}$ inside $\mathbb{A}%
_{4}$. This breaking may be directly checked from Eq. (\ref{sp}) showing that%
\begin{equation}
\begin{tabular}{lllllll}
$\mathcal{S}\left\langle \Omega \right\rangle $ & $=$ & $\left\langle \Omega
\right\rangle $ & $\qquad ,\qquad $ & $\mathcal{S}^{2}$ & $=$ & $I_{id}$ \\
$\mathcal{T}\left\langle \Omega \right\rangle $ & $\neq $ & $\left\langle
\Omega \right\rangle $ & $\qquad ,\qquad $ & $\mathcal{T}^{3}$ & $=$ & $%
I_{id}$ \\
$\mathcal{S}^{\prime }\left\langle \Omega \right\rangle $ & $\neq $ & $%
\left\langle \Omega \right\rangle $ & $\qquad ,\qquad $ & $\mathcal{S}%
^{\prime 2}$ & $=$ & $I_{id}$%
\end{tabular}%
\end{equation}%
\textrm{\ }with\textrm{\ }$\mathcal{S}$\textrm{\ }generating\textrm{\ }$%
\mathbb{Z}_{2}$. Notice that the breaking of $\mathbb{A}_{4}$ down to a $%
\mathbb{Z}_{2}$ can be also realized by the nontrivial $\mathbb{A}_{4}$-
singlet $S$ whose VEV $\upsilon _{S}$ preserves $\mathbb{Z}_{2}$ since,
according to Table \ref{C} based on the characters of $\mathcal{S}$ and $%
\mathcal{T}$, we have
\begin{equation}
\mathcal{S}\left\langle S\right\rangle =\left\langle S\right\rangle \qquad
,\qquad \mathcal{T}\left\langle S\right\rangle =e^{\frac{2i\pi }{3}%
}\left\langle S\right\rangle
\end{equation}%
In what follows, we study separately these two breaking patterns $\mathbb{A}%
_{4}\rightarrow \mathbb{Z}_{3}$ (for charged leptons) and $\mathbb{A}%
_{4}\rightarrow \mathbb{Z}_{2}$ (for neutrinos); they are respectively
realised by the VEVs of the flavon triplets $\Phi $ and $\Omega $. As an
ultimate goal of this study, we aim to describe with details the induced
domain walls in the\textrm{\ }flavored NMSSM with building blocks as in
Tables \ref{A}-\ref{C} and provide a solution for the neutrino sector where
DWs are inevitable.

\subsection{Domain walls in charged lepton sector}

In this subsection we study the domain walls in the charged lepton sector in
the\textrm{\ }flavored NMSSM which are induced by the breaking $\mathbb{A}%
_{4}\rightarrow \mathbb{Z}_{3}$. We will show that the DWs in this sector
are not problematic due to our estimation of the lower bound of $\upsilon
_{\Phi }$ VEV which happens to be greater than the inflationary scale which
is around 10$^{14}$ GeV. To that purpose, we first describe the properties
of the building blocks in our flavored NMSSM prototype after the $\mathbb{A}%
_{4}\rightarrow \mathbb{Z}_{3}$ breaking. Then, we analyse the properties of
the split degenerate vacua in the space $\mathfrak{F}$ of flavons as well as
the structure of the induced domain walls.

\subsubsection{Breaking pattern of $\mathbb{A}_{4}$ to $\mathbb{Z}_{3}$}

We start from the list of Eq. (\ref{a4}) concerning the discrete alternating
$\mathbb{A}_{4}$ group that acts on a given four- states system say $\left\{
\left\vert X_{1}\right\rangle ,..,\left\vert X_{4}\right\rangle \right\} $.
From this table, we learn that $\mathbb{A}_{4}$ may be spontaneously broken
down to any one of the subgroups in (\ref{a4}) by choosing appropriate VEV
directions. In the case where $\mathbb{A}_{4}$\emph{\ }is broken down to one
of\emph{\ }the four possible\emph{\ }$\mathbb{Z}_{3}$'s living inside of $%
\mathbb{A}_{4}$; say to the subgroup given by
\begin{equation}
\mathbb{Z}_{3}=\left\{ I_{id},\left( 123\right) ,\left( 132\right) \right\}
\end{equation}%
with point $\left\vert X_{4}\right\rangle $ fixed, only the $\mathcal{T}$
generator of the $\mathbb{Z}_{3}$ subsymmetry of $\mathbb{A}_{4}$ survives;
the other $\mathcal{S}$ generator gets broken. By taking $\mathcal{T}=\left(
132\right) $; then we have $\left( 123\right) =\mathcal{T}^{-1}=\mathcal{T}%
^{2}$ and so $\mathbb{Z}_{3}$ is just the set
\begin{equation}
\mathbb{Z}_{3}=\left\{ I,\mathcal{T},\mathcal{T}^{2}\right\}
\end{equation}%
with $\mathcal{T}^{3}=I$. In this picture, $\mathcal{T}$ can be represented
by a diagonal 3$\times $3 matrix with eigenvalues $\left( 1,\bar{\omega}%
,\omega \right) $ and its typical eigenvectors $X_{q}^{\prime }$ are
respectively given by
\begin{eqnarray}
X_{1}^{\prime } &=&X_{1}+X_{2}+X_{3}  \notag \\
X_{2}^{\prime } &=&X_{1}+\omega ^{2}X_{2}+\omega X_{3}  \label{pr} \\
X_{3}^{\prime } &=&X_{1}+\omega X_{2}+\omega ^{2}X_{3}  \notag
\end{eqnarray}%
For later use, notice the three following features: (a) In the $%
X_{i}^{\prime }$- basis, the generator $\mathcal{S}$ acts as $\left(
X_{1},X_{2},X_{3}\right) \rightarrow \left( X_{1},-X_{2},-X_{3}\right) $ and
it is represented by a non diagonal matrix given by Eq. (\ref{sp}); this
change can be explicitly checked by computing $\mathcal{S}:X_{i}^{\prime
}\rightarrow \tilde{X}_{i}^{\prime }=\mathcal{S}_{ij}X_{j}^{\prime }$. (b)
The analogue of the character relation (\ref{ch}) of $\mathbb{A}_{4}$ reads
in the $\mathbb{Z}_{3}$ case as follows
\begin{equation}
\mathbb{Z}_{3}:3=\left( 1\right) ^{2}+\left( 1^{\prime }\right) ^{2}+\left(
1^{\prime \prime }\right) ^{2}
\end{equation}%
showing that there are three kinds of one- dimensional representations
which, by using characters, can be denoted like $\mathbf{1}=\mathbf{1}_{1},$
$\mathbf{1}^{\prime }=\mathbf{1}_{\omega }$ and $\mathbf{1}^{\prime \prime }=%
\mathbf{1}_{\bar{\omega}}$. Notice moreover that by breaking $\mathbb{A}_{4}$%
\emph{\ }down to the above\emph{\ }$\mathbb{Z}_{3}$, the $\mathcal{S}$
generator is no longer a conserved symmetry. So, the triplet $\mathbf{3}%
_{\left( -1,0\right) }$ should be imagined as $\mathbf{3}_{0}$ where we have
kept only the character of $\mathcal{T}$; and then can be decomposed in
terms of irreducible representations like
\begin{equation}
\mathbf{3}_{0}=\mathbf{1}_{1}\oplus \mathbf{1}_{\omega }\oplus \mathbf{1}_{%
\bar{\omega}}
\end{equation}%
since all irreducible representations of $\mathbb{Z}_{3}$ are one-
dimensional. The broken part of the symmetry is given by $\mathbb{V}_{4}$.
The new quantum numbers of the different superfields get replaced, after the
breaking $\mathbb{A}_{4}\rightarrow \mathbb{Z}_{3}$, by the ones defined in
Tables \ref{E} and \ref{F}.
\begin{table}[h]
\centering\renewcommand{\arraystretch}{1.2}
\begin{tabular}{|c|c|c|c|c|c|}
\hline
Superfields & $L_{i}$ & $e^{c}$ & $\mu ^{c}$ & $\tau ^{c}$ & $N_{i}^{c}$ \\
\hline
$\mathbb{Z}_{3}$ & $\mathbf{1}_{1}\oplus \mathbf{1}_{\omega }\oplus \mathbf{1%
}_{\bar{\omega}}$ & $\mathbf{1}_{\bar{\omega}}$ & $1_{\omega }$ & $1_{1}$ & $%
\mathbf{1}_{1}\oplus \mathbf{1}_{\omega }\oplus \mathbf{1}_{\bar{\omega}}$
\\ \hline
\end{tabular}%
\caption{Lepton and right-handed neutrino superfields and their quantum
numbers under $\mathbb{Z}_{3}$.}
\label{E}
\end{table}
\begin{table}[h]
\centering\renewcommand{\arraystretch}{1.2}
\begin{tabular}{|c|c|c|c|c|c|c|}
\hline
Superfields & $H_{u}$ & $H_{d}$ & $\Phi _{i}$ & $\Omega _{i}$ & $S$ & $%
\mathbf{\chi }$ \\ \hline
$\mathbb{Z}_{3}$ & $\mathbf{1}_{1}$ & $\mathbf{1}_{\omega }$ & $\mathbf{1}%
_{1}\oplus \mathbf{1}_{\omega }\oplus \mathbf{1}_{\bar{\omega}}$ & $\mathbf{1%
}_{1}\oplus \mathbf{1}_{\omega }\oplus \mathbf{1}_{\bar{\omega}}$ & $\mathbf{%
1}_{\bar{\omega}}$ & $1_{1}$ \\ \hline
\end{tabular}%
\caption{Higgs and flavon superfields and their quantum numbers under $%
\mathbb{Z}_{3}$.}
\label{F}
\end{table}
As a check, the monomials $S\nu _{e}^{c}\nu _{\mu }^{c}$, $S\nu _{\tau
}^{c}\nu _{\tau }^{c},$ $\nu _{e}^{c}\nu _{e}^{c}\Omega _{1},$ $\nu _{\mu
}^{c}\nu _{\tau }^{c}\Omega _{1},$ $\nu _{\tau }^{c}\nu _{\tau }^{c}\Omega
_{2},$ $\nu _{e}^{c}\nu _{\mu }^{c}\Omega _{2}$, $\nu _{\mu }^{c}\nu _{\mu
}^{c}\Omega _{3},$ $\nu _{e}^{c}\nu _{\tau }^{c}\Omega _{3}$, $\nu
_{e}^{c}\nu _{e}^{c}\mathbf{\chi }$ and $\nu _{\mu }^{c}\nu _{\tau }^{c}%
\mathbf{\chi }$ \ involved in Eq. (\ref{1}) are invariant under discrete $%
\mathbb{Z}_{3}$.

\subsubsection{Vacua and unproblematic domain walls}

\begin{figure}[tbph]
\begin{center}
\includegraphics[scale=0.4]{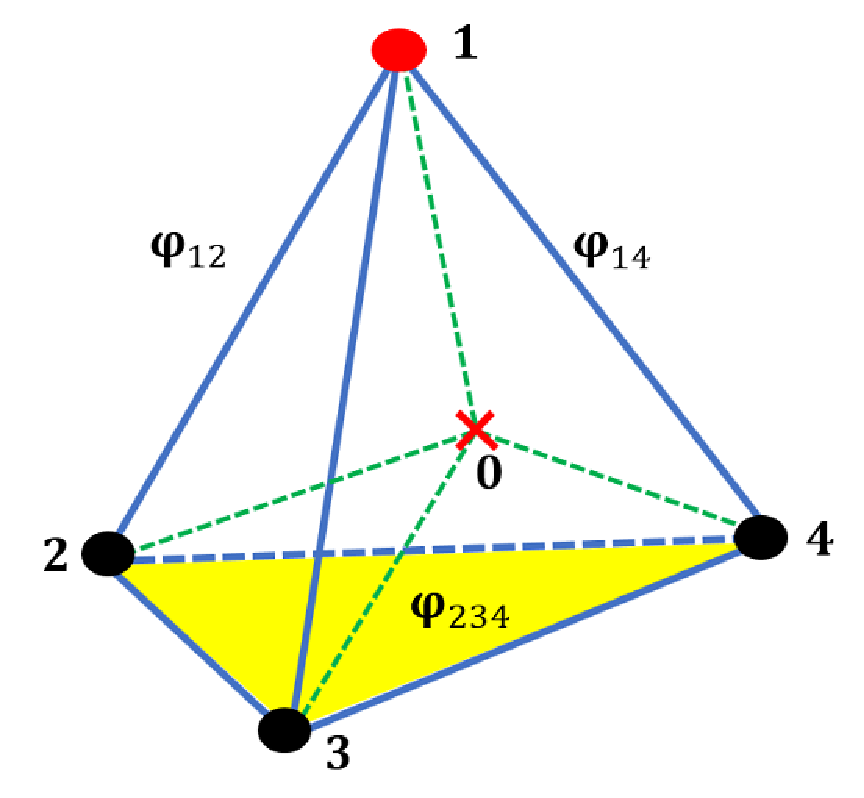}
\end{center}
\par
\vspace{-0.5cm}
\caption{Four vacua $\mathbf{\protect\varphi }_{a}$ related to each other by
$\mathbb{V}_{4}$ transformations. They define the four vertices of a
tetrahedron with $\mathbb{Z}_{3}$ axes $O\mathbf{\protect\varphi }_{a}$. It
is interpreted here as the DWs quiver in the flavon space $\mathfrak{F}$.
Dashed green lines represent the four $\mathbb{Z}_{3}$ subsymmetry axes of $%
\mathbb{A}_{4}$. }
\label{w1}
\end{figure}

After the breaking of the discrete $\mathbb{A}_{4}$ down to $\mathbb{Z}_{3}$%
, the flavon vacua $\left\langle \Phi \right\rangle $ sit in four degenerate
points $\mathbf{\varphi }_{1},\mathbf{\varphi }_{2},\mathbf{\varphi }_{3},%
\mathbf{\varphi }_{4}$ in the flavon space $\mathfrak{F}$. As the quantum
number of these vacua is given by $\mathbb{V}_{4}$;\textrm{\ }it follows
that the induced domain walls (\ref{dw}) are characterized by the Klein group%
\textrm{\ }$\mathbb{V}_{4}\sim \mathbb{Z}_{2}\times \mathbb{Z}_{2}^{\prime }$%
. One of the two $\mathbb{Z}_{2}$ factors is generated by the previous $%
\mathcal{S}$ while the other $\mathbb{Z}_{2}^{\prime }$ is generated by $%
\mathcal{S}^{\prime }$ whose expressions are given by (\ref{sp}). Thus, in
order to investigate the domain walls extending between the degenerate vacua
of the model created by the spontaneous breaking of the non-Abelian $\mathbb{%
A}_{4}$ group down to $\mathbb{Z}_{3}$, it is enough to use properties of
its Abelian subgroups $\mathbb{Z}_{2}$ and $\mathbb{Z}_{3}$. Indeed, by
thinking of $\mathbb{A}_{4}$ like $\mathbb{V}_{4}\rtimes \mathbb{Z}_{3}$ ,
the breaking of the flavor symmetry driven by the flavon triplet $\Phi $ in
the charged lepton sector may be expressed as%
\begin{equation}
\mathbb{V}_{4}\rtimes \mathbb{Z}_{3}\times \boldsymbol{Z}_{3}\qquad \overset{%
\left\langle \Phi \right\rangle }{\longrightarrow }\qquad \mathbb{Z}%
_{3}\times \boldsymbol{Z}_{3}.
\end{equation}%
So, the domain walls separating the $\mathbf{\varphi }_{1},\mathbf{\varphi }%
_{2},\mathbf{\varphi }_{3},\mathbf{\varphi }_{4}$ flavon vacua can be
completely characterized by the elements of $\mathbb{V}_{4}\simeq \mathbb{Z}%
_{2}\times \mathbb{Z}_{2}^{\prime }$. Being associated with $\mathbb{V}_{4}$%
, these vacua are then related to each other by $\mathbb{Z}_{2}\times
\mathbb{Z}_{2}^{\prime }$ transformations as sketched on the following
diagram%
\begin{equation}
\begin{tabular}{lllll}
&  & $\mathbf{\varphi }_{2}$ &  &  \\
& $\diagup $ &  & $\diagdown $ &  \\
$\mathbf{\varphi }_{1}$ &  & $\ 0$ &  & $\mathbf{\varphi }_{4}$ \\
& $\diagdown $ &  & $\diagup $ &  \\
&  & $\mathbf{\varphi }_{3}$ &  &
\end{tabular}
\label{ff}
\end{equation}%
where the centre $0$ refers to the singular situation where all four $%
\mathbf{\varphi }_{a}$'s vanish identically --- $\upsilon _{\Phi
}\rightarrow 0$. So, at this singular point lives the full $\mathbb{A}_{4}$
symmetry including its four $\mathbb{Z}_{3}$ subsymmetries. The $\mathbf{%
\varphi }_{1},\mathbf{\varphi }_{2},\mathbf{\varphi }_{3},\mathbf{\varphi }%
_{4}$ have same ground state energy and can be obtained by acting on the $%
\mathbb{Z}_{3}$- invariant vacuum (\ref{tv}) by the $\mathbb{V}_{4}$
generators $\mathcal{S}$ and $\mathcal{S}^{\prime }$. By taking $\mathbf{%
\varphi }_{1}=\left\langle \Phi \right\rangle $ as in Eq. (\ref{tv}), we
then have%
\begin{equation}
\begin{tabular}{lllll}
$\mathbf{\varphi }_{2}=\mathcal{S}\mathbf{\varphi }_{1}$ & $;$ & $\mathbf{%
\varphi }_{3}=\mathcal{S}^{\prime }\mathbf{\varphi }_{1}$ & $;$ & $\mathbf{%
\varphi }_{4}=\mathcal{S}^{\prime \prime }\mathbf{\varphi }_{1}$%
\end{tabular}
\label{fo}
\end{equation}%
where we have set $\mathcal{S}^{\prime \prime }=\mathcal{SS}^{\prime }=%
\mathcal{S}^{\prime }\mathcal{S}$. Notice that $\mathbf{\varphi }_{4}$ can
be also obtained from $\mathbf{\varphi }_{2}$ or from $\mathbf{\varphi }_{3}$%
; the relations are respectively given by $\mathbf{\varphi }_{4}=\mathcal{S}%
^{\prime }\mathbf{\varphi }_{2}$ and $\mathbf{\varphi }_{4}=\mathcal{S}%
\mathbf{\varphi }_{3}$. Using Eqs. (\ref{tv}) and (\ref{sp}), we can express
the four vacua explicitly in $\mathbb{C}^{3}$ as follows
\begin{equation}
\begin{tabular}{lllllll}
$\mathbf{\varphi }_{1}$ & $=$ & $\upsilon _{\Phi }\left(
\begin{array}{c}
1 \\
0 \\
0%
\end{array}%
\right) $ & $\quad \text{,\quad }$ & $\mathbf{\varphi }_{2}$ & $=$ & $\frac{%
\upsilon _{\Phi }}{3}\left(
\begin{array}{c}
-1 \\
2 \\
2%
\end{array}%
\right) $ \\
$\mathbf{\varphi }_{3}$ & $=$ & $\frac{\upsilon _{\Phi }}{3}\left(
\begin{array}{c}
-1 \\
2\bar{\omega} \\
2\omega%
\end{array}%
\right) $ & $\quad \text{,\quad }$ & $\mathbf{\varphi }_{4}$ & $=$ & $\frac{%
\upsilon _{\Phi }}{3}\left(
\begin{array}{c}
-1 \\
2\omega \\
2\bar{\omega}%
\end{array}%
\right) $%
\end{tabular}
\label{four}
\end{equation}%
with $\omega =-\frac{1}{2}+i\frac{\sqrt{3}}{2}$. In the vector basis (\ref%
{uv}), we have to use real variables; so the dimension of the above vectors
has to be doubled; for instance we have $\mathbf{\varphi }_{4}^{T}=\frac{%
\upsilon _{\Phi }}{3}\left( -1,0,-1,\sqrt{3},-1,-\sqrt{3}\right) $. \newline
Observe that the four $\mathbf{\varphi }_{a}$ vacua define four particular
vectors in the complex three- (real six-) dimensional space $\mathbb{C}%
^{3}\sim \mathbb{R}^{6}$ of possible vacua of the $\Phi $- flavon. As these
vectors are constrained like
\begin{equation}
\mathbf{\varphi }_{1}+\mathbf{\varphi }_{2}+\mathbf{\varphi }_{3}+\mathbf{%
\varphi }_{4}=0
\end{equation}%
they define a tetrahedron with Kahler modulus given by $\func{Re}\upsilon
_{\Phi }$. By using the $\left( X,Y,Z\right) $ basis vectors; we can express
the four $\mathbf{\varphi }_{a}$'s like $1\oplus 3$, the sum of a singlet
and a triplet with respect to the generator $\mathcal{T}$. The singlet is
given by
\begin{equation}
\mathbf{\varphi }_{1}=\upsilon _{\Phi }X\quad ,\quad \mathcal{T}\mathbf{%
\varphi }_{1}=\mathbf{\varphi }_{1}  \label{51}
\end{equation}%
and the remaining three others form a 3-cycle $\left( \mathbf{\varphi }_{2},%
\mathbf{\varphi }_{3},\mathbf{\varphi }_{4}\right) $ with components as
follows
\begin{eqnarray}
\mathbf{\varphi }_{2} &=&\frac{\upsilon _{\Phi }}{3}\left( -X+2Y+2Z\right)
\notag \\
\mathbf{\varphi }_{3} &=&\frac{\upsilon _{\Phi }}{3}\left( -X+2\bar{\omega}%
Y+2\omega Z\right)  \label{52} \\
\mathbf{\varphi }_{4} &=&\frac{\upsilon _{\Phi }}{3}\left( -X+2\omega Y+2%
\bar{\omega}Z\right) .  \notag
\end{eqnarray}%
Using the expression of $\mathcal{T}$ given by Eq. (\ref{sp}), one can
easily check that they obey indeed the cyclic property%
\begin{equation}
\mathcal{T}\mathbf{\varphi }_{2}=\mathbf{\varphi }_{3}\quad ,\quad \mathcal{T%
}\mathbf{\varphi }_{3}=\mathbf{\varphi }_{4}\quad ,\quad \mathcal{T}\mathbf{%
\varphi }_{4}=\mathbf{\varphi }_{2}.  \label{3c}
\end{equation}%
From the above relations, we learn that we indeed have the remarkable
identity $\sum_{i}\mathbf{\varphi }_{i}=0$ that leads in turns to $\mathbf{%
\varphi }_{1}=-\left( \mathbf{\varphi }_{2}+\mathbf{\varphi }_{3}+\mathbf{%
\varphi }_{4}\right) $. Observe also that for a fixed value of $\upsilon
_{\Phi }$, the four $\mathbf{\varphi }_{1},\mathbf{\varphi }_{2},\mathbf{%
\varphi }_{3},\mathbf{\varphi }_{4}$ are given by four special points in the
three dimensional $\mathfrak{F}\simeq \mathbb{C}^{3}$. These points define
the four vertices of a homogeneous tetrahedron given by the Fig. \ref{w1}.
Thus, from the view of the space of the flavon $\Phi $, the domain walls
separating the four vacua (i.e., vertices) are encoded in a nonregular 3d
geometry having: $\left( \mathbf{\alpha }\right) $ six 1- dimensional edges
that we denote like $\mathbf{\varphi }_{ab}=(\mathbf{\varphi }_{a},\mathbf{%
\varphi }_{b})$ with domain walls given by (\ref{dw}) namely%
\begin{equation}
l_{ab}=\pm \int_{\mathbf{\varphi }_{ab}}\varpi _{1}  \label{l}
\end{equation}%
with Hermitian 1-form $\varpi _{1}=\varpi _{1}\left( \xi _{\Phi },\bar{\xi}%
_{\Phi }\right) $ given by%
\begin{equation}
\varpi _{1}=\frac{d\varphi }{\sqrt{\mathcal{E}+\mathcal{\tilde{V}}\left( \xi
_{\Phi },\bar{\xi}_{\Phi }\right) }}\qquad ,\qquad d\varphi
^{2}=\sum_{i=1}^{3}d\bar{\xi}_{\Phi _{i}}d\xi _{\Phi _{i}}
\end{equation}%
$\left( \mathbf{\beta }\right) $ four triangular surfaces $\boldsymbol{T}%
_{abc}$ with vertices $(\mathbf{\varphi }_{a},\mathbf{\varphi }_{b},\mathbf{%
\varphi }_{c})$, one of them is given by $\mathbf{T}_{234}$---the face with
yellow color in \ref{w1}. In this case, the domain walls extending between $(%
\mathbf{\varphi }_{a},\mathbf{\varphi }_{b},\mathbf{\varphi }_{c})$ are
given by
\begin{equation}
\boldsymbol{S}_{abc}=\pm \dint\nolimits_{\boldsymbol{T}_{abc}}\varpi _{2}
\label{s}
\end{equation}%
where $\varpi _{2}=\varpi _{2}\left( \xi _{\Phi },\bar{\xi}_{\Phi }\right) $%
---generalizes the above 1- form $\varpi _{1}$---is a Hermitian 2-form
living on $\mathfrak{F}$ and dependent as well on the flavon scalar
potential $\mathcal{\tilde{V}}\left( \xi _{\Phi },\bar{\xi}_{\Phi }\right) $%
. $\left( \mathbf{\gamma }\right) $ the bulk of the tetrahedron $\boldsymbol{%
H}_{1234}$ with vertices $(\mathbf{\varphi }_{1},\mathbf{\varphi }_{2},%
\mathbf{\varphi }_{3},\mathbf{\varphi }_{4})$ leading to
\begin{equation}
\boldsymbol{V}_{1234}=\pm \dint\nolimits_{\boldsymbol{H}_{1234}}\varpi _{3}
\label{v}
\end{equation}%
where now $\varpi _{3}=\varpi _{3}\left( \xi _{\Phi },\bar{\xi}_{\Phi
}\right) $ is a Hermitian 3-form on $\mathfrak{F}$. Notice that for the
limit $\upsilon _{\Phi }\rightarrow 0,$ all four $\mathbf{\varphi }_{1},%
\mathbf{\varphi }_{2},\mathbf{\varphi }_{3},\mathbf{\varphi }_{4}$ vacua
merge at the origin $O$ of $\mathfrak{F}$---the centre of the
tetrahedron---where lives the full $\mathbb{A}_{4}$ symmetry. In this
singular limit, the DWs disappear. Notice also that on the $O\mathbf{\varphi
}_{1}$ axis normal to $\boldsymbol{T}_{234}$ face of the tetrahedron; it
lives a $\mathbb{Z}_{3}$ subsymmetry of $\mathbb{A}_{4}$. In fact, on each
of the four $O\mathbf{\varphi }_{a}$ axes, passing through the center $O$
and the vertices $\mathbf{\varphi }_{a}$, lives one of the four $\mathbb{Z}%
_{3}$ subsymmetries (\ref{a4}) of the discrete $\mathbb{A}_{4}$
characterizing the\textrm{\ }flavored NMSSM.
Notice moreover that from Eq. (\ref{ff}) we learn that one can also use the
elements $\left\{ I,\mathcal{S},\mathcal{S}^{\prime },\mathcal{SS}^{\prime
}\right\} $ of the $\mathbb{V}_{4}$ to describe the four degenerate vacua (%
\ref{four}) and then the quivers representing the DWs. Indeed, seen that the
Klein $\mathbb{V}_{4}$ is isomorphic to $\mathbb{Z}_{2}\times \mathbb{Z}_{2}$%
, the set $\left\{ I,\mathcal{S},\mathcal{S}^{\prime },\mathcal{SS}^{\prime
}\right\} $ can be then factorized in three manners as follows
\begin{eqnarray}
\mathbb{Z}_{2}\times \mathbb{Z}_{2}^{\prime } &=&\left\{ I,\mathcal{S}%
\right\} \times \left\{ I,\mathcal{S}^{\prime }\right\}  \notag \\
\mathbb{Z}_{2}\times \mathbb{Z}_{2}^{\prime \prime } &=&\left\{ I,\mathcal{S}%
\right\} \times \left\{ I,\mathcal{S}^{\prime \prime }\right\} \\
\mathbb{Z}_{2}^{\prime }\times \mathbb{Z}_{2}^{\prime \prime } &=&\left\{ I,%
\mathcal{S}^{\prime }\right\} \times \left\{ I,\mathcal{S}^{\prime \prime
}\right\}  \notag
\end{eqnarray}%
with $\mathcal{S}^{\prime \prime }=\mathcal{SS}^{\prime }$ and $\mathcal{S}%
^{2}=\left( \mathcal{S}^{\prime }\right) ^{^{2}}=\left( \mathcal{S}^{\prime
\prime }\right) ^{^{2}}=I$.\textrm{\ }Thus, we can represent the domain
walls extending between the vacua by a real quiver diagram similar to Fig. %
\ref{w1}; but with vertices given by the group elements of $\mathbb{V}%
_{4}=\left\{ I,\mathcal{S},\mathcal{S}^{\prime },\mathcal{SS}^{\prime
}\right\} $. In this case, the $\mathbb{Z}_{2}$ subgroups of $\mathbb{V}_{4}$
are associated with edges of the tetrahedron with a boundary point given by $%
I$ and the other boundary by one of the three other vertices. The four $%
\mathbb{Z}_{3}$ subsymmetries are associated with the four faces $%
\boldsymbol{T}_{abc}$ of the diamond. Indeed, we can write down an explicit
correspondence between the vacua $\mathbf{\varphi }_{a}$ and the elements of
$\mathbb{V}_{4}$. For example, for the three $\mathbf{\varphi }_{1a}$ edges,
we have the following
\begin{eqnarray}
\left( I,\mathcal{S}\right) \qquad &\leftrightarrow &\qquad \mathbf{\varphi }%
_{12}  \notag \\
\left( I,\mathcal{S}^{\prime }\right) \qquad &\leftrightarrow &\qquad
\mathbf{\varphi }_{13} \\
\left( I,\mathcal{S}^{\prime \prime }\right) \qquad &\leftrightarrow &\qquad
\mathbf{\varphi }_{14}  \notag
\end{eqnarray}%
By using (\ref{fo}), the vertex $\mathbf{\varphi }_{1}$ is represented by
the identity\textrm{\ }$I$\textrm{\ }and the vertex $\mathbf{\varphi }_{2}$
by the generator\textrm{\ }$\mathcal{S}$. A similar thing is valid for the
other $\mathbf{\varphi }_{13}$ and $\mathbf{\varphi }_{14}$ edges with one
vertex given by $\mathbf{\varphi }_{1}$\textrm{\ }and the other by $\mathbf{%
\varphi }_{3}$ or $\mathbf{\varphi }_{4}$.\ The three other remaining edges $%
\mathbf{\varphi }_{23},$ $\mathbf{\varphi }_{24},$ and $\mathbf{\varphi }%
_{34}$\ can be interpreted as follows
\begin{eqnarray}
\mathbf{\varphi }_{23}\qquad &\leftrightarrow &\qquad \left( \mathcal{S},%
\mathcal{S}^{\prime }\right) =\mathcal{S}\times \left( I,\mathcal{S}^{\prime
\prime }\right) \qquad \leftrightarrow \qquad \mathcal{S}\mathbf{\varphi }%
_{14}  \notag \\
\mathbf{\varphi }_{24}\qquad &\leftrightarrow &\qquad \left( \mathcal{S},%
\mathcal{S}^{\prime \prime }\right) =\mathcal{S}\times \left( I,\mathcal{S}%
^{\prime }\right) \qquad \leftrightarrow \qquad \mathcal{S}\mathbf{\varphi }%
_{13} \\
\mathbf{\varphi }_{34}\qquad &\leftrightarrow &\qquad \left( \mathcal{S}%
^{\prime },\mathcal{S}^{\prime \prime }\right) =\mathcal{S}^{\prime }\times
\left( I,\mathcal{S}\right) \qquad \leftrightarrow \qquad \mathcal{S}%
^{\prime }\mathbf{\varphi }_{12}  \notag
\end{eqnarray}%
where for instance, $\mathbf{\varphi }_{23}$ is just the transformation of $%
\mathbf{\varphi }_{14}$\ under $\mathcal{S}$. Notice also that the face $%
\boldsymbol{T}_{234}$ is associated with $\left( \mathcal{S},\mathcal{S}%
^{\prime },\mathcal{S}"\right) $ and is related with the $\mathbb{Z}_{3}$
subsymmetry of $\mathbb{A}_{4}$; so%
\begin{equation}
\boldsymbol{T}_{234}\qquad \leftrightarrow \qquad \mathbb{Z}_{3}=\left\{ I,%
\mathcal{T},\mathcal{T}^{2}\right\}
\end{equation}%
with $\mathcal{T}=\left( 234\right) $. The domain walls induced by the
breaking $\mathbb{A}_{4}\rightarrow \mathbb{Z}_{3}$ may be also described by
using characters of the Klein four- group $\mathbb{V}_{4}$ as in Fig. \ref%
{w3}. These characters are given by $\left( +,+\right) ,$ $\left( -,+\right)
,$ $\left( +,-\right) ,$ $\left( -,-\right) $ respectively associated with
the group elements $I,\mathcal{S},\mathcal{S}^{\prime },\mathcal{SS}^{\prime
}$.
\begin{figure}[tbph]
\begin{center}
\includegraphics[scale=0.6]{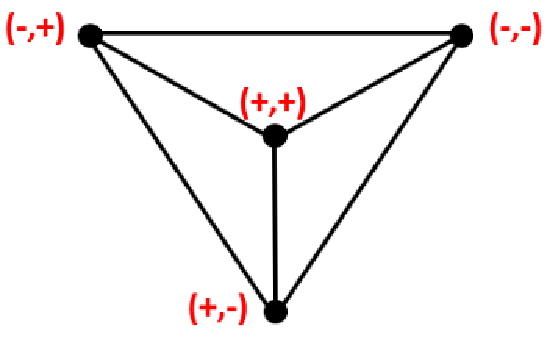}
\end{center}
\par
\vspace{-0.5cm}
\caption{A quiver diagram representing domain walls using characters of the
generators of the Klein group. \textrm{The four vertices correspond to the
four degenerate vacua, while the edges that connects the four vertices stand
for domain walls.}}
\label{w3}
\end{figure}
After this graphic description of DWs in the VEV space $\mathfrak{F}$, we
are now in position to address the question on whether these domain walls of
the \emph{FNMSSM} are problematic in the charged lepton sector or not. The
answer is fortunately no!; this is due to the very high energy scale where
the spontaneous breaking of the discrete group $\mathbb{A}_{4}\rightarrow
\mathbb{Z}_{3}$ takes place. In this regard, recall that the inflation based
scenario might be a nice approach to bypass DWs provided the inflationary
scale; say around $\mathcal{O}(10^{14})$\ $GeV$\textrm{\ \cite{R64,R65}, }is
lower than the flavor symmetry breaking scale. In this case, the DWs are
formed before the end of inflation at scales comparable to the GUT scale%
\textrm{\ \cite{R64}.} Hence, assuming that the cutoff scale $\Lambda $
associated to $\mathbb{A}_{4}$ breaking given in Eq. (\ref{cl}) is of order
of the SUSY-GUT scale $M_{GUT}\sim 2\times 10^{16}$ $\mathrm{GeV}$, we find
by using (\ref{cl}) the following lower bound of the flavon VEV
\begin{equation}
\upsilon _{\Phi }>1.4\times 10^{14}\text{ }\mathrm{GeV}
\end{equation}%
Therefore, the flavon triplet $\Phi $\ acquired its VEV before the end of
inflation; thus, in the charged lepton sector the discrete $\mathbb{A}_{4}$
symmetry is broken at a scale that is above the inflationary scale.
Consequently, the domain walls\emph{\ }produced in this case are inflated
away \textrm{\cite{R64}}; and one is left with the $\mathbb{Z}_{3}$
invariant ground state $\mathbf{\varphi }_{1}$.

\subsection{Domain walls in neutrino sector}

In this subsection we study the domain walls in the chargeless lepton sector
of the flavored NMSSM and show that they are really problematic. Recall that
in the neutrino sector, the breaking pattern corresponds to $\mathbb{A}_{4}$%
\emph{\ }$\rightarrow $\emph{\ }$\mathbb{Z}_{2}$; and can be driven either
by the VEV of the flavon $\Omega $; or by the VEV of the singlet $S$. This
feature may be observed from the properties of the quantum numbers given by
Table \ref{C}. So, we first study the breaking $\mathbb{A}_{4}\rightarrow
\mathbb{Z}_{2}$ by using a diagonal VEV in case of the flavon $\Omega $;
then we make comments on the breaking using $S$.

\subsubsection{Breaking $\mathbb{A}_{4}$ to $\mathbb{Z}_{2}$ by flavon $%
\Omega $ and stable domain walls}

\begin{figure}[tbph]
\begin{center}
\includegraphics[scale=0.65]{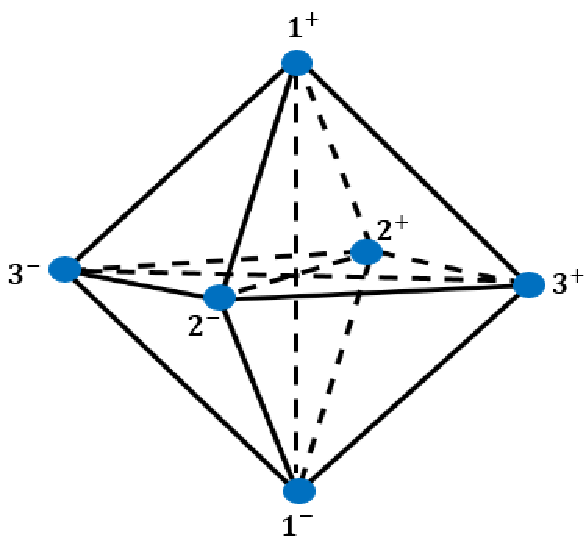}
\end{center}
\par
\vspace{-0.5cm}
\caption{The six vacua define the vertices of a homogeneous octahedron of
Kahler modulus $\protect\upsilon _{\Omega }$. \textrm{While the edges stand
for the domain walls interpolating between these six degenerate vacua.}}
\label{w4}
\end{figure}

By using the flavon $\Omega $, the breaking $\mathbb{A}_{4}\rightarrow
\mathbb{Z}_{2}$ is realised by Eq. (\ref{om}). Clearly, this particular
flavon triplet $\Omega $ vacuum aligned along the diagonal of the complex
space $\mathfrak{F}\simeq \mathbb{C}^{3}$ is not a symmetry of $\mathcal{T}$%
; but preserves the $\mathcal{S}$ generator,%
\begin{equation}
\mathcal{S}\left\langle \Omega \right\rangle =\left\langle \Omega
\right\rangle \quad ,\quad \mathcal{T}\left\langle \Omega \right\rangle \neq
\left\langle \Omega \right\rangle
\end{equation}%
These relations can be explicitly checked by using the matrix
representations of $\mathcal{T}$ and $\mathcal{S}$ given by Eq. (\ref{sp}).
As a consequence of this breaking of $\mathbb{A}_{4}$ down to $\mathbb{Z}%
_{2}=\left\{ I,\mathcal{S}\right\} $; the quantum charges of the
right-handed neutrino $N_{i}^{c}$\ and the lepton doublets $L_{i}$
superfields under the $\mathbb{Z}_{2}$ residual symmetry are summarized in
Tables \ref{V9} and \ref{V10}.
\begin{table}[h]
\centering\renewcommand{\arraystretch}{1.2}
\begin{tabular}{|c|c|c|}
\hline
Superfields & $L_{i}$ & $N_{i}^{c}$ \\ \hline
$\mathbb{Z}_{2}$ & $\mathbf{1}_{-}\oplus \mathbf{1}_{-}\oplus \mathbf{1}_{-}$
& $\mathbf{1}_{-}\oplus \mathbf{1}_{-}\oplus \mathbf{1}_{-}$ \\ \hline
\end{tabular}%
\caption{Lepton and right-handed neutrino superfields and their quantum
numbers under $\mathbb{Z}_{2}$.}
\label{V9}
\end{table}
\begin{table}[h]
\centering\renewcommand{\arraystretch}{1.2}
\begin{tabular}{|c|c|c|c|c|c|c|}
\hline
Superfields & $H_{u}$ & $H_{d}$ & $\Phi $ & $\Omega $ & $S$ & $\chi $ \\
\hline
$\mathbb{Z}_{2}$ & $\mathbf{1}_{+}$ & $\mathbf{1}_{+}$ & $\mathbf{1}%
_{-}\oplus \mathbf{1}_{-}\oplus \mathbf{1}_{-}$ & $\mathbf{1}_{+}\oplus
\mathbf{1}_{+}\oplus \mathbf{1}_{+}$ & $\mathbf{1}_{+}$ & $\mathbf{1}_{+}$
\\ \hline
\end{tabular}%
\caption{Higgs and flavon superfields and their quantum numbers under $%
\mathbb{Z}_{2}$.}
\label{V10}
\end{table}
\newline
Therefore, we have the following symmetry transformations
\begin{equation}
\begin{array}{ccccc}
\nu _{e}^{c}\rightarrow -\nu _{e}^{c} & \quad ,\quad & \nu _{\mu
}^{c}\rightarrow -\nu _{\mu }^{c} & \quad ,\quad & \nu _{\tau
}^{c}\rightarrow -\nu _{\tau }^{c} \\
L_{e}\rightarrow -L_{e} & , & L_{\mu }\rightarrow -L_{\mu } & , & L_{\tau
}\rightarrow -L_{\tau }%
\end{array}%
\end{equation}%
while the right-handed charged leptons are $\mathbb{Z}_{2}$ even, because
the singlet representations ($\mathbf{1}_{\left( 1,1\right) },$ $\mathbf{1}%
_{\left( 1,\omega \right) },$ $\mathbf{1}_{\left( 1,\omega ^{2}\right) }$)
transform trivially under $\mathcal{S}$, see Eq. (\ref{sr})\ of Appendix A.%
\newline
By substituting $\mathbb{A}_{4}$ by $\mathbb{V}_{4}\rtimes \mathbb{Z}_{3}$
and the four order group $\mathbb{V}_{4}$ by $\mathbb{Z}_{2}\times \mathbb{Z}%
_{2}^{\prime }$, the breaking pattern $\mathbb{A}_{4}\rightarrow \mathbb{Z}%
_{2}$ can be expressed like
\begin{equation}
\left( \mathbb{Z}_{2}\times \mathbb{Z}_{2}^{\prime }\right) \rtimes \mathbb{Z%
}_{3}\text{ }\quad \overset{\left\langle \Omega \right\rangle }{%
\longrightarrow }\quad \text{\ }\mathbb{Z}_{2}  \label{2p}
\end{equation}%
This breaking mode shows that the induced domain walls are characterized by
the broken part of the alternating $\mathbb{A}_{4}$ flavor symmetry which
may naively be presented as $\mathbb{Z}_{2}^{\prime }\rtimes \mathbb{Z}_{3}$%
. The $\mathbb{Z}_{2}$ and $\mathbb{Z}_{2}^{\prime }$ groups are subgroups
of the Klein four- group $\mathbb{V}_{4}$ respectively generated by $%
\mathcal{S}$ and $\mathcal{S}^{\prime }\mathcal{=TST}^{-1}$. The $\mathbb{Z}%
_{3}$ is generated by $\mathcal{T}$, it is one of the four kinds of $\mathbb{%
Z}_{3}$\ subgroups inside of $\mathbb{A}_{4}$. Under the breaking (\ref{2p}%
), the initial $\mathbb{A}_{4}$ invariant vacuum gets now split into six
vacua with same energy%
\begin{equation}
\left\langle \Omega \right\rangle _{i}\equiv \vartheta _{i}\qquad ,\qquad
i=1,...,6  \label{cv}
\end{equation}%
In the space $\mathfrak{F}$ of flavons, these vacua define the six vertices
of a homogeneous octahedron $\sum_{i=1}^{6}\vartheta _{i}=0$ embedded in $%
\mathbb{R}^{6}\sim \mathbb{C}^{3}$ as schematized in Fig. \ref{w4}.
Notice that one of the six vacua (\ref{cv}), say $\vartheta _{1}$, is just
the $\mathbb{Z}_{2}$- invariant vacuum Eq. (\ref{om}); that is $\vartheta
_{1}=\left\langle \Omega \right\rangle $. By using the basis vectors $X,Y,Z$%
, it can be expressed as follows
\begin{equation}
\vartheta _{1}=\upsilon _{\Omega }\left( X+Y+Z\right)
\end{equation}%
and points in the diagonal direction of the complex 3d (real 6-) space. To
get the remaining five other $\vartheta _{i}$ vacua, we act on this $%
\vartheta _{1}$ by the $\mathcal{S}^{\prime }$ and $\mathcal{T}$ generators
of $\mathbb{Z}_{2}^{\prime }\rtimes \mathbb{Z}_{3}$. As the $\mathcal{S}%
^{\prime }$ acts on $\vartheta _{1}$ like a reflection: $\vartheta
_{1}\rightarrow -\vartheta _{1}$, it is instructive to denote the six $%
\vartheta _{i}$ vacua like $\vartheta _{\alpha }^{\pm }$ with $\alpha
=1,2,3. $ Using the explicit expression of $\mathcal{S}^{\prime }$ and $%
\mathcal{T}$ , we obtain%
\begin{eqnarray}
\vartheta _{1}^{\pm } &=&\pm \upsilon _{\Omega }\left( X+Y+Z\right)  \notag
\\
\vartheta _{2}^{\pm } &=&\pm \upsilon _{\Omega }\left( X+\bar{\omega}%
Y+\omega Z\right)  \label{an} \\
\vartheta _{3}^{\pm } &=&\pm \upsilon _{\Omega }\left( X+\omega Y+\bar{\omega%
}Z\right)  \notag
\end{eqnarray}%
with%
\begin{equation}
\mathcal{S}\left\langle \vartheta _{\alpha }^{\pm }\right\rangle
=\left\langle \vartheta _{\alpha }^{\pm }\right\rangle
\end{equation}%
Explicitly, we have%
\begin{equation}
\vartheta _{1}^{\pm }=\pm \upsilon _{\Omega }\left(
\begin{array}{c}
1 \\
1 \\
1%
\end{array}%
\right) \ ,\,\,\quad \ \vartheta _{2}^{\pm }=\pm \upsilon _{\Omega }\left(
\begin{array}{c}
1 \\
\bar{\omega} \\
\omega%
\end{array}%
\right) ,\,\,\quad \ \vartheta _{3}^{\pm }=\pm \upsilon _{\Omega }\left(
\begin{array}{c}
1 \\
\omega \\
\bar{\omega}%
\end{array}%
\right) \   \label{voc1}
\end{equation}%
with the transformation relations%
\begin{equation}
\vartheta _{1}^{\pm }\quad ,\quad \vartheta _{2}^{\pm }=\mathcal{T}\vartheta
_{1}^{\pm }\quad ,\quad \vartheta _{3}^{\pm }=\mathcal{T}\vartheta _{2}^{\pm
}\quad ,\quad \mathcal{S}^{\prime }\vartheta _{2}^{\pm }=\vartheta _{2}^{\pm
}\quad ,\quad \mathcal{S}^{\prime }\vartheta _{1/3}^{\pm }=\vartheta
_{1/3}^{\mp }
\end{equation}%
with $\mathcal{S}^{\prime }=\mathcal{TST}^{-1}$. Notice the real VEV $%
\upsilon _{\Omega }$ of the chiral superfield $\Omega $, which breaks $%
\mathbb{A}_{4}$ as in (\ref{2p}), breaks as well the extra\textrm{\ }$%
\boldsymbol{Z}_{3}$\textrm{\ }symmetry\footnote{%
Notice that the domain walls created by this extra $\boldsymbol{Z}_{3}$\
symmetry during the breaking pattern Eq. (\ref{2p}) in the neutrino sector
will disappear together with the ones from the broken $\mathbb{A}_{4}$.}.
Notice also that contrary to the charged sector, domain walls induced by the
breaking pattern $\mathbb{A}_{4}\rightarrow \mathbb{Z}_{2}$ in the neutrino
sector\ are problematic.\ The couplings involved in the superpotential (\ref%
{ma1}) are at the renormalizable level where we expect a particular
hierarchy among the VEVs as\textrm{\ }$\upsilon _{S}<\upsilon _{\mathbf{\chi
}}\lesssim \upsilon _{\Omega }$. This hierarchy is reasonable since $%
\upsilon _{S}$\ is known to acquire a VEV of the order of\textrm{\ }$%
M_{SUSY} $ and it is also responsible for a small deviation form the TBM
pattern and thus should be smaller than the other VEVs while the VEVs of the
flavons\textrm{\ }$\mathbf{\chi }$\textrm{\ }and\textrm{\ }$\Omega $\textrm{%
\ }can be safely assumed to vary in the range $\left[ 10^{7}\rightarrow
10^{10}\right] $ GeV which will lead to right-handed neutrino masses $M_{i}$%
\ in Eq. (\ref{ev}) compatible with type-I seesaw mechanism.\textrm{\ }This
can be accomplished simply by choosing the right values for the free
parameters\textrm{\ }$Y_{0},\lambda ,\lambda ^{\prime }$ and\textrm{\ }$%
\beta $\textrm{\ }involved in $\upsilon _{u}=\upsilon \sin \beta $.
Therefore, these VEVs are then smaller than the inflationary scale and thus,
the domain walls in the neutrino sector of the flavored NMSSM are created
below the inflationary scale; they are inevitable and then inconsistent with
the standard cosmology \cite{R67}; they must be avoided. In order to
circumvent this domain wall problem, we break explicitly the $\mathbb{Z}%
_{2}^{\prime }\rtimes \mathbb{Z}_{3}$ permuting the six degenerate vacua.
This will be done by using higher order Planck-suppressed operators $%
\mathcal{O}(\frac{1}{M_{Pl}})$.

\subsection{Solving the DW problem in neutrino sector}

\begin{figure}[tbph]
\begin{center}
\includegraphics[scale=0.8]{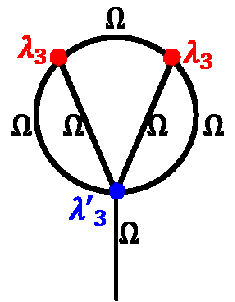}
\end{center}
\caption{The tadpole diagram for the nonrenormalizable term in Eq. (\protect
\ref{op}).}
\label{f20}
\end{figure}

To start, recall that several ways have been suggested in the literature to
overcome the domain wall problem induced by breaking discrete symmetries of
the degenerate vacua; the simplest one being inflation \cite{R9}.\emph{\ }A
second interesting method to solve the domain wall problem is that the
discrete symmetry connecting the vacua must be broken explicitly before the
spontaneous breaking takes place \cite{R68}. It was shown in Refs. \cite%
{R10,R69,A30,A32,A33,A34} that the last method can be achieved by
introducing higher dimensional operators $\frac{1}{M_{Pl}^{n}}\mathcal{O}%
_{n+3}$ suppressed by powers of the Planck scale $M_{Pl}$ leading to favor
one of the vacua over the others, and consequently no walls will be formed
and the domain wall problem is resolved; for NMSSM see \cite{R42}. Moreover,
an analysis has been performed in Refs.\ \cite{R71,A35} proves that the
relevant non-renormalizable operators $\mathcal{O}_{n+3}$\textrm{\ }that
solve the problem are the odd superpotential terms (n even integer).
However,\ we should notice that there are also dimensional operators of
order four (operators suppressed by one inverse power of the Planck mass)
that must be eliminated. We do not go into details here but we point out
that this can be realized by invoking an additional symmetry such as extra $%
\mathcal{Z}_{2}$ or $\mathcal{Z}_{5}$ R-symmetries; for details in this
direction see for instance Refs. \cite{R71,A35}.\textrm{\ }In the present
paper,\textrm{\ }we follow Ref. \cite{R70} that adopts the second approach;
concretely we break the $\mathbb{Z}_{2}^{\prime }\rtimes \mathbb{Z}%
_{3}\times \boldsymbol{Z}_{3}$\ part of the flavor symmetry explicitly via
higher dimensional operators---Planck scale operators---of order five; in
other words, we want to keep only one vacuum among the six $\vartheta
_{\alpha }^{\pm }$; for instance $\vartheta _{1}^{+}$. For our concern, the $%
\mathbb{A}_{4}\times \boldsymbol{Z}_{3}$ invariant superpotential $W_{scal}$
restricted to the Higgs doublet $H_{u,d}$, the gauge singlet $S$ and the
flavon superfields $\mathbf{\chi }$, $\Omega $, $\Phi $\ is as follows%
\begin{eqnarray}
W_{scal} &=&\lambda _{1}SH_{u}H_{d}+\lambda _{2}S^{3}+\lambda _{3}\Omega
^{3}+\lambda _{4}\Phi ^{3}+\lambda _{5}\mathbf{\chi }^{3}  \notag \\
&&+\lambda _{6}S\Omega ^{2}+\lambda _{7}\mathbf{\chi }\Omega ^{2}+\mu _{\Phi
}\Phi ^{2}  \label{sca}
\end{eqnarray}%
The non-renormalizable operators $\mathcal{W}_{NR}$ of order five
operators---suppressed by two inverse powers of the Planck mass---that break
the full discrete symmetry group of our model down to $\mathbb{Z}_{2}$ can
be chosen as given by a perturbation $\delta W_{scal}$ of one of the
trilinear couplings that already exist in the scalar superpotential (\ref%
{sca}). Thus, we are left with multiple choices for order five operators
where the most suitable one that can break explicitly $\mathbb{A}_{4}\times
\boldsymbol{Z}_{3}$ down to $\mathbb{Z}_{2}$ is given by the following
operator%
\begin{equation}
\mathcal{W}_{NR}=\frac{\lambda _{3}^{\prime }}{M_{Pl}^{2}}\left. \left(
\Omega ^{5}\right) \right\vert _{\left( 1,\omega \right) }  \label{op}
\end{equation}%
\textrm{\ } \textrm{\ }and, at quantum level, is represented by the Feynman
diagram in Fig. \ref{f20} contributing to the effective potential $\mathcal{V%
}_{eff}$. Higher dimensional operators of order five that can contribute to
the explicit breaking of the full flavor group are shown in Appendix C.%
\newline
The choice of\textrm{\ }above\textrm{\ }$\mathcal{W}_{NR}$\textrm{\ }has
been motivated by the fact that\textrm{\ }$\Omega $\ carries quantum charges
under both\textrm{\ }$\mathbb{A}_{4}$\textrm{\ }and $\boldsymbol{Z}_{3}$
symmetries\footnote{%
\ It is clear that the nonrenormalizable operator given in Eq. (\ref{op})
will have negligible effect on the low energy phenomenology, considering the
fact that this operator is suppressed by two power of the Planck mass.}.
Moreover, this triplet creates its own walls during the spontaneous breaking
of the full discrete symmetry $G_{\mathrm{f}}$. For the breaking of $\mathbb{%
A}_{4}$, the DWs are located on the boundaries of the six degenerate vacua
given in Eq. (\ref{voc1}) and Fig. \ref{w4}. We note that the other
five-dimensional operators (given in Appendix C) do not present any risk on
the phenomenology of the model, since their contributions at low energy
theory are generated through the same tadpole diagram presented in Fig. \ref%
{f20} with a factor difference depending on the coupling constant used in
the perturbation term.
By using the Feynman rules of supergraphs\footnote{%
\ For the detailed calculations that led to the term given in the effective
potential $\delta \mathcal{V}_{eff}$, see the analysis given in Appendix C.}
\cite{R72,A36,A37}, we obtain the following contribution to the effective
scalar potential%
\begin{equation}
\delta \mathcal{V}_{eff}=M_{W}^{3}(\eta \phi _{\Omega }+\bar{\eta}\bar{\phi}%
_{\Omega })+M_{W}^{2}(\eta \digamma _{\Omega }+\bar{\eta}\bar{\digamma}%
_{\Omega })
\end{equation}%
where $\phi _{\Omega }$ and $\digamma _{\Omega }$ are respectively the
scalar component and the F-term of the flavon superfield $\Omega =\phi
_{\Omega }+\theta \psi _{\Omega }+\theta ^{2}\digamma _{\Omega }$; the $%
M_{W} $ is the scale of the electroweak theory\ and $\eta =\frac{\lambda
_{3}^{2}\lambda _{3}^{\prime }}{(16\pi ^{2})^{3}}$. It is clear that this
operator induces a linear term in the soft SUSY breaking scalar
potential---the full soft SUSY breaking terms are given by Eq. (\ref{VS}) of
Appendix B---expressed as%
\begin{equation}
\mathcal{V}_{soft}\supset \eta M_{W}^{3}\phi _{\Omega }+H.c  \label{vs}
\end{equation}%
and since the triplet $\phi _{\Omega }$ transforms nontrivially under $%
\mathbb{A}_{4}\times \boldsymbol{Z}_{3}$ flavor symmetry, the term in Eq. (%
\ref{vs}) breaks explicitly $\mathbb{A}_{4}\times \boldsymbol{Z}_{3}$ down
to $\mathbb{Z}_{2}$. This contribution creates an energy gap among the
degenerate vacua of $\Omega $ (\ref{voc1}) with energetically dominant
vacuum given by the $\vartheta _{1}^{+}$ of (\ref{voc1}) whose region in
space start\ to expand and thus pushing the walls away.

\section{Summary and conclusion}

Motivated by the recent progress in describing neutrino masses and mixing
consistent with current data by using non-Abelian discrete flavor symmetries
as well as their possible implications in cosmology, we have explored in
this paper the neutrino phenomenology in\ a flavored NMSSM based on the
flavor symmetry $G_{\mathrm{f}}=\mathbb{A}_{4}\times \boldsymbol{Z}_{3}$\
and studied the domain wall problem related to the spontaneous breaking of $%
G_{\mathrm{f}}$. To perform this study, we have first studied the extension
of the usual NMSSM with $G_{\mathrm{f}}$ and three right-handed neutrino in
order to engineer appropriate neutrino masses and mixing. Then, we have
analyzed the phenomenological implications of this \emph{FNMSSM} including
the problem of domain walls and provided a solution to this issue by using
the breaking of the flavour symmetry and the effective scalar potential
approach.\newline
In our modeling, the non-Abelian alternating $\mathbb{A}_{4}$ group has been
motivated by its simplest and its straightforwardness to describe the $\mu
-\tau $ symmetry, while the extra $\boldsymbol{Z}_{3}$ discrete symmetry
plays several roles such as avoiding the communication between the charged
lepton and neutrino sectors, solving the usual $\mu $-problem of the MSSM
and eliminating unwanted terms in the superpotential. This\textrm{\ }\emph{%
FNMSSM} is also motivated by the several implications carried by the NMSSM
singlet superfield $S$ where in addition to its role of dynamically
generating the $\mu $-term and its contribution to the Higgs mass, here it
also contributes to the neutrino masses.\newline
Indeed, the first main objective of this paper is to study the neutrino
masses and mixing and their phenomenological implications. First, we showed
that the light neutrino mass matrix of the TBM type arises at the
renormalizable level through the type I seesaw mechanism; this is achieved
by using the $\mathbb{A}_{4}$ representation properties and only one $%
\mathbb{A}_{4}$ flavon triplet $\Omega $\ where we made use of the flavon
singlet $\mathbf{\chi }$ in the Majorana neutrinos mass matrix. Then, in
order to generate neutrino masses and mixing compatible with the
experimental data, a Majorana term arises in the superpotential from the
contribution of the singlet $S$\ plays the role of a deviation from the TBM
pattern. The adopted deviation leads to the well-known trimaximal mixing
predicting nonzero reactor angle $\theta _{13}$ as well as\ a near maximal
atmospheric angle $\theta _{23}$ is permitted. Next, we studied the
phenomenological implications of the neutrino sector where by using the $%
3\sigma $ allowed ranges of the oscillation parameters, we showed through
scatter plots the allowed ranges for the sum of neutrino masses $\sum m_{i}$%
, the effective Majorana neutrino mass $m_{ee}$, the effective\textrm{\ }%
electron neutrino mass $m_{\beta }$, and the Dirac $CPV$ phase $\delta _{CP}$%
. This study is done for both mass hierarchies where we found that numerical
results concerning the ranges of these parameters may be tested in current
and future experiments with a slight preference for the normal hierarchy
over the inverted one. We report in the following Table a summary of the
obtained ranges.%
\begin{equation*}
\begin{tabular}{||l|l|l||l|l|l||}
\hline\hline
\multicolumn{3}{||l||}{\ \ \ \ \ \ \ \ \ \ \ \ \ {\small \ \ \ \ \ \ \ \ \ \
\ \ \ \ \ \ \ \ \ \ Normal Hierarchy}} & \multicolumn{3}{|l||}{\small \ \ \
\ \ \ \ \ \ \ \ \ \ \ \ \ \ \ \ \ Inverted Hierarchy} \\ \hline\hline
$0.007271$ & $\ \ \lesssim m_{ee}(\mathrm{eV})\lesssim $ & $0.04202$ & $%
0.05519$ & $\ \lesssim m_{ee}(\mathrm{eV})\lesssim $ & $0.0641$ \\ \hline
$0.00959$ & $\ \ \lesssim m_{\nu _{e}}(\mathrm{eV)}\lesssim $ & $0.0439$ & $%
0.0554$ & $\ \lesssim m_{\nu _{e}}(\mathrm{eV)}\lesssim $ & $0.0638$ \\
\hline
$0.063$ & $\lesssim \sum m_{i}(\mathrm{eV})\lesssim $ & $0.153$ & $0.14$ & $%
\ \lesssim \sum m_{i}(\mathrm{eV})\lesssim $ & $0.17$ \\ \hline
$0.0043$ & $\lesssim m_{lightest}(\mathrm{eV})\lesssim $ & $0.043$ & $0.0282$
& $\lesssim m_{lightest}(\mathrm{eV})\lesssim $ & $0.041$ \\ \hline
$0.87\pi \left( \text{\&}1.04\pi \right) $ & $\ \ \ \ \ \lesssim \delta
_{CP}\lesssim $ & $0.96\pi \left( \text{\&}1.94\pi \right) $ & $1.12\pi $ & $%
\ \ \ \ \ \lesssim \delta _{CP}\lesssim $ & $1.93\pi $ \\ \hline\hline
\end{tabular}%
\end{equation*}%
On a related note, lepton flavor violating decays generally gives strong
constraints on off-diagonal couplings and scalar masses which might in turns
have an impact on the phenomenology of our model; the discussion of these
processes as well as the dangerous contribution to flavor changing neutral
currents (FCNCs) are left as future work.\newline
The second main goal of this paper concerns the domain wall problem induced
by the spontaneous breaking of the\textrm{\ }$\mathbb{A}_{4}\times
\boldsymbol{Z}_{3}$\textrm{\ }discrete flavour group. To discuss this issue,
we first explained the different breaking patterns encountered in the
charged leptons and neutrino sectors. In the former, the spontaneous
breaking is generated by the flavon triplet $\Phi $ that leads to the
formation of domain walls among the degenerate VEVs of $\Phi $ up to $%
\mathbb{Z}_{2}\times \mathbb{Z}_{2}$ transformations, however, we showed
that these walls do not present any danger since this kind of breaking took
place at the inflationary scale.\textrm{\ }This is not the case for the
breaking pattern in the neutrino sector driven by the flavon triplet $\Omega
$\ and leading to the creation of its own domain walls up to $\mathbb{Z}%
_{2}\rtimes \mathbb{Z}_{3}\times \boldsymbol{Z}_{3}$ charges that are
problematic since this breaking occurs far below the inflationary scale.%
\textrm{\ }To deal with this problem, we added a nonrenormalizable term to
the superpotential---suppressed by two units of the Planck scale---that
breaks explicitly the\textrm{\ }$\mathbb{Z}_{2}\rtimes \mathbb{Z}_{3}\times
\boldsymbol{Z}_{3}$\textrm{\ }discrete symmetry.\textrm{\ }Normally, there
are too many choices for the nonrenormalizable term, and we have chosen one
of them to develop its contribution to the effective potential in details
while we reported the rest of the terms---which lead to a similar
contribution to the one we discussed in details---in Appendix C.

\section{Appendices}

\renewcommand{\thesubsection}{\Alph{subsection}} We give four appendices
where we report some technical details.

\subsection{THE DISCRETE ALTERNATING $\mathbb{A}_{4}$ GROUP}

The alternating group $\mathbb{A}_{4}$ has two noncommuting generators $%
\mathcal{S}$ and $\mathcal{T}$ satisfying the following relation: $\mathcal{S%
}^{2}=\mathcal{T}^{3}=\left( \mathcal{ST}\right) ^{3}=I$. Because of their
noncommutativity---$\mathcal{ST}\neq \mathcal{TS}$---only one of these
generators can be chosen diagonal. In the present paper we have worked in
the Altarelli-Feruglio (AF) basis where the generator $\mathcal{T}$ is
diagonal \cite{R43}. Recall that the discrete group $\mathbb{A}_{4}$ has
four irreducible representations; one irreducible triplet $\mathbf{3}%
_{\left( -1,0\right) }$ and three different singlets $\mathbf{1}_{\left(
1,1\right) },$ $\mathbf{1}_{\left( 1,\omega \right) },$ $\mathbf{1}_{\left(
1,\bar{\omega}\right) }$ distinguished by their representation characters.
In the AF basis, the three-dimensional unitary representation of the
generators $\mathcal{T}$ and $\mathcal{S}$ is given by%
\begin{equation}
\begin{array}{cccc}
\mathbf{3}_{\left( -1,0\right) }: & \mathcal{S}=\frac{1}{3}\left(
\begin{array}{ccc}
-1 & 2 & 2 \\
2 & -1 & 2 \\
2 & 2 & -1%
\end{array}%
\right) & , & \mathcal{T}=\left(
\begin{array}{ccc}
1 & 0 & 0 \\
0 & \bar{\omega} & 0 \\
0 & 0 & \omega%
\end{array}%
\right)%
\end{array}
\label{st}
\end{equation}%
$\allowbreak $ where $\omega =e^{\frac{2\pi i}{3}}$, while on the
one-dimensional representations, these generators are represented by%
\begin{equation}
\begin{array}{ccccc}
\mathbf{1}_{(1,1)}:\mathcal{S}=1~,~\mathcal{T}=1 & ; & \mathbf{1}_{(1,\omega
)}:\mathcal{S}=1~,~\mathcal{T}=\bar{\omega} & ; & \mathbf{1}_{(1,\bar{\omega}%
)}:\mathcal{S}=1~,~\mathcal{T}=\omega%
\end{array}
\label{sr}
\end{equation}%
On the other hand, the $\mathbb{A}_{4}$ representations which are denoted by
their basis characters obey the following tensor product algebra%
\begin{equation}
\begin{tabular}{lll}
$\mathbf{3}_{\left( -1,0\right) }\times \mathbf{3}_{\left( -1,0\right) }$ & $%
=$ & $\mathbf{1}_{\left( 1,1\right) }+\mathbf{1}_{\left( 1,\omega \right) }+%
\mathbf{1}_{\left( 1,\omega ^{2}\right) }+\mathbf{3}_{\left( -1,0\right) }+%
\mathbf{3}_{\left( -1,0\right) }$ \\
$\mathbf{3}_{\left( -1,0\right) }\times \mathbf{1}_{\left( 1,\omega
^{r}\right) }$ & $=$ & $\mathbf{3}_{\left( -1,0\right) }$ \\
$\mathbf{1}_{\left( 1,\omega ^{r}\right) }\times \mathbf{1}_{\left( 1,\omega
^{s}\right) }$ & $=$ & $\mathbf{1}_{\left( 1,\omega ^{r+s}\right) }$%
\end{tabular}
\label{re}
\end{equation}%
where $r$ and $s$ take values $0,1,2$. Here we give useful tools for the
computation of the tensor product of $\mathbb{A}_{4}$ triplets. For the case
of two $\mathbb{A}_{4}$ triplets taken as $\mathbf{a}=(a_{1},a_{2},a_{3})$
and $\mathbf{b}=(b_{1},b_{2},b_{3})$, their tensor product is reducible with
irreducible components given by the first decomposition relation in Eq. (\ref%
{re}). These irreducible components are given in a $\mathcal{T}$-diagonal
basis by%
\begin{equation}
\begin{array}{c}
\left. \left( \mathbf{a\otimes b}\right) \right\vert
_{1}=a_{1}b_{1}+a_{2}b_{3}+a_{3}b_{2}\quad ,\quad \left. \left( \mathbf{%
a\otimes b}\right) \right\vert _{1^{\prime
}}=a_{3}b_{3}+a_{2}b_{1}+a_{1}b_{2} \\
\left. \left( \mathbf{a\otimes b}\right) \right\vert _{1^{\prime \prime
}}=a_{2}b_{2}+a_{1}b_{3}+a_{3}b_{1}%
\end{array}%
\end{equation}%
and%
\begin{equation}
\left. \left( \mathbf{a\otimes b}\right) \right\vert _{3_{S}}=\frac{1}{3}%
\left(
\begin{array}{c}
2a_{1}b_{1}-a_{2}b_{3}-a_{3}b_{2} \\
2a_{3}b_{3}-a_{2}b_{1}-a_{1}b_{2} \\
2a_{2}b_{2}-a_{1}b_{3}-a_{3}b_{1}%
\end{array}%
\right) \quad ,\quad \left. \left( \mathbf{a\otimes b}\right) \right\vert
_{3_{A}}=\frac{1}{2}\left(
\begin{array}{c}
a_{2}b_{3}-a_{3}b_{2} \\
a_{1}b_{2}-a_{2}b_{1} \\
a_{3}b_{1}-a_{1}b_{3}%
\end{array}%
\right) .  \label{tp}
\end{equation}%
Recall that the $\mathbb{A}_{4}$ group has four conjugacy classes $\mathcal{C%
}_{1},\mathcal{C}_{2},\mathcal{C}_{3},\mathcal{C}_{4}$ given by%
\begin{eqnarray*}
\mathcal{C}_{1} &=&\left\{ e\right\} \quad ,\quad \mathcal{C}_{2}=\left\{
\mathcal{S},\mathcal{TST}^{-1},\mathcal{T}^{-1}\mathcal{ST}\right\} \\
\mathcal{C}_{3} &=&\left\{ \mathcal{T},\mathcal{TS},\mathcal{ST},\mathcal{STS%
}\right\} \quad ,\quad \mathcal{C}_{4}=\left\{ \mathcal{T}^{2},\mathcal{ST}%
^{2},\mathcal{T}^{2}\mathcal{S},\mathcal{TST}\right\}
\end{eqnarray*}%
and they are used in building the character Table $\chi _{_{R_{i}}}\left(
\mathcal{C}_{j}\right) =\chi _{ij}$ of the group $\mathbb{A}_{4}$ which
reads as follows%
\begin{equation}
\begin{tabular}{|l|l|l|l|l|}
\hline
$\chi _{ij}\left( \mathbb{A}_{4}\right) $ & $\boldsymbol{R}_{1}$ & $%
\boldsymbol{R}_{1^{\prime }}$ & $\boldsymbol{R}_{1^{\prime \prime }}$ & $%
\boldsymbol{R}_{3}$ \\ \hline
$\mathcal{C}_{1}$ & $1$ & $1$ & $1$ & $3$ \\ \hline
$\mathcal{C}_{2}$ & $1$ & $1$ & $1$ & $-1$ \\ \hline
$\mathcal{C}_{3}$ & $1$ & $\omega $ & $\omega ^{2}$ & $0$ \\ \hline
$\mathcal{C}_{4}$ & $1$ & $\omega ^{2}$ & $\omega $ & $0$ \\ \hline
\end{tabular}%
\ \   \label{c5}
\end{equation}%
To get these numbers, one uses the formula $\chi _{\varrho }\left( g\right)
=tr\left( \varrho _{g}\right) $ with $\varrho $ an irreducible
representation and $g$ a group element. Because of the trace cyclicity
property, this formula is invariant under group conjugation $hgh^{-1}$. For
instance, we have for $\chi _{\boldsymbol{R}_{3}}\left( e\right) =tr\left(
I_{3}\right) =3$, and for $\chi _{\boldsymbol{R}_{3}}\left( \mathcal{S}%
\right) =tr\left( \varrho _{\mathcal{S}}\right) =-1$ while for $\chi _{%
\boldsymbol{R}_{3}}\left( \mathcal{T}\right) =tr\left( \varrho _{\mathcal{T}%
}\right) =0$ due to the identity $1+\omega +\omega ^{2}=0$ for $\omega $
third root of identity.

\subsection{SCALAR SECTOR AND THE VACUUM ALIGNMENT}

In this appendix, we examine the vacuum alignment of the flavon triplets $%
\Phi $ and $\Omega $ used in the charged lepton and neutrino sectors
respectively. As we already mentioned, to achieve the desired vacuum
alignments of $\Phi $ and $\Omega $---the VEV structures that produced the
mass matrices \ref{mm} and \ref{mr1}---we need to avoid any couplings
between these flavons triplets. Indeed, this is one of the reasons for
adding the $\boldsymbol{Z}_{3}$ symmetry under which the coupling $\Phi
\otimes \Omega $ transform as $Q^{2}$. Moreover, it is easy to check from
our scalar superfield content in Table \ref{C} that all trilinear scalar
couplings involving $\Phi \otimes \Omega $ are not invariant under the $%
\boldsymbol{Z}_{3}$ symmetry. The scalar potential of our proposal contains
the usual F, D and soft mass terms of the MSSM but this time with the
singlet $S$ and the flavons $\mathbf{\chi }$, $\Omega $, $\Phi $\ included.
Using Eq. (\ref{sca}), the F-term contribution $\left\vert
F_{scal}\right\vert ^{2}$\ to the scalar potential is given by%
\begin{equation}
\begin{array}{ccc}
\mathcal{V}_{F} & = & \left\vert F_{H_{u}}\right\vert ^{2}+\left\vert
F_{H_{d}}\right\vert ^{2}+\left\vert F_{S}\right\vert ^{2}+\left\vert
F_{\Phi }\right\vert ^{2}+\left\vert F_{\Omega }\right\vert ^{2}+\left\vert
F_{\mathbf{\chi }}\right\vert ^{2}\text{ \ \ \ \ \ \ \ \ } \\
& = & \left\vert \lambda _{1}SH_{d}\right\vert ^{2}+\left\vert \lambda
_{1}SH_{u}\right\vert ^{2}+\left\vert \lambda _{1}H_{u}H_{d}+\lambda
_{2}S^{2}+\lambda _{6}\Omega ^{2}\right\vert ^{2}\text{ \ \ \ } \\
& + & \left\vert \lambda _{4}\Phi ^{2}+\mu _{\Phi }\Phi \right\vert
^{2}+\left\vert \lambda _{3}\Omega ^{2}+\lambda _{6}S\Omega \right\vert
^{2}+\left\vert \lambda _{5}\mathbf{\chi }^{2}+\lambda _{7}\Omega
^{2}\right\vert ^{2}.%
\end{array}%
\end{equation}%
On the other hand, as the singlet $S$ and the flavon superfields are gauge
singlets, they do not have any D-term contributions to the scalar potential.
This restrict the $D$-terms to be exactly the same as in the usual MSSM. The
final contributions to the scalar potential arising from the soft SUSY
breaking terms are given explicitly as%
\begin{eqnarray}
\mathcal{V}_{soft} &=&m_{H_{u}}^{2}\left\vert H_{u}\right\vert
^{2}+m_{H_{d}}^{2}\left\vert H_{d}\right\vert ^{2}+m_{s}^{2}\left\vert
S\right\vert ^{2}+m_{\Phi }^{2}\left\vert \Phi \right\vert ^{2}+m_{\Omega
}^{2}\left\vert \Omega \right\vert ^{2}+m_{\mathbf{\chi }}^{2}\left\vert
\mathbf{\chi }\right\vert ^{2}  \notag \\
&&+[b_{\Phi }\Phi ^{2}+t_{1}SH_{u}H_{d}+t_{2}S^{3}+t_{3}\Omega
^{3}+t_{4}\Phi ^{3}+t_{5}\mathbf{\chi }^{3}+t_{6}S\Omega ^{2}+t_{7}\mathbf{%
\chi }\Omega ^{2}+H.c.]  \label{VS}
\end{eqnarray}%
where $m_{H_{u}}^{2}$, $m_{H_{d}}^{2}$, $m_{s}^{2}$, $m_{\Phi }^{2}$, $%
m_{\Omega }^{2}$, and $m_{\mathbf{\chi }}^{2}$ are soft supersymmetry
breaking masses, $b_{\Phi }$\ is a bilinear soft mass parameter and $t_{i}$
stand for the trilinear coupling between different scalar fields. Now, to
study the vacuum alignment of the flavon triplets $\Phi =\varphi _{_{\Phi
}}+\theta \psi _{_{\Phi }}+\theta ^{2}F_{_{\Phi }}$ and $\Omega =\varphi
_{_{\Omega }}+\theta \psi _{_{\Omega }}+\theta ^{2}F_{_{\Omega }}$, we
denote by $\left\langle \Phi \right\rangle =(\upsilon _{\Phi _{1}},\upsilon
_{\Phi _{2}},\upsilon _{\Phi _{3}})$ and $\left\langle \Omega \right\rangle
=(\upsilon _{\Omega _{1}},\upsilon _{\Omega _{2}},\upsilon _{\Omega _{3}})$
their VEVs solving the minimum conditions
\begin{equation}
\frac{\partial \mathcal{V}}{\partial \varphi _{_{\Phi _{i}}}}=0\qquad
,\qquad \frac{\partial \mathcal{V}}{\partial \varphi _{_{\Omega _{i}}}}=0
\label{mc}
\end{equation}%
with $\mathcal{V}=\mathcal{V}_{F}+\mathcal{V}_{soft}$. We start by
minimizing the scalar potential $\mathcal{V}$ with respect to $\varphi
_{\Phi }$ where\ we obtain the following three equations%
\begin{eqnarray}
\left. \frac{\partial \mathcal{V(}\Phi \mathcal{)}}{\partial \varphi
_{_{\Phi _{1}}}}\right\vert _{\left\langle \Phi _{i}\right\rangle =\upsilon
_{_{\Phi _{i}}}}\text{ \ } &=&\lambda _{4}^{2}\left( \frac{52}{9}\upsilon
_{\Phi _{1}}^{3}+24\upsilon _{\Phi _{1}}\upsilon _{\Phi _{2}}\upsilon _{\Phi
_{3}}+\frac{28}{9}\upsilon _{\Phi _{2}}^{3}+\frac{28}{9}\upsilon _{\Phi
_{3}}^{3}\right) +4\lambda _{4}b_{\Phi }\upsilon _{\Phi _{1}}^{2}  \notag \\
&&-4\lambda _{4}b_{\Phi }\upsilon _{\Phi _{2}}\upsilon _{\Phi
_{3}}+2\upsilon _{\Phi _{1}}\left( \mu _{\Phi }^{2}+m_{\Phi }^{2}+2b_{\Phi
}\right) +2t_{4}\left( \upsilon _{\Phi _{1}}^{2}-\upsilon _{\Phi
_{2}}\upsilon _{\Phi _{3}}\right)
\end{eqnarray}%
\begin{eqnarray}
\left. \frac{\partial \mathcal{V(}\Phi \mathcal{)}}{\partial \varphi
_{_{\Phi _{2}}}}\right\vert _{\left\langle \Phi _{i}\right\rangle =\upsilon
_{_{\Phi _{i}}}} &=&\lambda _{4}^{2}\left( 12\upsilon _{\Phi
_{1}}^{2}\upsilon _{\Phi _{3}}+\frac{28}{3}\upsilon _{\Phi _{1}}\upsilon
_{\Phi _{2}}^{2}+\frac{44}{3}\upsilon _{\Phi _{2}}\upsilon _{\Phi
_{3}}^{2}\right) +4\lambda _{4}b_{\Phi }\upsilon _{\Phi _{2}}^{2}  \notag \\
&&-4\lambda _{4}b_{\Phi }\upsilon _{\Phi _{1}}\upsilon _{\Phi
_{3}}+2\upsilon _{\Phi _{3}}\left( \mu _{\Phi }^{2}+m_{\Phi }^{2}+2b_{\Phi
}\right) +2t_{4}\left( \upsilon _{\Phi _{2}}^{2}-\upsilon _{\Phi
_{1}}\upsilon _{\Phi _{3}}\right)
\end{eqnarray}%
\begin{eqnarray}
\left. \frac{\partial \mathcal{V(}\Phi \mathcal{)}}{\partial \varphi
_{_{\Phi _{3}}}}\right\vert _{\left\langle \Phi _{i}\right\rangle =\upsilon
_{_{\Phi _{i}}}} &=&\lambda _{4}^{2}\left( 12\upsilon _{\Phi
_{1}}^{2}\upsilon _{\Phi _{2}}+\frac{28}{3}\upsilon _{\Phi _{1}}\upsilon
_{\Phi _{3}}^{2}+\frac{44}{3}\upsilon _{\Phi _{2}}^{2}\upsilon _{\Phi
_{3}}\right) +4\lambda _{4}b_{\Phi }\upsilon _{\Phi _{3}}^{2}  \notag \\
&&-4\lambda _{4}b_{\Phi }\upsilon _{\Phi _{1}}\upsilon _{\Phi
_{2}}+2\upsilon _{\Phi _{2}}\left( \mu _{\Phi }^{2}+m_{\Phi }^{2}+2b_{\Phi
}\right) +2t_{4}\left( \upsilon _{\Phi _{3}}^{2}-\upsilon _{\Phi
_{1}}\upsilon _{\Phi _{2}}\right)
\end{eqnarray}%
We can approximate the solution of the minimum conditions of $\mathcal{V}$
for the $\mathbb{A}_{4}$ triplet $\Phi $ through the relations%
\begin{equation}
\begin{tabular}{lllll}
$R_{1}$ & $=$ & $\upsilon _{\Phi _{2}}\frac{\partial \mathcal{V}}{\partial
\upsilon _{_{\Phi _{1}}}}-\upsilon _{\Phi _{1}}\frac{\partial \mathcal{V}}{%
\partial \upsilon _{_{\Phi _{2}}}}$ & $=$ & $0$ \\
$R_{2}$ & $=$ & $\upsilon _{\Phi _{3}}\frac{\partial \mathcal{V}}{\partial
\upsilon _{_{\Phi _{1}}}}-\upsilon _{\Phi _{1}}\frac{\partial \mathcal{V}}{%
\partial \upsilon _{_{\Phi _{3}}}}$ & $=$ & $0$ \\
$R_{3}$ & $=$ & $\upsilon _{\Phi _{3}}\frac{\partial \mathcal{V}}{\partial
\upsilon _{_{\Phi _{2}}}}-\upsilon _{\Phi _{2}}\frac{\partial \mathcal{V}}{%
\partial \upsilon _{_{\Phi _{3}}}}$ & $=$ & $0$%
\end{tabular}%
\end{equation}%
where we find%
\begin{eqnarray}
R_{1} &=&4\lambda _{4}^{2}\left[ \frac{13}{9}\upsilon _{\Phi
_{1}}^{3}\upsilon _{\Phi _{2}}+\upsilon _{\Phi _{3}}\left( 6\upsilon _{\Phi
_{1}}\upsilon _{\Phi _{2}}^{2}-3\upsilon _{\Phi _{1}}^{3}-\frac{11}{3}%
\upsilon _{\Phi _{1}}\upsilon _{\Phi _{2}}\upsilon _{\Phi _{3}}\right) %
\right]  \notag \\
&&+\frac{28}{9}\lambda _{4}^{2}\upsilon _{\Phi _{2}}\left( \upsilon _{\Phi
_{2}}^{3}+\upsilon _{\Phi _{3}}^{3}-\upsilon _{\Phi _{1}}^{2}\upsilon _{\Phi
_{2}}\right) +4\lambda _{4}b_{\Phi }\left( \upsilon _{\Phi _{1}}-\upsilon
_{\Phi _{2}}\right) \upsilon _{\Phi _{1}}\upsilon _{\Phi _{2}}  \notag \\
&&+4\lambda _{4}b_{\Phi }\left( \upsilon _{\Phi _{1}}^{2}-\upsilon _{\Phi
_{2}}^{2}\right) \upsilon _{\Phi _{3}}+2t_{4}\left( \upsilon _{\Phi
_{1}}^{2}-\upsilon _{\Phi _{1}}\upsilon _{\Phi _{2}}\right) \upsilon _{\Phi
_{2}} \\
&&+2\upsilon _{\Phi _{1}}\left( \upsilon _{\Phi _{2}}-\upsilon _{\Phi
_{3}}\right) \left( \mu _{\Phi }^{2}+m_{\Phi }^{2}+2b_{\Phi }\right)
+2t_{4}\left( \upsilon _{\Phi _{1}}^{2}-\upsilon _{\Phi _{2}}^{2}\right)
\upsilon _{\Phi _{3}}  \notag
\end{eqnarray}%
\begin{eqnarray}
R_{2} &=&\frac{4}{9}\lambda _{4}^{2}\left[ 13\upsilon _{\Phi
_{1}}^{3}\upsilon _{\Phi _{3}}-\upsilon _{\Phi _{2}}\left( 27\upsilon _{\Phi
_{1}}^{3}+7\upsilon _{\Phi _{2}}^{2}\upsilon _{\Phi _{3}}-33\upsilon _{\Phi
_{2}}\upsilon _{\Phi _{1}}\upsilon _{\Phi _{3}}\right) \right]  \notag \\
&&+\frac{4}{9}\lambda _{4}^{2}\upsilon _{\Phi _{3}}^{2}\left( 7\upsilon
_{\Phi _{3}}^{2}+54\upsilon _{\Phi _{1}}\upsilon _{\Phi _{2}}-21\upsilon
_{\Phi _{1}}^{2}\right) +4\lambda _{4}b_{\Phi }\left( \upsilon _{\Phi
_{1}}-\upsilon _{\Phi _{3}}\right) \upsilon _{\Phi _{1}}\upsilon _{\Phi _{3}}
\notag \\
&&+4\lambda _{4}b_{\Phi }\left( \upsilon _{\Phi _{1}}^{2}-\upsilon _{\Phi
_{3}}^{2}\right) \upsilon _{\Phi _{2}}+\allowbreak 2t_{4}\left( \upsilon
_{\Phi _{1}}^{2}-\upsilon _{\Phi _{3}}^{2}\right) \upsilon _{\Phi _{2}} \\
&&-2\upsilon _{\Phi _{1}}\left( \upsilon _{\Phi _{2}}-\upsilon _{\Phi
_{3}}\right) \left( \mu _{\Phi }^{2}+m_{\Phi }^{2}+2b_{\Phi }\right)
+\allowbreak 2t_{4}\left( \upsilon _{\Phi _{1}}^{2}-\upsilon _{\Phi
_{1}}\upsilon _{\Phi _{3}}\right) \upsilon _{\Phi _{3}}  \notag
\end{eqnarray}%
\begin{eqnarray}
R_{3} &=&\frac{4}{3}\lambda _{4}^{2}\left[ 9\upsilon _{\Phi _{1}}^{2}\left(
\upsilon _{\Phi _{3}}^{2}-\upsilon _{\Phi _{2}}^{2}\right) -\upsilon _{\Phi
_{2}}^{2}\left( 11\upsilon _{\Phi _{2}}\upsilon _{\Phi _{3}}+7\upsilon
_{\Phi _{1}}\upsilon _{\Phi _{3}}\right) \right]  \notag \\
&&+\frac{4}{3}\lambda _{4}^{2}\upsilon _{\Phi _{3}}^{2}\left( 11\upsilon
_{\Phi _{2}}\upsilon _{\Phi _{3}}-7\upsilon _{\Phi _{1}}\upsilon _{\Phi
_{2}}\right) +4\lambda _{4}b_{\Phi }\left( \upsilon _{\Phi _{1}}\upsilon
_{\Phi _{2}}-\upsilon _{\Phi _{3}}^{2}\right) \upsilon _{\Phi _{2}}  \notag
\\
&&+4\lambda _{4}b_{\Phi }\left( \upsilon _{\Phi _{2}}^{2}-\upsilon _{\Phi
_{1}}\upsilon _{\Phi _{3}}\right) \upsilon _{\Phi _{3}}+2t_{4}\left(
\upsilon _{\Phi _{1}}\upsilon _{\Phi _{2}}-\upsilon _{\Phi _{3}}^{2}\right)
\upsilon _{\Phi _{2}} \\
&&-2\left( \upsilon _{\Phi _{2}}^{2}-\upsilon _{\Phi _{3}}^{2}\right) \left(
\mu _{\Phi }^{2}+m_{\Phi }^{2}+2b_{\Phi }\right) +2t_{4}\left( \upsilon
_{\Phi _{2}}^{2}-\upsilon _{\Phi _{1}}\upsilon _{\Phi _{3}}\right) \upsilon
_{\Phi _{3}}  \notag
\end{eqnarray}%
Clearly, the only solution for these three equations is the one we have
chosen to produce the charged leptons masses (\ref{lm}), namely $\upsilon
_{\Phi _{1}}\neq 0$ and $\upsilon _{\Phi _{2}}=\upsilon _{\Phi _{3}}=0$.
Analogously, we perform the same calculations for the minimum conditions
coming from the triplet $\Omega $; we have%
\begin{eqnarray}
\left. \frac{\partial \mathcal{V(}\Omega \mathcal{)}}{\partial \varphi
_{_{\Omega _{1}}}}\right\vert _{\left\langle \Omega _{i}\right\rangle
=\upsilon _{\Omega _{i}}} &=&(4\lambda _{2}\lambda _{6}\upsilon
_{S}^{2}+4\lambda _{1}\lambda _{6}^{2}\upsilon _{d}\upsilon _{u}+2\lambda
_{6}^{2}\upsilon _{S}^{2}+4\lambda _{5}\lambda _{7}\upsilon _{\varphi
}^{2}+2m_{\Omega }^{2}+2t_{6}\upsilon _{S}+2t_{7}\upsilon _{\varphi
})\upsilon _{\Omega _{1}}  \notag \\
&&+(\lambda _{6}^{2}+\lambda _{3}^{2}+\lambda _{7}^{2})(\frac{52}{9}\upsilon
_{\Omega _{1}}^{3}+24\upsilon _{\Omega _{1}}\upsilon _{\Omega _{2}}\upsilon
_{\Omega _{3}}+\frac{28}{9}\upsilon _{\Omega _{2}}^{3}+\frac{28}{9}\upsilon
_{\Omega _{3}}^{3})  \notag \\
&&+(2\lambda _{3}\lambda _{6}\upsilon _{S}+t_{3})(2\upsilon _{\Omega
_{1}}^{2}-2\upsilon _{\Omega _{2}}\upsilon _{\Omega _{3}})
\end{eqnarray}%
\begin{eqnarray}
\left. \frac{\partial \mathcal{V(}\Omega \mathcal{)}}{\partial \varphi
_{_{\Omega _{2}}}}\right\vert _{\left\langle \Omega _{i}\right\rangle
=\upsilon _{\Omega _{i}}} &=&2(2\lambda _{2}\lambda _{6}\upsilon
_{S}^{2}+2\lambda _{1}\lambda _{6}^{2}\upsilon _{d}\upsilon _{u}+\lambda
_{6}^{2}\upsilon _{S}^{2}+2\lambda _{5}\lambda _{7}\upsilon _{\varphi
}^{2}+m_{\Omega }^{2}+t_{6}\upsilon _{S}+t_{7}\upsilon _{\varphi })\upsilon
_{\Omega _{3}}  \notag \\
&&+(\lambda _{6}^{2}+\lambda _{3}^{2}+\lambda _{7}^{2})(12\upsilon _{\Omega
_{1}}^{2}\upsilon _{\Omega _{3}}+\frac{28}{3}\upsilon _{\Omega _{1}}\upsilon
_{\Omega 2}^{2}+\frac{44}{3}\upsilon _{\Omega _{2}}\upsilon _{\Omega
_{3}}^{2})  \notag \\
&&+(2\lambda _{3}\lambda _{6}\upsilon _{S}+t_{3})(2\upsilon _{\Omega
_{2}}^{2}-2\upsilon _{\Omega _{1}}\upsilon _{\Omega _{3}})
\end{eqnarray}%
\begin{eqnarray}
\left. \frac{\partial \mathcal{V(}\Omega \mathcal{)}}{\partial \varphi
_{_{\Omega _{3}}}}\right\vert _{\left\langle \Omega _{i}\right\rangle
=\upsilon _{\Omega _{i}}} &=&2(2\lambda _{2}\lambda _{6}\upsilon
_{S}^{2}+2\lambda _{1}\lambda _{6}^{2}\upsilon _{d}\upsilon _{u}+\lambda
_{6}^{2}\upsilon _{S}^{2}+2\lambda _{5}\lambda _{7}\upsilon _{\varphi
}^{2}+m_{\Omega }^{2}+t_{6}\upsilon _{S}+t_{7}\upsilon _{\varphi })\upsilon
_{\Omega _{2}}  \notag \\
&&+(\lambda _{6}^{2}+\lambda _{3}^{2}+\lambda _{7}^{2})(12\upsilon _{\Omega
_{1}}^{2}\upsilon _{\Omega _{2}}+\frac{28}{3}\upsilon _{\Omega _{1}}\upsilon
_{\Omega 3}^{2}+\frac{44}{3}\upsilon _{\Omega _{2}}^{2}\upsilon _{\Omega
_{3}})  \notag \\
&&+(2\lambda _{3}\lambda _{6}\upsilon _{S}+t_{3})(2\upsilon _{\Omega
_{3}}^{2}-2\upsilon _{\Omega _{1}}\upsilon _{\Omega _{2}})
\end{eqnarray}%
Similarly to the case of the flavon $\Phi $, we can approach the solution of
the minimum conditions of $\mathcal{V}$ for the $\mathbb{A}_{4}$ triplet $%
\Omega $ through the following equations%
\begin{eqnarray}
\left( \upsilon _{\Omega _{3}}\frac{\partial \mathcal{V(}\Omega \mathcal{)}}{%
\partial \varphi _{_{\Omega _{1}}}}-\upsilon _{\Omega _{1}}\frac{\partial
\mathcal{V(}\Omega \mathcal{)}}{\partial \varphi _{_{\Omega _{3}}}}\right)
_{\left\langle \Omega _{i}\right\rangle =\upsilon _{\Omega _{i}}} &=&0
\notag \\
\left( \upsilon _{\Omega _{2}}\frac{\partial \mathcal{V(}\Omega \mathcal{)}}{%
\partial \varphi _{_{\Omega _{1}}}}-\upsilon _{\Omega _{1}}\frac{\partial
\mathcal{V(}\Omega \mathcal{)}}{\partial \varphi _{_{\Omega _{2}}}}\right)
_{\left\langle \Omega _{i}\right\rangle =\upsilon _{\Omega _{i}}} &=&0
\label{va} \\
\left( \upsilon _{\Omega _{3}}\frac{\partial \mathcal{V(}\Omega \mathcal{)}}{%
\partial \varphi _{_{\Omega _{2}}}}-\upsilon _{\Omega _{2}}\frac{\partial
\mathcal{V(}\Omega \mathcal{)}}{\partial \varphi _{_{\Omega _{3}}}}\right)
_{\left\langle \Omega _{i}\right\rangle =\upsilon _{\Omega _{i}}} &=&0
\notag
\end{eqnarray}%
leading to%
\begin{equation}
\begin{tabular}{lll}
$\mathbf{T}_{1}(\upsilon _{\Omega _{3}}-\upsilon _{\Omega _{2}})+\mathbf{T}%
_{2}(\upsilon _{\Omega _{1}}-\upsilon _{\Omega _{3}})+\mathbf{T}_{3}$ & $=$
& $0$ \\
$\mathbf{T}_{1}(\upsilon _{\Omega _{2}}-\upsilon _{\Omega _{3}})+\mathbf{T}%
_{4}(\upsilon _{\Omega _{1}}-\upsilon _{\Omega _{2}})+\mathbf{T}_{5}$ & $=$
& $0$ \\
$\mathbf{T}_{6}(\upsilon _{\Omega _{3}}^{2}-\upsilon _{\Omega _{2}}^{2})+%
\mathbf{T}_{7}(\upsilon _{\Omega _{2}}-\upsilon _{\Omega _{3}})$ & $=$ & $0$%
\end{tabular}%
\end{equation}%
with%
\begin{equation}
\begin{tabular}{lll}
$\mathbf{T}_{1}$ & $=$ & $2\upsilon _{\Omega _{1}}(2\lambda _{2}\lambda
_{6}\upsilon _{S}^{2}+2\lambda _{1}\lambda _{6}^{2}\upsilon _{d}\upsilon
_{u}+\lambda _{6}^{2}\upsilon _{S}^{2}+2\lambda _{5}\lambda _{7}\upsilon
_{\varphi }^{2}+m_{\Omega }^{2}+t_{6}\upsilon _{S}+t_{7}\upsilon _{\varphi
}) $ \\
$\mathbf{T}_{2}$ & $=$ & $2(2\lambda _{3}\lambda _{6}\upsilon
_{S}+t_{3})(\upsilon _{\Omega _{1}}\upsilon _{\Omega _{3}}+\upsilon _{\Omega
_{2}}(\upsilon _{\Omega _{1}}+\upsilon _{\Omega _{3}}))$ \\
$\mathbf{T}_{3}$ & $=$ & $\frac{4}{9}(\lambda _{6}^{2}+\lambda
_{3}^{2}+\lambda _{7}^{2})(7\upsilon _{\Omega _{3}}^{4}-21\upsilon _{\Omega
_{1}}^{2}\upsilon _{\Omega _{3}}^{2}-27\upsilon _{\Omega _{1}}^{3}\upsilon
_{\Omega _{2}}+13\upsilon _{\Omega _{1}}^{3}\upsilon _{\Omega _{3}}$ \\
&  & $+7\upsilon _{\Omega _{2}}^{3}\upsilon _{\Omega _{3}}-33\upsilon
_{\Omega _{2}}^{2}\upsilon _{\Omega _{1}}\upsilon _{\Omega _{3}}+54\upsilon
_{\Omega _{3}}^{2}\upsilon _{\Omega _{1}}\upsilon _{\Omega _{2}})$ \\
$\mathbf{T}_{4}$ & $=$ & $2(2\lambda _{3}\lambda _{6}\upsilon
_{S}+t_{3})(\upsilon _{\Omega _{1}}\upsilon _{\Omega _{2}}+\upsilon _{\Omega
_{3}}(\upsilon _{\Omega _{1}}+\upsilon _{\Omega _{2}}))$%
\end{tabular}%
\end{equation}%
and%
\begin{equation}
\begin{tabular}{lll}
$\mathbf{T}_{5}$ & $=$ & $\frac{4}{9}(\lambda _{6}^{2}+\lambda
_{3}^{2}+\lambda _{7}^{2})(7\upsilon _{\Omega _{2}}^{4}-21\upsilon _{\Omega
_{1}}^{2}\upsilon _{\Omega _{2}}^{2}+13\upsilon _{\Omega _{1}}^{3}\upsilon
_{\Omega _{2}}-27\upsilon _{\Omega _{1}}^{3}\upsilon _{\Omega _{3}}$ \\
&  & $+7\upsilon _{\Omega _{3}}^{3}\upsilon _{\Omega _{2}}-33\upsilon
_{\Omega _{3}}^{2}\upsilon _{\Omega _{1}}\upsilon _{\Omega _{2}}+54\upsilon
_{\Omega _{2}}^{2}\upsilon _{\Omega _{1}}\upsilon _{\Omega _{3}})$ \\
$\mathbf{T}_{6}$ & $=$ & $2[2\lambda _{2}\lambda _{6}\upsilon
_{S}^{2}+2\lambda _{1}\lambda _{6}^{2}\upsilon _{d}\upsilon _{u}+\lambda
_{6}^{2}\upsilon _{S}^{2}+2\lambda _{5}\lambda _{7}\upsilon _{\varphi
}^{2}+m_{\Omega }^{2}+t_{6}\upsilon _{S}+t_{7}\upsilon _{\varphi }$ \\
&  & $+2\lambda _{3}\lambda _{6}\upsilon _{S}\upsilon _{\Omega
_{1}}+t_{3}\upsilon _{\Omega _{1}}+(\lambda _{6}^{2}+\lambda
_{3}^{2}+\lambda _{7}^{2})(12\upsilon _{\Omega _{1}}^{2}+\frac{44}{3}%
\upsilon _{\Omega _{2}}\upsilon _{\Omega _{3}})]$ \\
$\mathbf{T}_{7}$ & $=$ & $\upsilon _{\Omega _{2}}\upsilon _{\Omega
_{3}}[4\lambda _{3}\lambda _{6}\upsilon _{S}+2t_{3}+\frac{28}{3}(\lambda
_{6}^{2}+\lambda _{3}^{2}+\lambda _{7}^{2})\upsilon _{\Omega _{1}}].$%
\end{tabular}%
\end{equation}%
It is clear that the solution for Eq. (\ref{va}) is indeed the VEV structure
we have chosen to produce the right-handed Majorana neutrinos mass matrix (%
\ref{mr1}), namely $\upsilon _{\Omega _{1}}=\upsilon _{\Omega _{2}}=\upsilon
_{\Omega _{3}}=\upsilon _{\Omega }$.

\subsection{MORE ON FLAVOR SYMMETRY BREAKING OPERATORS}

In this appendix, we provide a simple list of the five dimensional operators
$\mathcal{O}_{5}$ that can be used to break explicitly the discrete $G_{%
\mathrm{f}}=\mathbb{A}_{4}\times \boldsymbol{Z}_{3}$ flavor group of our
model. As before, we focus on the flavor invariant trilinear couplings $%
\mathcal{W}_{R}=\sum \lambda _{abc}\Upsilon _{a}\Upsilon _{b}\Upsilon _{c}$
that includes the superfields relevant to the neutrino sector namely
\begin{equation}
\begin{tabular}{lllllllllllll}
$\Upsilon _{a}\Upsilon _{b}\Upsilon _{c}$ & $\text{\ \ }\sim \text{ \ \ }$ &
$S^{3}$ & $,$ & $\Omega ^{3}$ & $\,,$ & $\mathbf{\chi }^{3}$ & $,$ & $%
S\Omega ^{2}$ & $,$ & $\mathbf{\chi }\Omega ^{2}$ & $,$ & $SH_{u}H_{d}$%
\end{tabular}%
\end{equation}%
and we would like to look for a chiral superpotential $\mathcal{W}%
_{NR}^{\prime }$ made of gauge invariant five dimensional operators type $%
\frac{1}{M_{Pl}^{2}}\mathcal{O}_{5}$ that breaks explicitly the full or a
part of the flavor symmetry $G_{\mathrm{f}}$. Clearly one can write down a
big list of such gauge symmetric operators; a remarkable family of gauge
invariant five dimensional operators $\mathcal{O}_{5}$ that breaks the
flavor symmetry $G_{\mathrm{f}}$ is obtained by thinking of $\mathcal{O}_{5}$
as follows%
\begin{equation}
\mathcal{O}_{5}\text{ \ \ }\sim \text{ \ \ }\mathcal{O}_{2}\times \Upsilon
_{a}\Upsilon _{b}\Upsilon _{c}
\end{equation}%
with $\mathcal{O}_{2}$ breaking $G_{\mathrm{f}}$ partially or completely.
Moreover, seen that $G_{\mathrm{f}}$ is the product of two factors, the
breaking may be carried either by one of the two factors; that is $\mathbb{A}%
_{4}$ or by $\boldsymbol{Z}_{3}$; or by both of them\footnote{%
For the contributions that arise from the combination between two different
fields, namely $S$, $\Omega $ and $\chi $, they can be eliminated by the
same R-symmetries usually used to avoid the existence of terms of order four
\cite{R71,A35}.}. For example, candidates for $\mathcal{O}_{2}$ breaking $%
\boldsymbol{Z}_{3}$ are directly read from the Table \ref{C}; they are given
by the following quadratic chiral superfield monomials
\begin{equation}
\mathcal{O}_{2}:\text{ \ \ \ \ }S^{2}\quad ,\,\quad \Omega ^{2}\quad
,\,\quad \mathbf{\chi }^{2}\quad ,\,\quad H_{u}H_{d}
\end{equation}%
From these four quadratic candidates, only $S^{2}$ breaks both $\mathbb{A}%
_{4}$ and $\boldsymbol{Z}_{3}$ symmetries. Hence, a simple candidate form
for the superpotential $\mathcal{W}_{NR}^{\prime }$ reads therefore as
follows%
\begin{equation}
\mathcal{W}_{NR}^{\prime }=\frac{1}{M_{Pl}^{2}}\left[ h_{i}H_{u}H_{d}+\text{
}h_{i}^{\prime }S^{2}+l_{i}\Omega ^{2}+l_{i}^{\prime }\mathbf{\chi }^{2}%
\right] \times \mathcal{W}_{R}
\end{equation}%
where $\mathcal{W}_{R}=\sum \lambda _{abc}\Upsilon _{a}\Upsilon _{b}\Upsilon
_{c}$ and where $l_{3}\equiv \lambda _{3}^{\prime }$ according to Eq. (\ref%
{op}). From this $\mathcal{W}_{NR}^{\prime }$, one can compute the
contribution to the effective scalar potential $\mathcal{V}$ that breaks the
flavor symmetry. In what follows, we illustrate the calculation leading to
the scalar potential for the very particular and simple example where $%
\mathcal{W}_{NR}^{\prime }$ is restricted to the operator of Eq. (\ref{op})
namely%
\begin{equation}
\mathcal{W}_{NR}=\frac{\lambda _{3}^{\prime }}{M_{Pl}^{2}}\left. \left(
\Omega ^{5}\right) \right\vert _{\left( 1,\omega \right) }
\end{equation}%
By using the Feynman rules for supergraphs of chiral superfields; in
particular for the\textrm{\ chiral} $S=S_{s}+\theta S_{\psi }+\theta
^{2}S_{F}$, the contribution of the supergraph in Fig. \ref{f20} to the
effective action is given by \cite{A34,R74,A38,A39}%
\begin{equation}
\begin{array}{ccc}
\delta S & \approx  & \frac{\lambda _{3}^{2}\lambda _{3}^{\prime }}{%
M_{Pl}^{2}}\dint d^{4}x_{1}d^{4}x_{2}d^{4}x_{3}d^{4}\theta _{1}d^{4}\theta
_{2}d^{4}\theta _{3}\Omega (x_{1},\theta _{1})\frac{1}{\mathbf{\bar{\phi}}(%
\bar{\theta}_{1})}e^{\frac{2K_{(12)}}{3}}e^{\frac{K_{(23)}}{3}}e^{\frac{%
2K_{(13)}}{3}} \\
&  & \times \left( \frac{\overline{D}_{1}^{2}}{4\square _{1}}\delta
_{12}\right) \left( \frac{D_{2}^{2}\overline{D}_{2}^{2}}{16\square _{2}}%
\delta _{12}\right) \left( \frac{D_{2}^{2}}{4\square _{2}}\delta
_{23}\right) \left( \frac{\overline{D}_{3}^{2}}{4\square _{3}}\delta
_{13}\right) \left( \frac{\overline{D}_{3}^{2}D_{3}^{2}}{16\square _{3}}%
\delta _{13}\right)
\end{array}
\label{in}
\end{equation}%
where
\begin{equation}
\delta _{ij}=\delta ^{4}(x_{i}-x_{j})\delta ^{4}(\theta _{i}-\theta _{j}),
\end{equation}%
the compensator superfield is given by $\mathbf{\bar{\phi}}(\bar{\theta}%
_{1})=1+\bar{\theta}_{1}^{2}\frac{M_{s}^{2}}{M_{Pl}}$ and $K_{(ij)}=K(\theta
_{i},\bar{\theta}_{j})$ which is the minimal Kahler potential\emph{\ }of the
model defined as $K=\Sigma \bar{\Sigma}$ where $\Sigma $ stands for chiral
superfields of the model. One can evaluate this expression by integrating by
parts and use the factors of $\delta ^{4}(\theta _{i}-\theta _{j})$\ to
eliminate $\theta $\ integrals. Moreover, by using the algebra of the
supersymmetric covariant derivatives we can remove $D^{2}\overline{D}^{2}$
due to properties \cite{A34,R70,R74,A33,A39} like%
\begin{equation}
D^{2}\overline{D}^{2}D^{2}=16\square D^{2}\quad ;\quad \overline{D}^{2}D^{2}%
\overline{D}^{2}=16\square \overline{D}^{2}
\end{equation}%
and%
\begin{equation}
\dint d^{2}\theta _{2}\delta ^{4}(\theta _{2}-\theta _{1})D^{2}\overline{D}%
^{2}\delta ^{4}(\theta _{2}-\theta _{1})=16
\end{equation}%
In doing so, the integral (\ref{in}) gets reduced to a single integral over $%
\theta _{1}$ of the form%
\begin{equation}
\delta S\approx \frac{\lambda _{3}^{2}\lambda _{3}^{\prime }}{M_{Pl}^{2}}%
\dint d^{4}x_{1}d^{4}\theta _{1}\Omega (x_{1},\theta _{1})\frac{1}{\mathbf{%
\bar{\phi}}(\bar{\theta}_{1})}(\frac{\bar{D}^{2}}{4})e^{\frac{5K_{(11)}}{3}%
}\times \mathcal{I}_{3}
\end{equation}%
with%
\begin{equation}
\mathcal{I}_{3}=\dint \frac{d^{4}k_{1}}{(2\pi )^{4}}\frac{d^{4}k_{2}}{(2\pi
)^{4}}\frac{d^{4}k_{3}}{(2\pi )^{4}}\frac{1}{%
k_{1}^{2}k_{2}^{2}k_{3}^{2}(k_{1}-k_{2})^{2}(k_{1}-k_{3})^{2}}\sim \mathcal{O%
}(\frac{M_{P}^{2}}{(16\pi ^{2})^{3}})
\end{equation}%
Using the following relations \cite{A34,R74,A38,A39}%
\begin{equation}
\exp \left( \frac{2K}{3M_{Pl}^{2}}\right) \approx 1+\theta ^{2}\frac{%
M_{s}^{2}}{M_{Pl}}+\bar{\theta}^{2}\frac{M_{s}^{2}}{M_{Pl}}+\theta ^{2}\bar{%
\theta}^{2}\frac{M_{s}^{4}}{M_{Pl}^{2}}~\text{and }\mathbf{\phi }\approx
1+\theta ^{2}\frac{M_{s}^{2}}{M_{Pl}}
\end{equation}%
which are respectively the classical VEVs for the Kahler potential and the
nonpropagating compensator superfield, we get an additional contribution to
the effective potential; it is given by%
\begin{equation}
\delta \mathcal{V}_{eff}=\frac{\lambda _{3}^{2}\lambda _{3}^{\prime
}M_{W}^{3}}{(16\pi ^{2})^{3}}(\phi _{\Omega }+\bar{\phi}_{\Omega })+\frac{%
\lambda _{2}^{2}\lambda _{5}^{\prime }M_{W}^{2}}{(16\pi ^{2})^{3}}(\digamma
_{\Omega }+\bar{\digamma}_{\Omega })
\end{equation}%
where the linear term proportional to $M_{W}^{3}$\ is the contribution used
to break explicitly the full flavor symmetry $G_{\mathrm{f}}=\mathbb{A}%
_{4}\times \mathbf{Z}_{3}$ of our model.

\subsection{DOMAIN WALLS FROM BREAKING $\mathbb{A}_{4}$ BY $\left\langle
S\right\rangle $}

\begin{figure}[tbph]
\begin{center}
\includegraphics[scale=0.5]{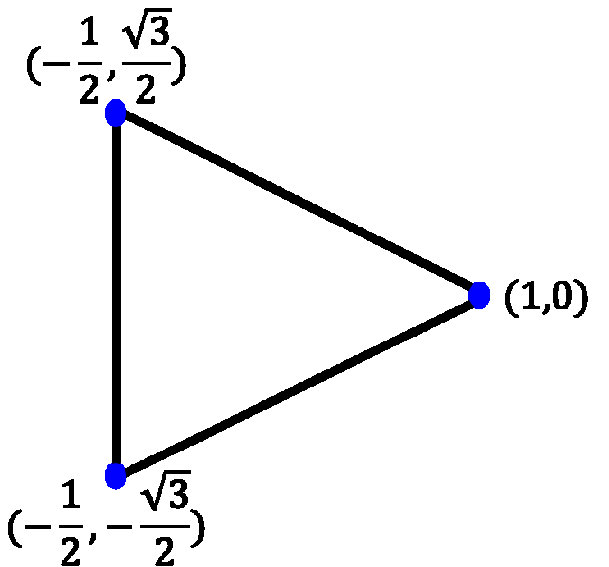}
\end{center}
\par
\vspace{-0.5cm}
\caption{A quiver diagram representing domain walls using characters of the
generator of $\mathbb{Z}_{3}$. \textrm{The vertices correspond to the three
vacua, while the domain walls are represented by line segments (edges)
interpolating between the vacua.}}
\label{w5}
\end{figure}

In this appendix,\textrm{\ }we describe the case where the breaking pattern
in the neutrino sector is driven by the VEV\ of the NMSSM singlet $S$. Since
$S$ is invariant under the actions of $\mathcal{S}$, $\mathcal{S}^{\prime }$%
, and $\mathcal{S}^{\prime \prime }$, $\left\langle S\right\rangle $ breaks $%
G_{\mathrm{f}}$ down to its Klein subgroup $\mathbb{V}_{4}$ with the broken
part is given by $\mathbb{Z}_{3}$. The analogue of the $\vartheta _{\alpha
}^{\pm }$- vacua (\ref{an}) are now given by three degenerate $\left\langle
S\right\rangle _{i}$ rotated by $\mathbb{Z}_{3}$ that we denote like%
\begin{equation}
\zeta _{1}=\upsilon _{S}X_{4}\quad ,\quad \zeta _{2}=\omega \upsilon
_{S}X_{4}\quad ,\quad \zeta _{3}=\bar{\omega}\upsilon _{S}X_{4}  \label{ksi}
\end{equation}%
where $\omega =e^{\frac{2i\pi }{3}}$ and where $X_{4}$\ is as in (\ref{fl}).
These vacua can be obtained by starting from $\zeta _{1}$ and acting by $%
\mathcal{T}$ generator of $\mathbb{Z}_{3}$. They obey the property
\begin{equation}
\sum_{\alpha =1}^{3}\zeta _{\alpha }=0
\end{equation}%
and so can be represented by a 2d triangle. Indeed, by using the real basis
vector (\ref{r8}); we can express the $X_{4}^{T}$ and $i\otimes X_{4}^{T}$
directions respectively like $\left( 1,0\right) $ and $\left( 0,1\right) $;
then substituting back into the above vacua, we obtain after setting $\vec{%
\zeta}_{i}=\left( \func{Re}\zeta _{i},\func{Im}\zeta _{i}\right) $ the
following real VEVs
\begin{equation}
\begin{tabular}{lllll}
$\vec{\zeta}_{1}=\left(
\begin{array}{c}
\upsilon _{S} \\
0%
\end{array}%
\right) $ & $\quad ,\quad $ & $\vec{\zeta}_{2}=\left(
\begin{array}{c}
-\frac{\upsilon _{S}}{2} \\
\frac{\sqrt{3}}{2}\upsilon _{S}%
\end{array}%
\right) $ & $\quad ,\quad $ & $\vec{\zeta}_{3}=\left(
\begin{array}{c}
-\frac{\upsilon _{S}}{2} \\
-\frac{\sqrt{3}}{2}\upsilon _{S}%
\end{array}%
\right) $%
\end{tabular}%
\end{equation}%
defining the vertices of a triangle quiver diagram given by Fig. \ref{w5}
representing as well the DWs extending between the three vertices.
The nonrenormalizable operators $\mathcal{W}_{NR}$ of order five operators
that can break explicitly the full $\mathbb{A}_{4}\times \boldsymbol{Z}_{3}$
flavor symmetry down to $\mathbb{V}_{4}$ is given by the following operator%
\begin{equation}
\mathcal{W}_{NR}=\frac{\lambda _{5}^{\prime }}{M_{Pl}^{2}}S^{5}.
\end{equation}%
\begin{equation*}
\end{equation*}

\end{document}